\documentclass[12pt,english,nofootinbib]{revtex4}
\usepackage[T1]{fontenc}
\pdfoutput=1
\usepackage[latin9]{inputenc}
\usepackage{amsmath}
\usepackage{graphicx}
\usepackage{amssymb}
\usepackage{esint}
\usepackage{epstopdf}
\usepackage{color}
\makeatletter
\@ifundefined{textcolor}{}
{%
 \definecolor{BLACK}{gray}{0}
 \definecolor{WHITE}{gray}{1}
 \definecolor{RED}{rgb}{1,0,0}
 \definecolor{GREEN}{rgb}{0,1,0}

 \definecolor{BLUE}{rgb}{0,0,1}
 \definecolor{CYAN}{cmyk}{1,0,0,0}
 \definecolor{MAGENTA}{cmyk}{0,1,0,0}
 \definecolor{YELLOW}{cmyk}{0,0,1,0}
 }
 \definecolor{RED}{rgb}{1,0,0}
\usepackage{amsfonts}\usepackage{amstext}\usepackage{amsbsy}\usepackage{amscd}
\usepackage{amsopn}

\makeatother

\usepackage{babel}

\makeatother

\usepackage{babel}
\usepackage{setspace}

\makeatother

\usepackage{babel}

\begin{document}
\singlespacing

\title{Classes of fast and specific search mechanisms for proteins on DNA}

\author{M. Sheinman$^{1,2}$, O. B\'enichou$^{3}$, Y. Kafri$^{2}$, and R. Voituriez$^{3}$}
\affiliation{$1$ Department of Physics and Astronomy, Vrije
Universiteit, Amsterdam, The Netherlands} \affiliation{$2$
Department of Physics, Technion, Haifa 32000, Israel.}
\affiliation{$3$ UMR 7600, Universit\'e Pierre et Marie
Curie/CNRS, 4 Place Jussieu, 75255 Paris Cedex 05 France.}

\date{\today}

\begin{abstract}
Problems of search and recognition appear over different scales in
biological systems. In this review we focus on the challenges
posed by interactions between proteins, in particular transcription factors, and DNA
and possible mechanisms which allow for a fast and selective target location.
Initially we argue that DNA-binding proteins can be classified, broadly, into three distinct classes which we illustrate
 using experimental data.
Each class calls for a different search process
 and we discuss the possible application of different
search mechanisms proposed over the years to each class. The main thrust of this review is a new mechanism which
is based on barrier discrimination. We introduce the model and analyze in detail its consequences.
It is shown that this mechanism applies to all classes of transcription factors and can lead to a fast
and specific search. Moreover, it is shown that the mechanism has interesting transient features which
allow for stability at the target despite rapid binding and unbinding of the transcription factor from the target.

\tableofcontents{}
\end{abstract}
\maketitle

\section{Introduction\label{Introduction}}

Many biochemical processes require both an appropriate speed and
a high specificity for proper biological functions to occur -- a fast desirable process should not be accompanied by a
significant acceleration of undesirable ones. With typical
energy scales of a few $k_{B}T$, where $k_{B}$ is the Boltzmann
constant and $T$ is the temperature, evolution has devised many
efficient mechanisms which overcome the noisy environment and the
speed requirements. These range from mechanisms which rely on the
consumption of chemical energy, such as kinetic proofreading
\cite{Hopfield74}, to cooperativity, such as in the specific regulation of the
hemoglobin oxygen concentration \cite{Hill36,Fersht}.
Unraveling these mechanisms is an important step towards
understanding how cells function.

Being based on biopolymers, specificity in biological systems
implies that two (or more) well defined subsequences of two given polymers attach to
each other, but not to other subsequences of the same polymers
or to  other polymers. The two polymers can be proteins
(for example, in enzymes \cite{Fersht}), RNA molecules (for
example, in ribosomal action \cite{Schimmel98,Cech2000}),
a single-stranded and a double stranded DNA (for example, in
the homologous recombination \cite{Kupiec2008}) or a transcription factor (TF) and  a DNA molecule. The last example
highlights the challenges which a biological system faces.

Consider, for example, a prokaryotic cell (throughout the review we focus on these simpler systems).
Its typical DNA length is $N\simeq10^{7}$ basepairs. In a particularly
simple case a TF has to bind to a specific subsequence (target) of
a length of about $10-20$ basepairs on the DNA. The typical binding
energy between a protein and the DNA subsequence is of the order of
tens of $k_{B}T$, about one $k_{B}T$ per base-pair. Without using chemical
energy (which is true for almost all transcription factors) this
gives rise to a classical conflict between entropy and energy which puts a hamper on the
stability of the TF at the target\footnote{Chemical energy could lead to directed motion. This scenario is discussed in \cite{LBMV2008}).
}. Specifically, the entropy associated with the protein bound to non-target DNA is $k_{B}\ln N\simeq16k_{B}$
and therefore its contribution to the free energy is of the same order as the
binding energy. Unless the TF is designed to have a binding energy
at the target that is much lower than to the rest of the sequence the probability of finding it on the target
site will be very low. Of course, the copy number of a TF, which  in a
cell typically ranges  from about tens
to thousands \cite{Guptasarma95,RMC98,AbundanceDatabase,AbundanceDatabase2}, can increase the occupation probability of
the target site to a desired level (see, for instance, \cite{CspA}). This, however, comes at a cost
of producing many proteins and possibly activating or repressing
unwanted genes and loosing specificity, meaning that the TF is
likely to occupy nonspecific sites (below this argument in presented in a quantitative manner).

Following this line of thought early works \cite{BWH81,HB89,HM2004}
considered designed targets with a gapped binding energy which
is much lower than the rest of the DNA sequence. A sufficiently large energy gap at the target can then yield an arbitrarily large occupation probability of the target site even for one TF. When this is assumed the interesting question becomes that of the speed of the search. To address this question
various mechanisms, collectively called
\textit{facilitated diffusion}, were suggested. These combine one
dimensional diffusion along the DNA with three-dimensional
diffusion or intersegmental transfers. The combination of the
various search modes has been observed experimentally
\cite{BB76,winter2,WBH81,JWSP96,JP98,BGZY99,S99,SSMH2000,W2005,GWH2005,WAC2006,ELX2007,Bonnet2008rp,Fok2008,Loverdo:2009a,Fok2009,hopUL42} and shown
theoretically to be capable of decreasing the search time significantly
\cite{AD68,BB76,BWH81,LKWCJ93,GMH2002,B2002,HM2004,SM2004,CBVM2004,BZ2004,K2005,Z2005,LAM2005,HGS2006,HS2007,Oshanin2007a,Eliazar2007,kolo2008,Lomholt2009,Florescu2009,Florescu2009a,Givaty2009,Vuzman2010a,Vuzman2010b,Vuzman2010c,Rosa2010,benichou2011,Kolomeisky2011}. More recently the influence of facilitated diffusion on the noise level in gene regulation was analyzed in \cite{Bialek2009,Tamari2011}.

However, as realized early \cite{Hippe86} the assumption of a designed target
is far from obvious. In an alphabet of four letters a target sequence
of length $12$, quite common in TFs, will occur with essentially
probability one in a random sequence of length $\simeq10^{7}$.
Therefore, for target sequences shorter than $12$ bases, identical
and almost identical sequences will occur on the DNA. These competing sites
can easily ruin the stability of the target site. Furthermore, as
discussed in detail below, these almost identical sequences act as
traps \cite{JAWMP94} that hinder the search process and lead to  an antagonism
between the stability of the TF at the target site and the speed
of the target location. This problem, raised in \cite{WBH81}, is
commonly referred to as the \textit{speed-stability paradox}.

Recently, motivated by new experiments there has been renewed
interest in this rather old problem. To date there are now several
reviews (some very recent) which cover different aspects of the problem \cite{HB89,HM2004,pccp2008,Mirny2009,kolo2010,rmp2011}.
We believe that this review complements these and presents the problem using
a somewhat new angle. To this end we give an overview of the current status of the
speed-stability paradox and its implications on regulation
dynamics. We present the problem using both theoretical
considerations and experimental data. As we argue it is
preposterous to group all TFs in a single class
\cite{Wunderlich2009}. Different search mechanisms are likely to
apply to different proteins grouping them into different classes.
We show that three broad classes can be defined, which we term
gapped, marginally gapped and non-gapped transcription factors. The applicability
of previously suggested search mechanisms to each of the groups is analyzed in
some detail. Using this we turn to discuss in detail a recently proposed {\it barrier
controlled} search mechanism \cite{My3} which can in principle
resolve the speed-stability paradox for all classes of proteins. The possibility of such a mechanism
suggests that experiments should also probe activation barriers and not, as
commonly done, binding energies (see discussion below). Moreover, this mechanism allows for a
rich transient behavior and for transcription factors which are
efficient despite binding and unbinding rapidly from the target.

The structure of the review is as follows: In Section II we discuss
in detail the energetics associated with protein-DNA interaction.
We argue for the classification of transcription factors into the
three classes defined above. The classification is illustrated
using experimental data. In Section III we review the kinetics of
simple search mechanisms which have been discussed in the
literature. In Section IV we introduce the speed-stability paradox
and its possible resolution for each class of TFs. In Section V we
introduce and analyze in detail the barrier controlled search
mechanism. In Section VI an
effective model for the barrier controlled search mechanism is introduced and used to study transient
behaviors. We summarize the results in Sec. VII.

\section{Protein-DNA energetics \label{Protein-DNA energetics}}

Due to the sequences heterogeneity of the non-target DNA the binding
energy of a protein to a DNA is location dependent. The structure
of this \textit{disordered}, non-specific, energy landscape is crucial for understanding the stability of a TF at its target site and which search strategies
can or cannot be efficient. To this end, in this Section we consider
the energy landscape both from a theoretical point of view and by
looking at experimental data. Throughout what follows we use units
where $k_{B}T=1$.

\begin{figure}[ptb]
\begin{centering}
\includegraphics[width=15cm]{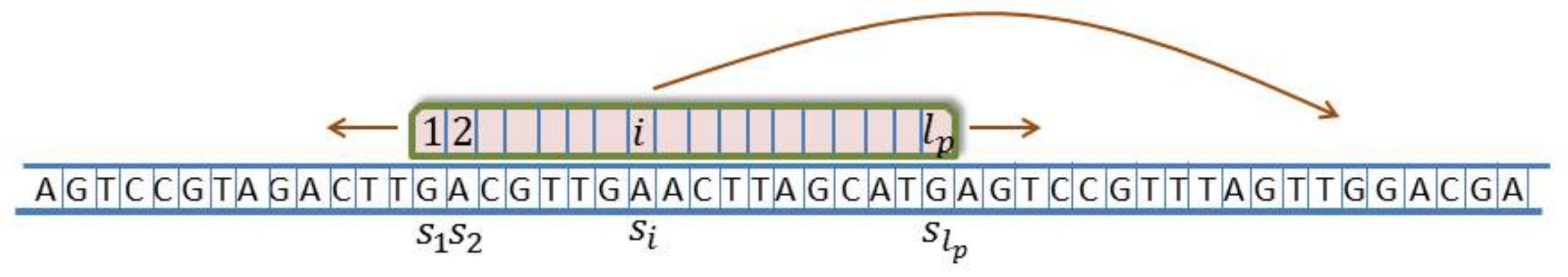}
\par\end{centering}
\caption{In this cartoon the interaction between the transcription
factor of length $l_p$ and the DNA sequence $\textbf{s}$ is illustrated.}
\label{ProteinCartoon}
\end{figure}

\begin{figure}[ptb]

\begin{centering}
\includegraphics[width=15cm]{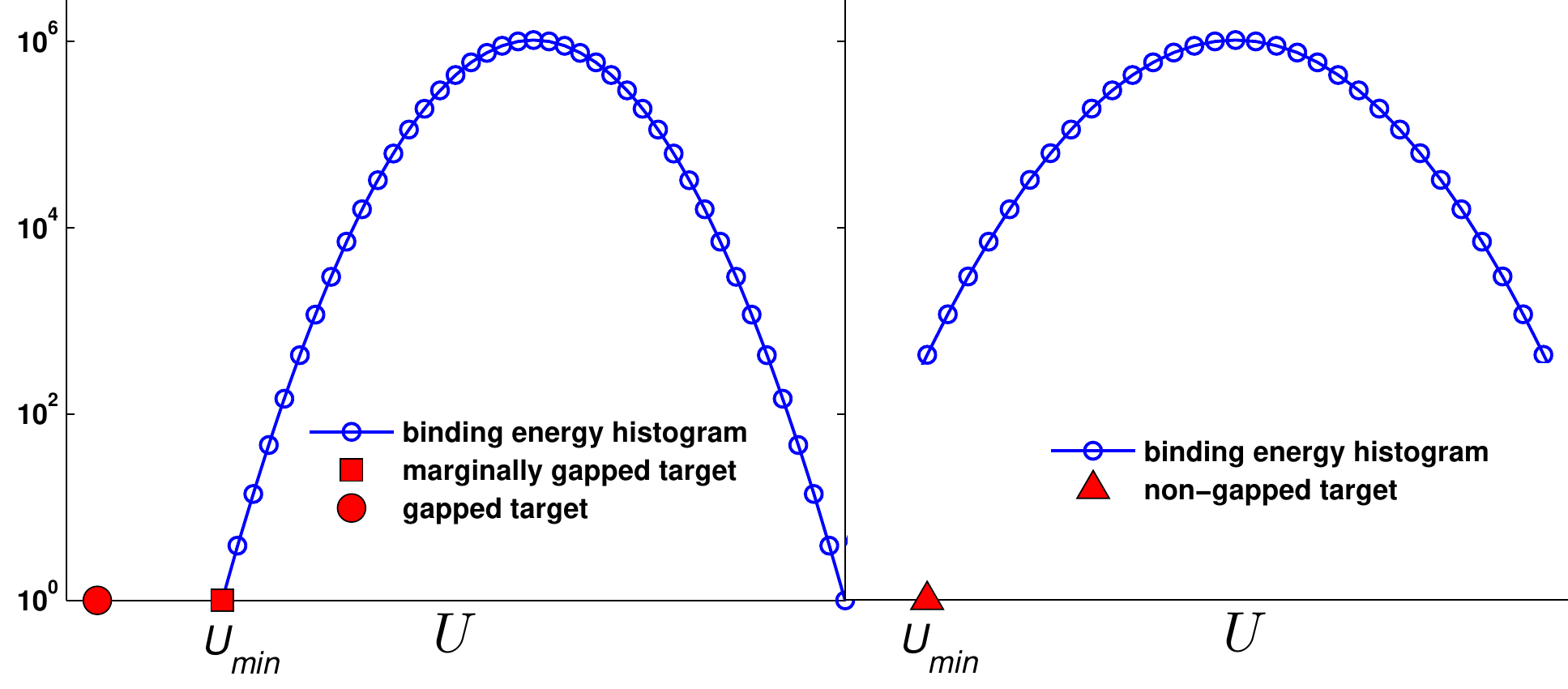}
\par\end{centering}

\caption{In this schematic plot three different types of a target are shown for a given
binding energy histogram (blue curve).}

\label{Targets}
\end{figure}

Equilibrium measurements \cite{SF98} reveal that to a good approximation
the binding energy, $U\left({\bf s}\right)$, of a transcription factor
which binds to  a sequence of $l_{p}$ bases ${\bf s}=\left(s_{1},s_{2},...,s_{l_{p}}\right)$
on the DNA is given by \cite{GMH2002}
\begin{equation}
U\left({\bf s}\right)=\underset{i=1}{\overset{l_{p}}{\sum}}\mathcal{E}\left(s_{i},i\right)\;.\label{Ueps}
\end{equation}
 Here $s_{i}=A,T,C,G$ is the nucleotide type on the $i$th binding
location of the protein and $l_{p}$ is the number of binding sites
on the protein (see Fig. \ref{ProteinCartoon}). The binding
energies $\mathcal{E}\left(s,i\right)$ are usually estimated
experimentally by measuring the probability,
$\Pr\left(s,i\right)$, that a nucleotide $s$ is bound to a
location $i$ on the protein in equilibrium \textit{in vitro
}experiments. Namely, one uses
\begin{equation}
\Pr\left(s,i\right)=\frac{e^{-\mathcal{E}\left(s,i\right)}}{Z_{i}}\;\;\;\;{\rm
where}\;\;\;\;\; Z_{i}=\underset{s^{\prime}=\left\{
A,T,C,G\right\}
}{\sum}e^{-\mathcal{E}\left(s^{\prime},i\right)}.\label{PrCI}
\end{equation}
The matrix $\Pr\left(s,i\right)$ has $4\times l_{p}$ elements and
is called the weight matrix (also known as Position-Specific
Scoring Matrix (PSSM) or \textquotedbl{}profile\textquotedbl{}).
It is important to note that these probabilities are measured only
for sequences which are close in structure to the target site\footnote{Since the binding probability is measured only in places close to the target sequence {\it on a finite sample} there are cases where one or more of the letters does not appear. To correct for this the probability of a letter to appear at a given site is derived from $(n_s+1/4)/(1+\sum\limits_{s} n_s)$, where $n_s$ is the number of occurrences of the letter $s$. This, standard procedure, ensured that when no measurements are made the probability is $1/4$.}. The
reason for this lies in the existence of other conformations of
the protein-DNA complex which we will allude to later \cite{Quake2007}. In Fig. \ref{Targets} we illustrate a sample
binding energy probability distribution for several \textit{E.
coli} proteins.

The structure of the binding energy implies that it can be
described by three parameters instead of the $4^{l_{p}}$ entries.
Specifically, the energy is a sum of contributions (see Eq.
(\ref{Ueps})) which can be assumed independent, if the DNA sequence
is uncorrelated, and can therefore be modeled to a good
approximation by a Gaussian random variable. (The assumption that
the DNA sequence is uncorrelated is believed to be true for coding
DNA and in particular for prokaryotic DNA\footnote{Algebraic
correlations have been claimed to be observed in non-coding DNA
\cite{HTWG98,PBGHSSS92} }.) The validity of this approximation is
illustrated for several proteins in Fig. \ref{EnHistFigure}. As can be
seen it holds for energies above the target energy,
$U_{\mathcal{T}}$, which is defined as the lowest possible binding
energy of the TF to any sequence. Explicitly, the probability
density of finding a given binding energy $U$ for non-target
sequences is well approximated by
\begin{equation}
P(U)\simeq\left\{ \begin{array}{cc}
{\cal N}^{-1}e^{-\frac{U^{2}}{2\sigma_{U}^{2}}} & U>U_{\mathcal{T}}\\
0 & U<U_{\mathcal{T}}\end{array}\right.,\label{Pr(U)}
\end{equation}
 where ${\cal N}$ is a normalization factor and the variance
\begin{equation}
\sigma_{U}^{2}=\underset{i=1}{\overset{l_{p}}{\sum}}\left\{ \underset{s_{i}=\left\{ A,T,G,C\right\} }{\frac{1}{4}\sum}\mathcal{E}^{2}\left(s_{i},i\right)-\left[\underset{s_{i}=\left\{ A,T,G,C\right\} }{\frac{1}{4}\sum}\mathcal{E}\left(s_{i},i\right)\right]^{2}\right\}\;.\label{Variance}
\end{equation}
The target energy is given by: \begin{equation}
U_{\mathcal{T}}=\underset{i=1}{\overset{l_{p}}{\sum}}\min\left[\mathcal{E}\left(A,i\right),\mathcal{E}\left(T,i\right),\mathcal{E}\left(C,i\right),\mathcal{E}\left(G,i\right)\right].\end{equation}
 The statistical properties of the binding energy are now encoded
by $\sigma_{U}$ and $U_{\mathcal{T}}$ and the mean binding energy
which we set to be zero. Note, that the Gaussian form is unchanged
even if one allows for corrections to the weight matrix which
depend, say, on near-neighbor configurations, as suggested in
\cite{Stormo86,Zhang93,Ponomarenko99,Bulyk2002}. The assumption
that the DNA sequence is uncorrelated also implies that the
binding energies $U_{i}$ and $U_{j}$ at different sites $i$ and
$j$ are independent. Strictly speaking this holds only for
$|i-j|>l_{p}$. In what follows we neglect these, unimportant, short
range correlations.

Another quantity which is important for understanding the binding
is the minimal energy, $U_{min}  \geq U_{\cal T}$, which occurs randomly on a typical
DNA sequence among the non-target sites. This site competes most strongly with
the target site. In a sequence of $N\gg1$ uncorrelated
base pairs, it is narrowly distributed (with a variance scaling as $1/\ln N$) and well approximated by \cite{HF2006}
\begin{equation}
\int_{-\infty}^{U_{min}}P(U)\simeq\frac{1}{N}
\end{equation}
or
\begin{equation}
U_{min}\simeq\max\left(-\sigma_{U}\sqrt{2}\operatorname{erfc}^{-1}\left(\frac{2}{N}\right),U_{\mathcal{T}}\right)\simeq\max\left(-\sigma_{U}\sqrt{2\ln
N},U_{\mathcal{T}}\right). \label{lowerLimit}
\end{equation}

For a given DNA length, $N$, $U_{min}$, $U_{\cal T}$ and $\sigma_{U}$ characterize the
binding properties of a TF. This naturally leads to three classes
of transcription factors (see Fig. \ref{Targets} for a schematic
illustration).

\paragraph{Gapped transcription factors.}

In this case there is a significant gap between the lowest non-target
energy, $U_{min}$, and the target energy, $U_{{\cal T}}$.
Namely,

\begin{equation}
U_{min}\simeq-\sigma_{U}\sqrt{2\ln N}
\end{equation}
and

\begin{equation}
U_{\mathcal{T}}<-\sigma_{U}\sqrt{2\ln N}.\label{gapcond}\end{equation}

\paragraph{Marginally gapped transcription factors.}

Here there is no energetic gap between the target and the rest of
the DNA but the number of sites with an energy close to $U_{{\cal T}}$
is small (of the order of one). This happens when

\begin{equation}
U_{min}\simeq U_{\mathcal{T}}\simeq-\sigma_{U}\sqrt{2\ln N}.\label{marggapcond}\end{equation}

\paragraph{Non-gapped transcription factors.}

In this case there is no energetic gap between the target and the
rest of the DNA and the number of sites with an energy close to the
target one is large. This happens when

\begin{equation}
U_{min}\simeq U_{\mathcal{T}}>-\sigma_{U}\sqrt{2\ln N}.
\end{equation}

Note that within the additive binding energy model, Eq. (\ref{Ueps}),
the possible existence of a gapped TF is directly related to its length.
In that case \begin{equation}
U_{{\cal \mathcal{T}}}=l_{p}E_{c}\label{Ut_lp}\end{equation}
 where $E_{c}<0$ is the average lowest binding energy per base and
\begin{equation}
\sigma_{U}^{2}=\frac{l_{p}E_{c}^{2}}{3}.\end{equation}
Here we assumed that each base appears with equal probability along the DNA. Then Eq. (\ref{gapcond}) implies that to produce an energetic gap between $U_{\mathcal{T}}$ and $U_{min}$ a TF has to be long enough. Namely, one finds
\begin{equation}
l_{p}>\frac{2}{3}\ln N\;.\label{l_p}
\end{equation}
 This has a particularly simple interpretation. It is equivalent to
demanding that on a DNA sequence, of length $N$, sites which are
identical to the target site do not appear randomly so that $1/4^{l_{p}}<1/N$.
For a typical bacterial DNA length, $N=10^{7}$, this gives $l_{p}>11$.
The argument can be refined using information theoretic arguments
(see Appendix \ref{An information theoretic approach} and for a similar line of reasoning \cite{Wunderlich2009}) to give a
stronger bound of $l_{p}>22$.

As we discuss below, the structure of the energy landscape, gap existence
and the properties of the target have important consequences on the
equilibrium probability of finding the protein on the target and the
search time. Interestingly, as we show below, experimental data suggests
that there are transcription factors which belong to each of the above
categories.

\subsection{Target occupation probability in equilibrium}\label{sec:targetoccupation}

Next, we turn to consider the probability of a TF to be at the
target, $P^{\mathcal{T}}$, in equilibrium. For TFs which appear in
small numbers (as believed to be the case in many examples
\cite{Guptasarma95}) this quantity has to be of the order of one
for proper control over gene expression. Otherwise, assuming
equilibration (we discuss other scenarios later), the TF has to be
present in a large copy number. Naively $P^{\mathcal{T}}$ will be
of the order of one as long as the TF is gapped. As we now show
this is not guaranteed and we outline the conditions for this to
occur. We ignore the free-energy contribution from configurations
where the protein is off the DNA. These can only hamper the stability at the target.

In equilibrium to ensure $P^{\mathcal{T}}$ close to one the partition
function has to be dominated by the target energy. Namely, for stability
we require \begin{equation}
Z=\sum_{i=1}^{N}e^{-U_{i}}\simeq e^{-U_{\mathcal{T}}}.\label{Stability-1}\end{equation}
 The typical partition function can be approximated, using Eq. (\ref{Pr(U)}),  by
  \begin{equation}
Z\simeq e^{-U_{{\cal \mathcal{T}}}}+N\frac{\underset{U_{min}}{\overset{\infty}{\int}}e^{-\frac{U^{2}}{2\sigma_{U}^{2}}}e^{-U}dU}{\underset{U_{min}}{\overset{\infty}{\int}}e^{-\frac{U^{2}}{2\sigma_{U}^{2}}}dU}.\end{equation}
 Note, that as standard in disordered systems, this can be different
from the average partition function which is obtained by setting the
lower bound of the integrations on the right hand side to $-\infty$.
This gives in the large $N$ limit
\begin{align}\label{Zapprox}
Z & \simeq\left\{ \begin{array}{cc}
e^{-U_{{\cal \mathcal{T}}}}+e^{\sigma_{U}\sqrt{2\ln N}} & \ {\rm for}\ \sigma_{U}\gg\sqrt{2\ln N}\\
e^{-U_{{\cal \mathcal{T}}}}+Ne^{\frac{\sigma_{U}^{2}}{2}} & \ {\rm
for}\ \sigma_{U}\ll\sqrt{2\ln N}\end{array}\right..\end{align} We therefore
identify two regimes: large disorder strength
$\sigma_{U}\gg\sqrt{2\ln N}$ and small disorder strength
$\sigma_{U}\ll\sqrt{2\ln N}$. Note, that the physics is very close
to that of the Random Energy Model (REM) \cite{Derrida1980}.

For large disorder strength $\sigma_{U}\gg\sqrt{2\ln N}$, which
corresponds to the frozen phase of the REM, gapped TFs or marginally
gapped TFs are stable on the target. Together with the definitions
(\ref{gapcond})-(\ref{marggapcond}), this condition reads
\begin{equation}
U_{{\cal \mathcal{T}}}\le-\sigma_{U}\sqrt{2\ln N}\;.\label{largedisorder}
\end{equation}
 To satisfy the stability requirement in the small disorder case,
which corresponds to a system above the freezing point of the REM,
it is required that
\begin{equation}
U_{{\cal \mathcal{T}}}\le-\ln N  - \sigma_U^2/2\;,\label{smalldisorder}
\end{equation}
 so that only gapped TFs can be stable on the target. Using the additive
binding model, so that $U_{{\cal \mathcal{T}}}=l_{p}E_{c}$ and $\sigma_{U}^{2}=\frac{l_{p}E_{c}^{2}}{3}$
implies that the small disorder regime corresponds to $l_{p}\ll\frac{6\ln N}{E_{c}^{2}}$
and the stability condition translates in this case to the constraint
\begin{equation}
l_{p}\geq\frac{\ln N}{-E_{c}(1+E_c/6)}.
\end{equation}
 This is possible only for $-E_{c}<6$. As expected the
bound on $l_{p}$ grows when $E_{c}$ approaches zero.

Note that for both large and small disorder strengths, the larger
$N$, the more stringent the condition on $U_{{\cal \mathcal{T}}}$.
With $E_{c}$
of the order of $-1$ the above conditions give $l_{p}\geq16$ for
small disorder and $l_{p}\geq32$ for large disorder.
We comment, that in principle a simple way to satisfy the conditions
(\ref{largedisorder}) or (\ref{smalldisorder}), is for example to
introduce large enough cooperative interactions between different
TF's binding domains. In this case the binding energy is not additive
so that Eq. (\ref{Ueps}) is not valid. These can single out the target
and generate an arbitrarily large gap between the target and the rest of
DNA sites.

In summary, TFs with non-gapped targets cannot be stabilized on their
targets. Marginally gapped TFs can be stabilized on their targets if
the disorder strength is large enough. Below, we show that this requirement
gives rise to a conflict with the speed of the target location. A gapped
TF is stable on its target when the disorder strength is large, or
in the small disorder regime if it is large enough (or if cooperative
effects are present). Without any cooperative interaction between
different TF's parts, such a gap may be achieved in both small and large
disorder regimes for reasonable TF's length (for a biochemically reasonable
energy scale of about $1$). Below we show that combining these requirements
with another set of constraints related to the speed of the search gives
much more stringent conditions on the length of the protein.

\subsection{Experimental data \label{Experimental data}}

\begin{figure}[ptb]

\begin{centering}
\includegraphics[width=12cm]{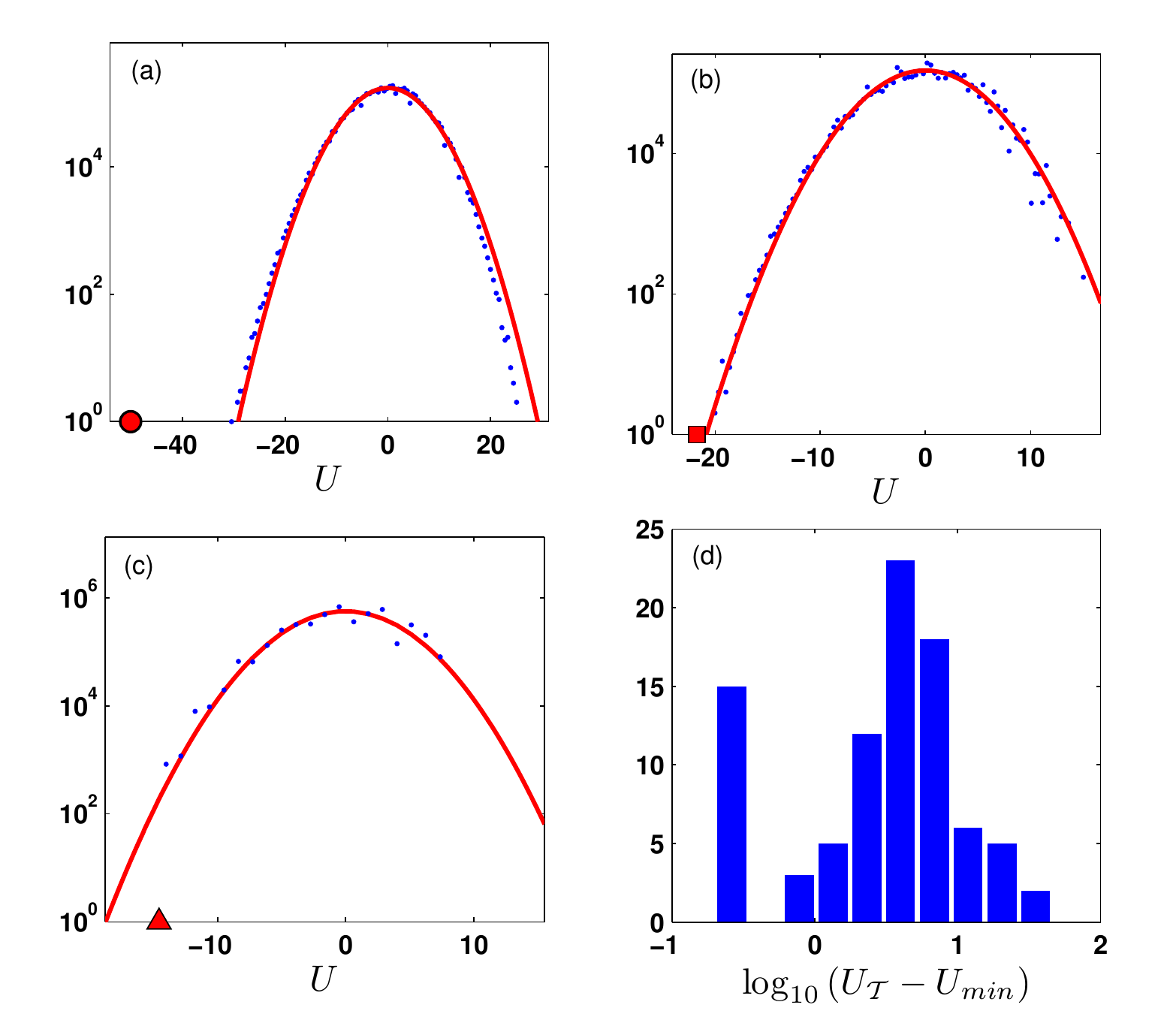}
\par\end{centering}

\caption{Here a histogram of the binding energy is presented for three different
TFs. (a) BaeR, $l_{p}=29$, $\sigma=6.86$, $U_{{\cal \mathcal{T}}}=-50.16$,
$E_{min}=-33.5$, $P^{\mathcal{T}}=1$. (b) DcuR, $l_{p}=15$, $\sigma=4.93$,
$U_{{\cal \mathcal{T}}}=-21.76$, $E_{min}=-21.75$, $P^{\mathcal{T}}=0.074$.
(c) AscG, $l_{p}=7$, $\sigma=4.2$, $U_{{\cal \mathcal{T}}}=-14.55$,
$E_{min}=-14.55$, $P^{\mathcal{T}}=0.0006$. The red lines are Gaussian approximations to the distributions using the measured variance calculated from Eq. (\ref{Variance}). (d) A histogram of
the estimated gap values. The data is based on 89
weight matrices of \textit{E. coli} DNA-binding proteins and was taken from the RegulonDB database \cite{regulonDB}. }

\label{EnHistFigure}
\end{figure}

In recent years much experimental data has been accumulated.
Specifically the weight matrix has been measured for many TFs. We
now use data from RegulonDB \cite{regulonDB} which contains 89
weight matrices to try and single out the different classes of proteins
discussed theoretically above. As we proceed to show,  the three classes can be identified in the
data. Three examples are shown in Fig.
\ref{EnHistFigure}. These correspond to a gapped (Fig.
\ref{EnHistFigure}(a)), marginally gapped (Fig.
\ref{EnHistFigure}(b)) and non-gapped (Fig.
\ref{EnHistFigure}(c)) proteins.

To analyze the stability of all the proteins in the database we look at several quantities. (i)
Their minimal possible binding energy $U_{{\cal \mathcal{T}}}=U({\bf s}^{*})$,
where ${\bf s}^{*}$ is defined to be the target of the protein. (ii)
The minimal binding energy on a typical disordered sequence of length
$N$, $U_{min}=U({\bf s}^{\dagger})$, where ${\bf s}^{\dagger}$
is the strongest binder on the sequence. (iii) The standard deviation  $\sigma_{U}$
for the different proteins and finally (iv) the occupation
probability at the target, $P^{\mathcal{T}}$. Some of the results presented below are demonstrated in Appendix
\ref{An information theoretic approach} using the language of information
theory (for a related discussion see \cite{Wunderlich2009}).

\begin{figure}[ptb]
\begin{centering}
\includegraphics[width=12cm]{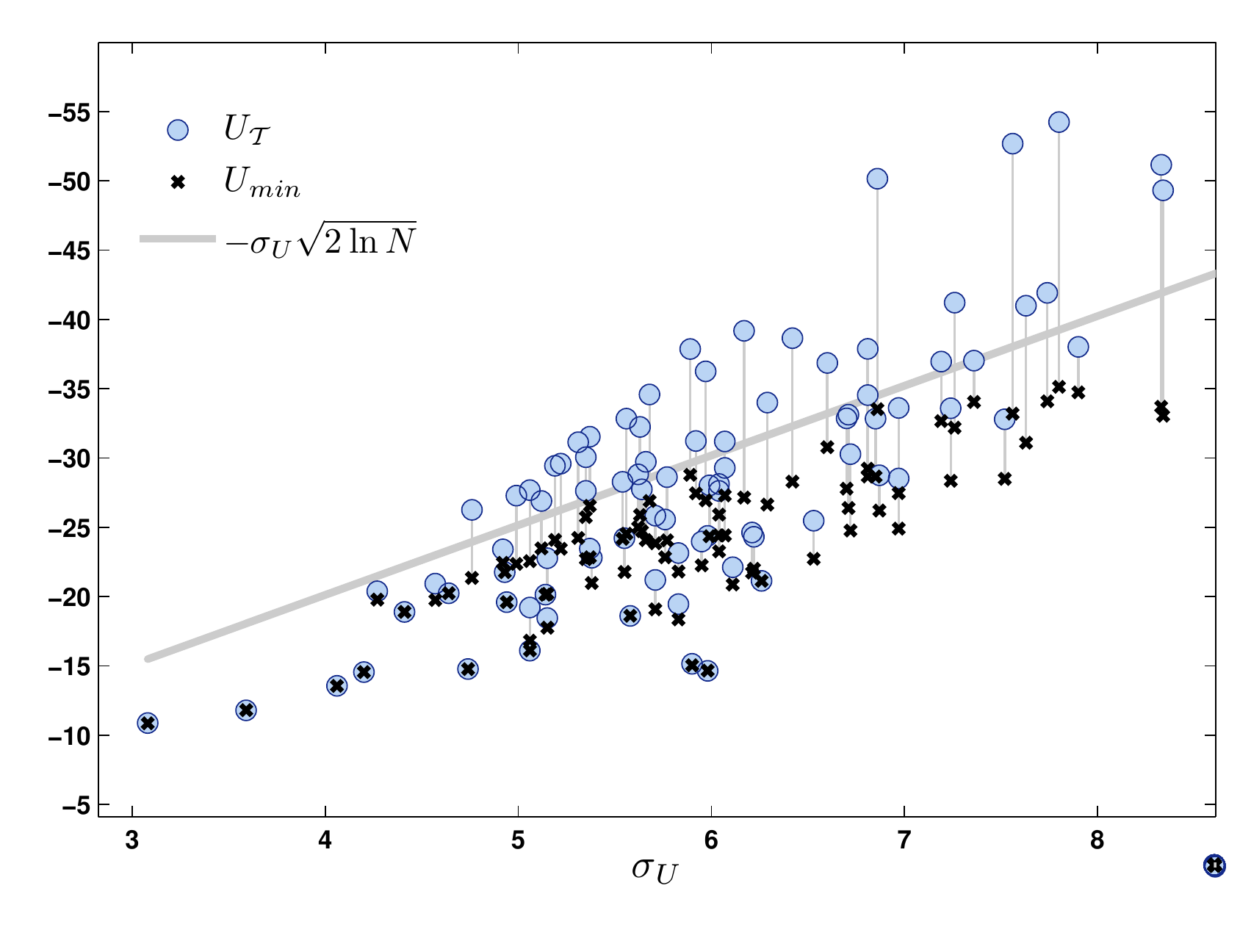}
\par\end{centering}

\caption{On this figure a comparison between (gray, thick, solid line)
the analytic upper limit for a minimal non-designed binding energy
Eq. (\ref{lowerLimit}), (black crosses) the estimated minimal
non-designed binding energy, $U_{min}$ and (blue circles) the
estimated binding energy of a perfectly designed full consensus
sequence, $U_{{\cal \mathcal{T}}}$. Each $\sigma_{U}$ corresponds
to a different protein. The data is based on $89$ weight matrices of \textit{E. coli}
DNA-binding proteins and was taken from the RegulonDB database
\cite{regulonDB}.}

\label{EminPlot}
\end{figure}

It is useful to present that data by plotting $U_{min}$ and
$U_{{\cal \mathcal{T}}}$ as a function of $\sigma_{U}$ (see Fig. \ref{EminPlot}). Each protein
on the graph is represented by two points with the same abscissa.
The graph shows several interesting features.

\begin{figure}[ptb]
\begin{centering}
\includegraphics[width=12cm]{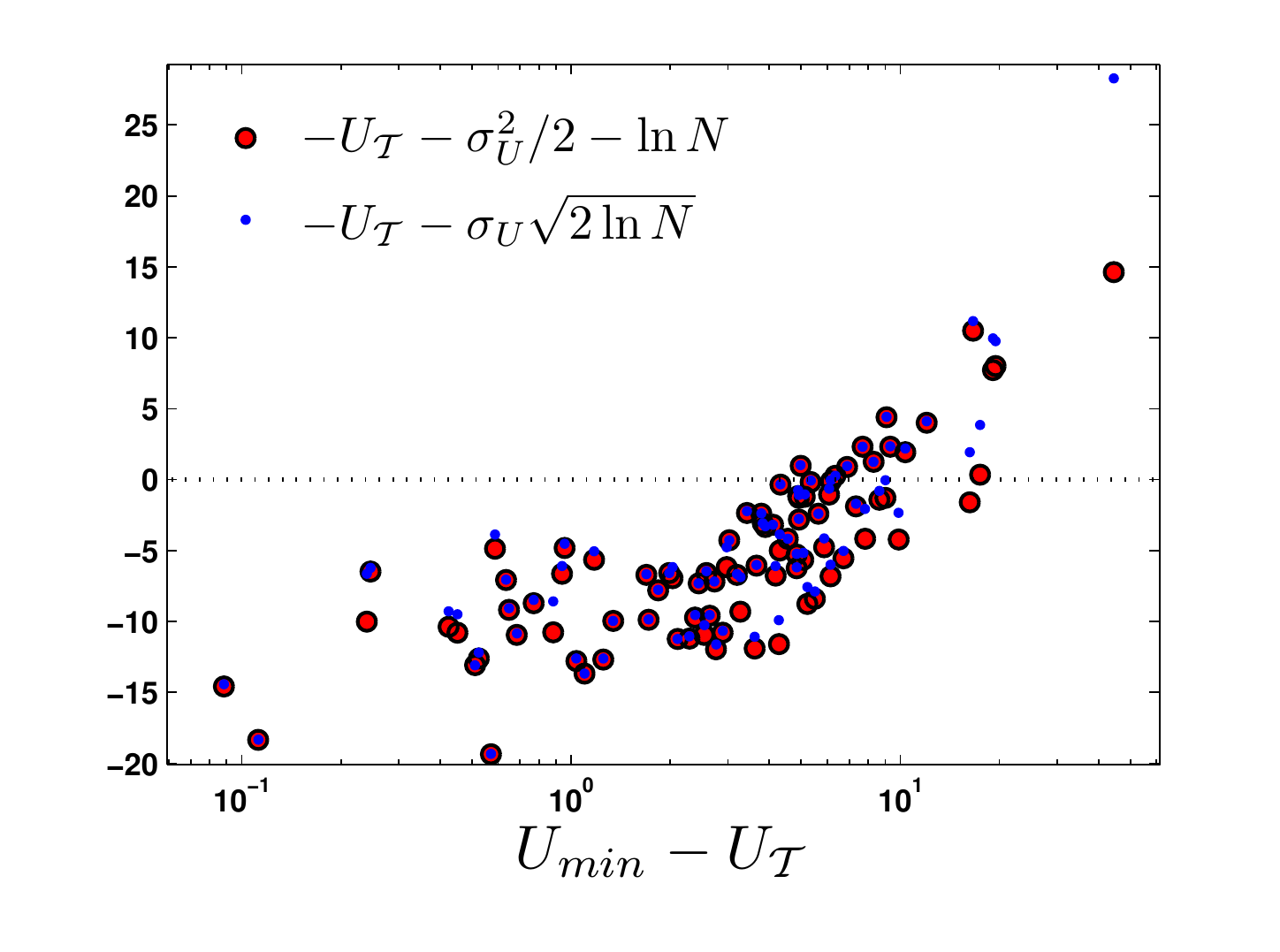}
\par\end{centering}

\caption{The stability criterions in the small (Eq. (\ref{smalldisorder}), large red dots) and large (Eq. (\ref{largedisorder}), small blue dots) disorder regimes. The data
is based on $89$ weight matrices of \textit{E. coli} DNA-binding proteins
and was taken from the RegulonDB database \cite{regulonDB}.}

\label{criter}
\end{figure}

(i) First, as expected, a significant part (about three fourth) of
the TFs are gapped with a gap size ranging from a few $k_BT$ to about $20k_BT$. A histogram of the gap size is shown in Fig.  \ref{EnHistFigure}(d). %Notice, that for about one fourth of the proteins the gap is smaller than $1 k_BT$.
As stated above such gapped proteins are stable only when the gap is large enough, see Eqs. (\ref{smalldisorder}) and (\ref{largedisorder}).
For an \textit{E. coli} DNA length this requires $U_{\cal T} < -15$ in the small disorder regime ($\sigma_U \ll \sqrt{2\ln N} \simeq 5.5$) and $U_{\cal T}< -30$ in the large disorder regime ($\sigma_U \gg 5.5$). Note  that indeed for $\sigma \geq 5.5$ a large fraction of the values of $U_{\cal T}$ are below $-30$ and therefore correspond to stable TFs. The stability criterions for both small (Eq. (\ref{smalldisorder})) and large (Eq. (\ref{largedisorder})) disorder strengths are shown  in Fig. \ref{criter} and indicate that most proteins with a large gap are stable. Note also that
the theoretical prediction for $U_{min}$ (shown in  Fig. \ref{EminPlot}) fits
reasonably well with the experimental results.

\begin{figure}[ptb]

\begin{centering}
\includegraphics[width=8cm]{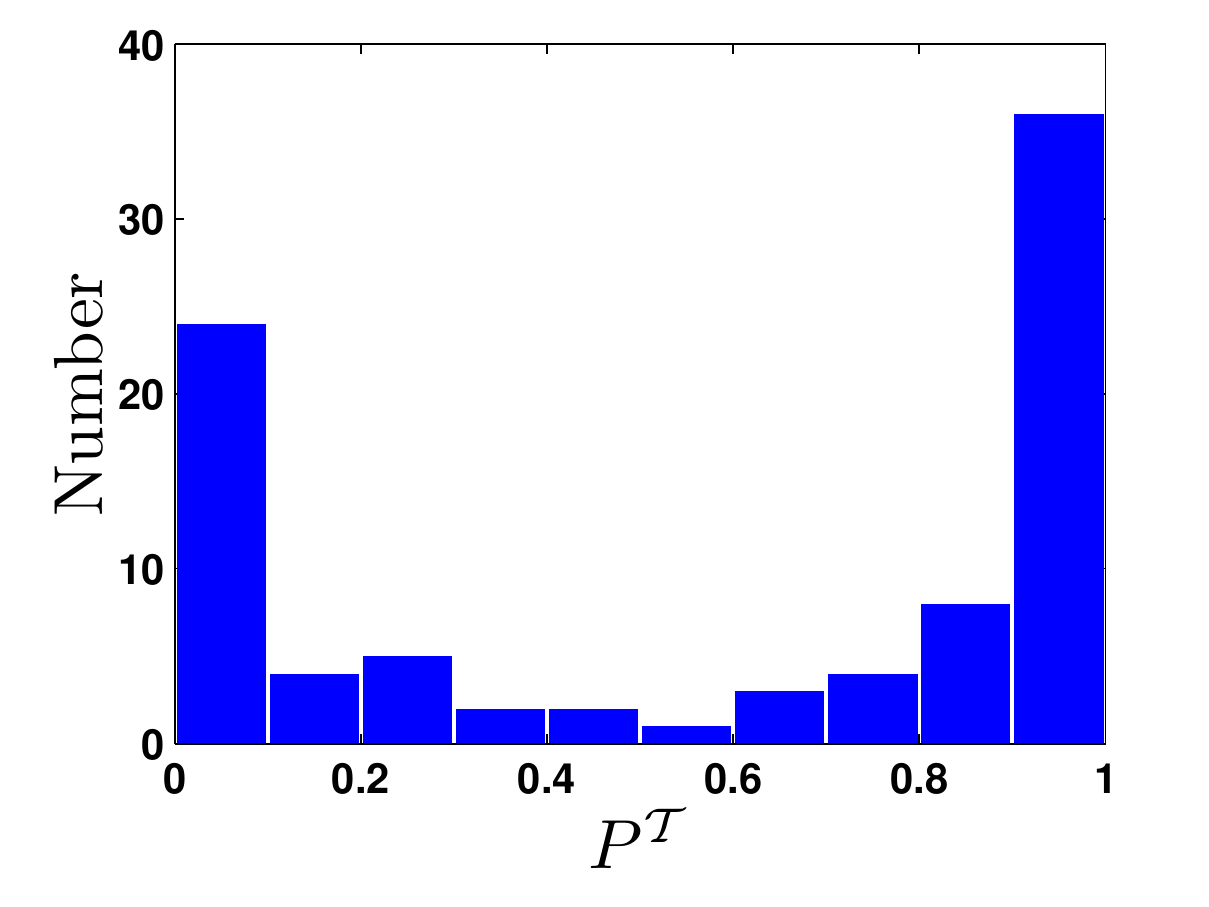}
\par\end{centering}

\caption{Here a histogram of the occupation probability of a
target, $P^{\mathcal{T}}$, is presented. The bulk term was not
taken into account such that the presented data slightly
overestimate $P^{\mathcal{T}}$. The data is based on $89$ weight
matrices of \textit{E. coli} DNA-binding proteins and was taken
from the RegulonDB database \cite{regulonDB}.}

\label{PbFigure}
\end{figure}

(ii) Second, for about one fourth of the  TFs $U_{\cal T} \simeq U_{min}$.
This indicates that they are either non-gapped or marginally gapped.
Recall that for such proteins a minimal criterion for being stable
at the target is that the disorder is large ($\sigma_{U}\ge\sqrt{2\ln N} \simeq 5.5$). This does not seem to be satisfied for most of the marginally gapped proteins. Therefore,  Fig. \ref{EminPlot} hints that most of the non gapped and marginally gapped TFs are actually unstable on the target.
This is more clearly illustrated in Fig. \ref{PbFigure} which shows that indeed
about one quarter of the proteins have a very small probability (less than
$10^{-1}$ with about half of them with a probability less than $10^{-2}$) for being on the target. %About half of these have a probability of less than $10^{-2}$ to be on the target.
This indicates that non gapped and marginally gapped  TFs seem to break the stability requirement. We return to these proteins later and suggest that either non-equilibrium effects or large copy numbers could stabilize them
on the target.

\begin{figure}[ptb]
\begin{centering}
\includegraphics[width=15cm]{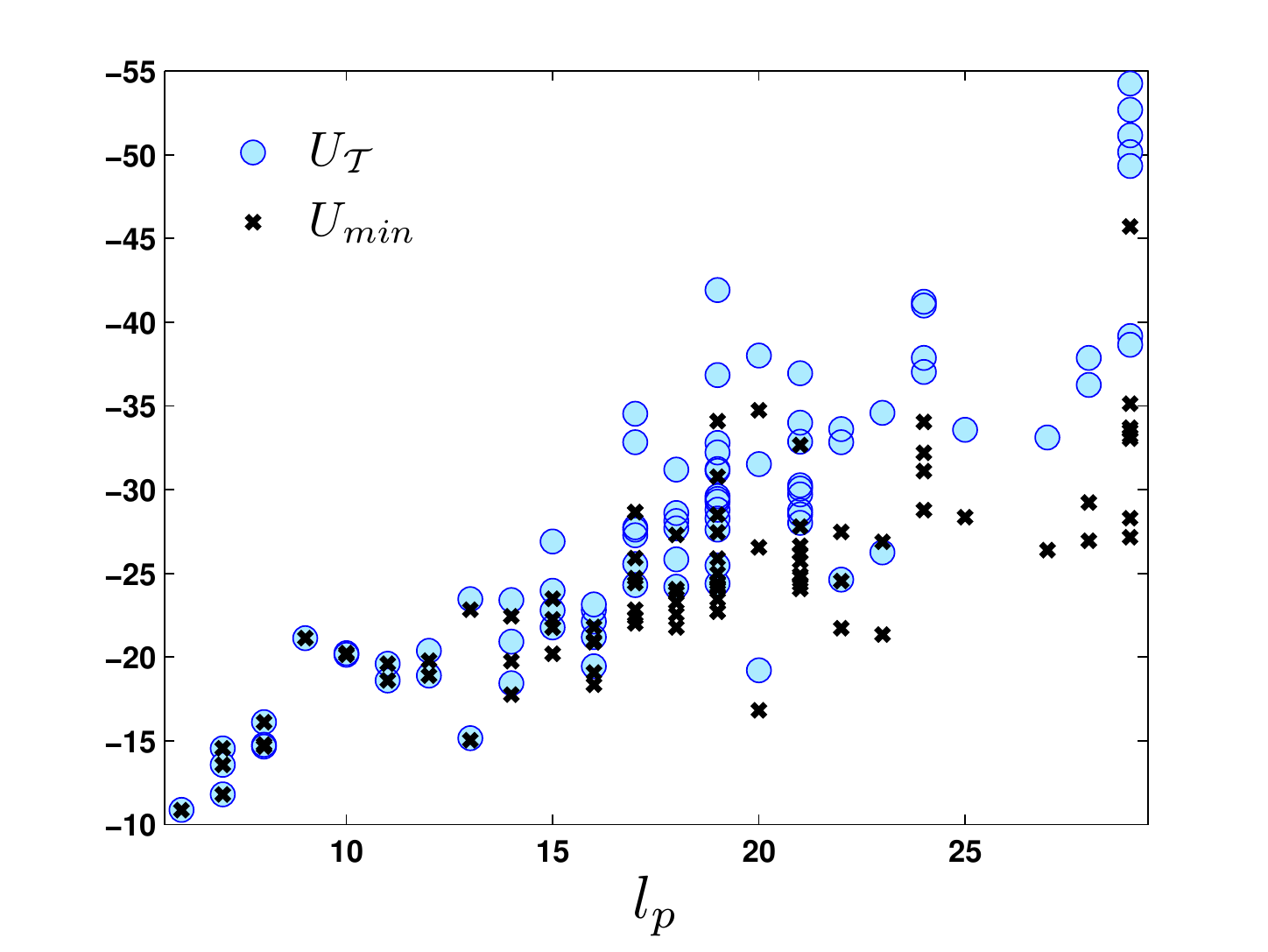}
\par\end{centering}

\caption{A comparison between (black crosses) the
estimated minimal non-designed binding energy, $U_{min}$ and (blue
circles) the estimated binding energy of a perfectly designed full
consensus sequence, $U_{{\cal \mathcal{T}}}$ as a function of the protein's length. The data is based on $89$ weight matrices of
\textit{E. coli} DNA-binding proteins and was taken from the
RegulonDB database \cite{regulonDB}.}

\label{Eminlp}
\end{figure}

\begin{figure}[ptb]
\begin{centering}
\includegraphics[width=10cm]{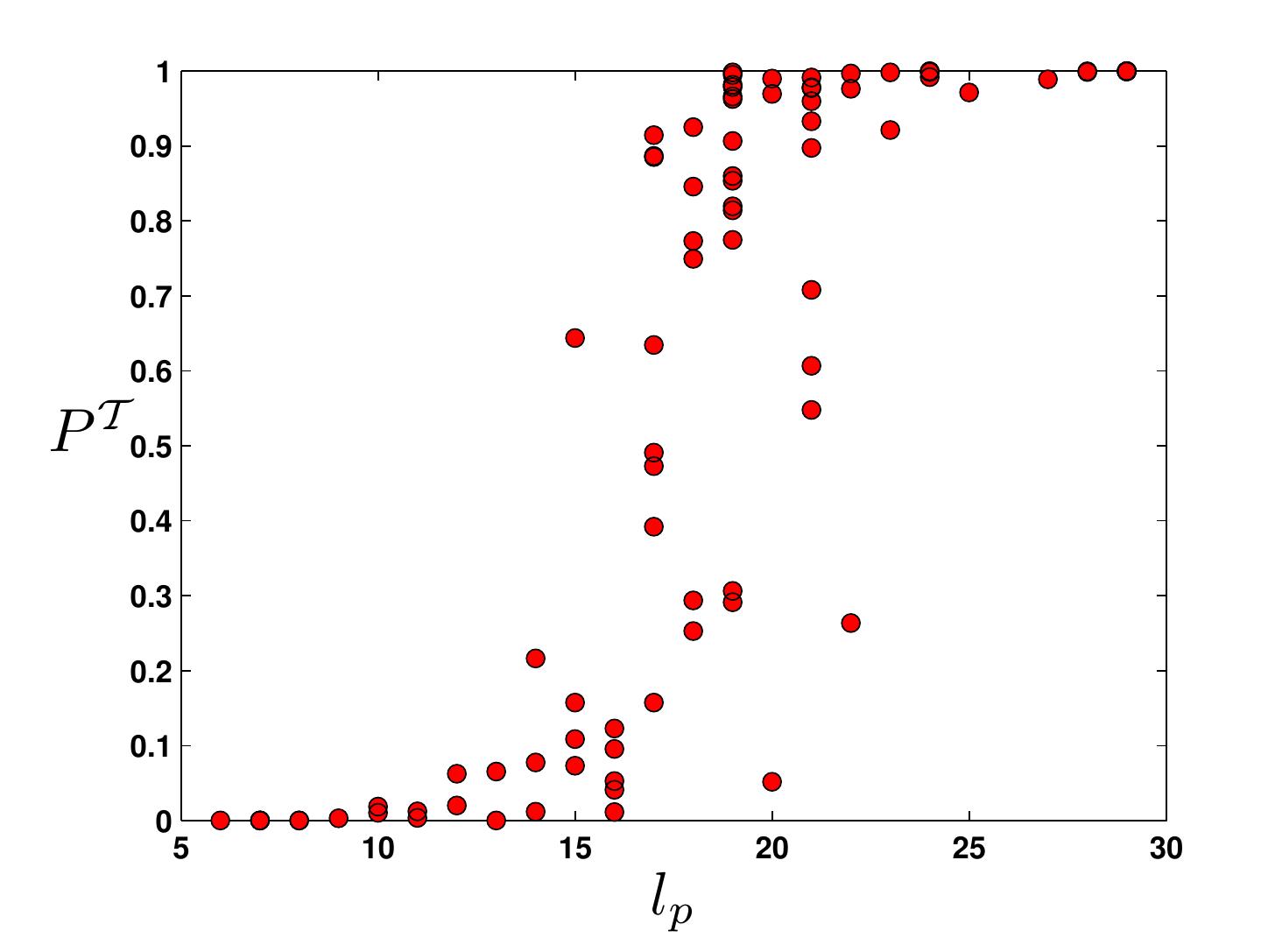}
\par\end{centering}

\caption{The occupation probability of a target,
$P^{\mathcal{T}}$, is presented as a function of a protein's
length, $l_p$. The bulk term was not taken into account such that
the presented data slightly overestimate $P^{\mathcal{T}}$. The
data is based on $89$ weight matrices of \textit{E. coli}
DNA-binding proteins and was taken from the RegulonDB database
\cite{regulonDB}.}

\label{Ptlp}
\end{figure}

It is interesting to present the same data, but instead of as a
function of $\sigma_U$, as a function of $l_p$. This is shown in
Figs. \ref{Eminlp} and \ref{Ptlp}. As is clearly seen there is a
close relation between the existence of a gap and $l_p$ being
large enough. In fact, in agreement with our simple arguments, a
gap begins to form at $l_p \simeq 13$. The data for $P^{\cal T}$
as a function of $l_p$ is even more striking. Essentially all
proteins with a binding site of size $l_p \simeq 13$ or smaller are unstable
on the target while those with $l_p \simeq 16$ or larger are
mostly stable at the target. The close correspondence between
$l_p$ and the gap is a direct result of a similar binding energy
per base for all TFs.

The above discussion focused on the stability characteristics. We
identified several distinct classes of TFs based on their stability
properties. An important question for transcription factors is their
speed of operation. The discussion above suggests that different TFs
could have different search strategies. Before attempting to map these out in what follows we first review
the different possible reactive pathways which have been suggested in the literature.

\section{The search dynamics}

Before discussing the reactive pathways it is useful to have a simple picture
of DNA packing in prokaryotic cells. In typical systems the DNA
has a total length of $L\sim10^{6}nm$, a persistence length $L_{0}\sim50nm$,
a cross section radius $\rho\sim1nm$, and is contained in a volume
of $\Lambda^{3}\sim10^{9}nm^{3}$. The typical distance between segments
of DNA of length $L_{0}$ is therefore much smaller than $L_{0}$:
$\frac{\Lambda^{3}}{L/L_{0}}\ll L_{0}^{3}$. Under these conditions,
using $\Lambda\gg L_{0}$, it is easy to check that the radius of
gyration of \emph{free} DNA, which is of the order of $L_{0}\sqrt{\frac{L}{L_{0}}}$
is much larger than the cell size $\Lambda$ - the DNA is densely
packed even though its fractional volume in the container $L\rho^{2}/\Lambda^{3}$,
is small (about one percent). By way of comparison, typical protein
sizes are in the range $R\mathbf{\sim}1-10nm$, much smaller than
the DNA's persistence length.

To quantify the search process one needs to estimate the time it takes the protein, from its initial production,
to activate (or repress) its target site. Early works considered a perfectly reactive target. In this case the search
efficiency can be quantified by studying the statistical properties of the first-passage time to the target  \cite{R2001,nature2007,natchem2010}. In this section we focus on the mean first-passage time. Later, we will discuss
the potential importance of other time scales in the problem.

For a cell to properly function the search process has to,
typically, be of the order of seconds. In principle, when the
target is perfectly reactive this can be achieved by a search
which is driven by pure three dimensional diffusion. However,
driven by experimental results, mostly on the Lac repressor
\cite{RSB70,RBC70}, which seem to give search times that are
faster than three-dimensional diffusion, various search strategies
were suggested. We now give simple arguments that quantify these
different search strategies. For a similar discussion see
\cite{SM2004,HGS2006}.

\subsection{Searching with three-dimensional diffusion}

Naively, one might expect the protein to search for its target (or,
equivalently, its specific binding site on the DNA) using only three-dimensional
diffusion. Neglecting interactions of the protein with the environment
and the DNA (apart from the target site), one then finds, using results
first obtained by Smoluchowski \cite{S17} or by dimensional analysis,
that the search time, $t^{search}$, defined as the mean first-passage time at the target, is given by:
\begin{equation}
t^{search}\sim\frac{\Lambda^{3}}{D_{3}r}.\label{t3D}
\end{equation}
Here $D_{3}$ is the three-dimensional diffusion constant of the
protein, $r$ is the target size, and $\Lambda^{3}$ is the volume
that needs to be searched. Assuming a target size of the order of
a base-pair $r\approx0.34nm$, a typical nucleus (or bacterium) size
as above and using the measured three-dimensional
diffusion coefficient for a GFP protein \textit{in vivo,} $D_{3}\sim10^{7}{nm}^{2}/s$
\cite{ESWSL99}, one finds $t^{search}$ of the order of hundreds
of seconds. We comment that $r$ can be increased significantly by changing the electrostatic
interactions between the protein and its target, for example, by changing the salt concentration.

These long time scales can be easily reduced if several proteins are
searching for the target. Namely, if $n_{p}$ proteins are searching
for the same target the average search time is given by %
\footnote{The relation between the search time $t^{search}$ for one protein
and search time $t_{n_{p}}^{search}$ for $n_{p}$ proteins remains
unchanged throughout the paper. In the next Section is shown that
in the case of wide distributions of the search time the dependence
on $n_{p}$ is more sensitive. %
} $t_{n_{p}}^{search}\simeq t^{search}/n_{p}$. This suggests that
about $10$ proteins could find a target in reasonable time for cells
to function properly. As we discuss below this simple relation between the search time
of one protein and $n_p$ proteins can fail in some cases.

\subsection{Searching with one-dimensional diffusion}

In real systems, due to the interactions of proteins with non-specific
DNA sequences and the environment \cite{LR72}, the picture is more
complex. Indeed, \textit{in vitro} experiments have suggested that
mechanisms other than three-dimensional diffusion are used by many
proteins to locate their targets. The simplest extension of the pure three-dimensional
diffusive search is using
three dimensional diffusion to reach the DNA and then scan it using
one-dimensional diffusion along its contour. This follows closely
ideas of Delbruck and Adam \cite{AD68}, introduced in a different
context. If the DNA is very long the search time is clearly controlled
by the one-dimensional diffusion along the DNA which is given by
\begin{equation}
t^{search}\sim\frac{L^{2}}{D_{1}}\sim O\left(hours\right) \;.
\end{equation}
Here $L\sim10^{6}nm$ is the genome length and $D_{1}$ is the
one-dimensional diffusion coefficient that was measured indirectly
\cite{BB76} and directly \cite{WAC2006,ELX2007} to be much smaller
than the three-dimensional diffusion coefficient
$D_{3}\sim10^{7} nm^{2}/s$ \cite{ESWSL99}. Effects of
disorder can be incorporated into an effective value of $D_{1}$
\cite{SM2004} (see discussion below). The above result renders
this search strategy useless for long DNA. However, if the
sequence scanned is short then it is easy to see that the search
time is given by
\begin{equation}
t^{search}\sim\frac{L^{2}}{D_{1}}+\frac{\Lambda^{3}}{D_{3}L}\;.
\end{equation}
Using the numbers cited above it is easy to check that search times
of the order of a $100$ sec (so that about 10 proteins can find the
target within seconds) can be obtain as long as $L$, the length of
the sequence scanned is smaller than $10^{4}$ nm, about $30$ kilobases
long. The results are mildly modified if the sequence has a globular
shape.

\subsection{Facilitated diffusion}
\label{facdif}

\begin{figure}[ptb]
\begin{centering}
\includegraphics[width=8cm]{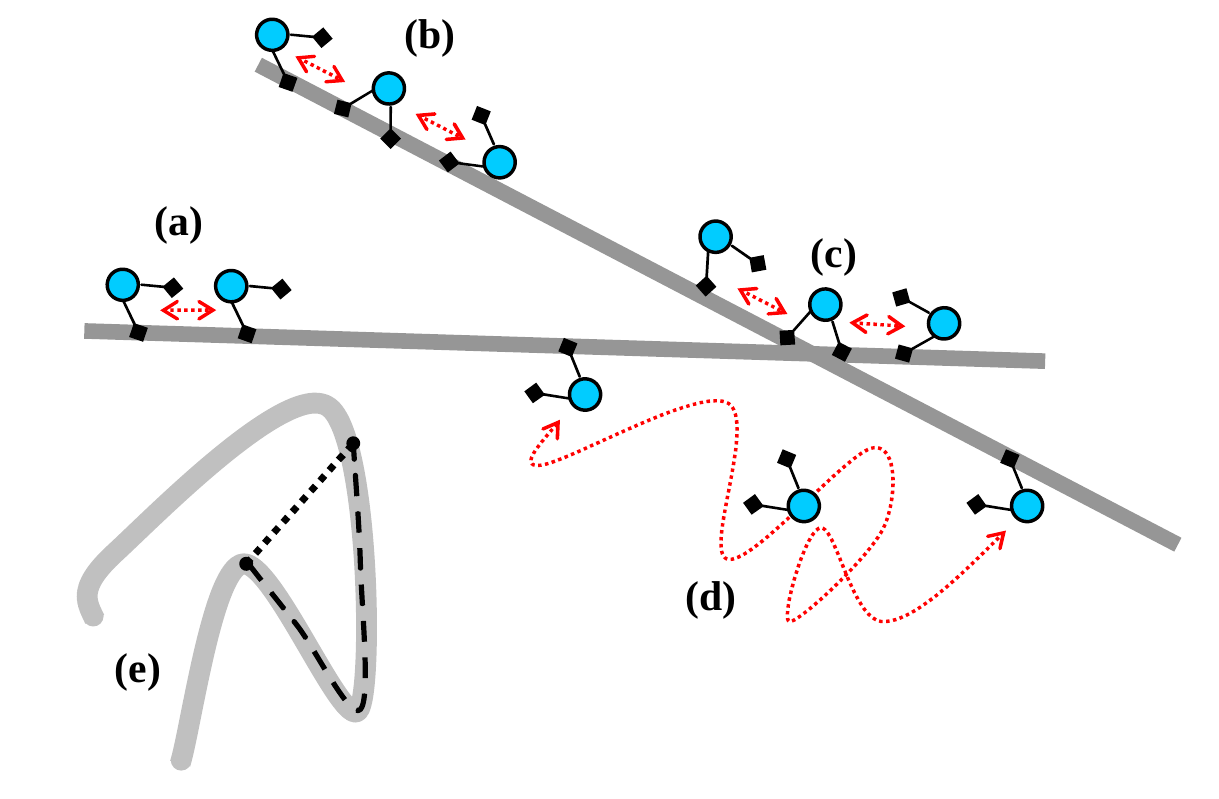}
\par\end{centering}

\caption{Schematic plots illustrating the different mechanisms that can participate
in the facilitated diffusion process. Here dashed arrows represent
different protein moves, the solid curve represents the DNA and a
small circle with two legs indicates a protein with two binding domains.
The figure shows (a) sliding, (b) a correlated intersegmental transfer,
(c) an uncorrelated intersegmental transfer, (d) jumping. (e) The dashed (dotted) line represents a one-dimensional (three-dimensional)
distance.}

\label{fig1ChIT}
\end{figure}

Motivated by experiments \cite{RSB70,RBC70}  an extension of the
Delbruck and Adam model was suggested in \cite{BWH81}. The model
combines one-dimensional diffusion (sliding) along the DNA which
is interrupted by periods of three dimensional diffusion
(typically called jumping or hopping in this context). This
combined strategy, called facilitated diffusion, has been studied
and debated extensively both in the context of \textit{in vivo}
\cite{BWH81,SM2004,HGS2006,HS2007,ELX2007} and \textit{in vitro} systems
\cite{BB76,BWH81,Terry85,HM2004,BZ2004,Z2005,LAM2005,HGS2006,Blainey2006}.
There is now a large body of evidence that such a mechanism plays
an important role for several TFs. It is illustrated in Fig.
\ref{fig1ChIT} and is believed to speed the search process.

Each of the individual search mechanisms described above, when applied alone, has shortcomings
and advantages over the other. When using only three-dimensional diffusion,
the number of \textit{distinct} three dimensional positions probed grows
linearly in time but the protein spends much time probing sites where
there is no DNA present. In contrast, during a one-dimensional diffusion
the protein is constantly bound to the DNA but suffers from a slow
increase in the number of \textit{distinct} positions probed as a function
of time ($\sim t^{1/2}$, where $t$ denotes time) \cite{H95Volume1}.
It is known that by intertwining one and three dimensional search strategies
and \textit{tuning the properties of both} one can in fact decrease
the search time significantly \cite{BWH81}. %

The discussion below follows Refs. \cite{HM2004} and \cite{SM2004}
closely. We imagine a single protein searching for a single target
located on the DNA. The search is composed of a series of
intervals of one-dimensional diffusion along the DNA (sliding) and
three-dimensional diffusion in the solution (jumping). The mean
time of each is denoted by $\tau_{1}$ and $\tau_{3}$ respectively.
Following a jump, the protein is assumed to associate on a new
randomly chosen location along the DNA. Note that one might be
worried if the structure of the packed DNA molecule invalidates
this approach. Numerics on typical frozen DNA conformations
indicate that as long as average search times are considered the
structure can be ignored \cite{SK2009}. Nonetheless much more
complicated structures  may arise in nature (for
example, in eukaryotic cells \cite{crumpleddna,HGS2006,Bancaud2009,Lieberman-Aiden:2009fk}) and these are ignored in the
discussion below.

Under the above assumptions, during each sliding event the protein
covers a typical length $l$, where $l\sim\sqrt{D_{1}\tau_{1}}$
(often called the \textit{antenna} size) \cite{H95Volume1}. To
complete the search process
\begin{equation}
N_{r}\sim\frac{L}{l}\label{NrGen}
\end{equation}
rounds of sliding and jumping are needed on average. While
this can be intuitively understood since the correlations between
the locations of the protein before and after the jump are neglected
the exact nature of the relation is in fact somewhat more subtle.
As shown in \cite{CBVM2004} the average length scanned before the
target is reached is half the total length. Nonetheless for the average
search time the expression is exact in the large $L$ limit. The
total time needed to find a specific site is then:
\begin{equation}
t^{search}=N_{r}\tau_{r},\label{SM}
\end{equation}
 where $\tau_{r}=\tau_{1}+\tau_{3}$ is the typical time of a round. Using
Eqs. (\ref{NrGen}) and (\ref{SM}) one obtains
\begin{equation}
t^{search}\sim\frac{L}{l}(\tau_{1}+\tau_{3})\sim\frac{L}{\sqrt{2D_{1}}}\left(\sqrt{\tau_{1}}+\frac{\tau_{3}}{\sqrt{\tau_{1}}}\right).\label{ts1}
\end{equation}
Furthermore, from dimensional analysis it is easy to argue that
\begin{equation}
\tau_{3}\sim\frac{\Lambda^{3}}{D_{3}L}.\label{t3}
\end{equation}
As shown in \cite{nature2007} this result holds up to a logarithmic correction
which diverges as the DNAs cross section, $\rho$, vanishes. The analysis leads to three
distinct regimes (i) For $\tau_{1}\ll\tau_{3}$ there is no dependence
on $L$ and the search time is given to a good approximation by Eq.
(\ref{t3D}). (ii) For $\frac{L^{2}}{D_{1}}\gg\tau_{1}\gg\tau_{3}$
the dependence on the DNA\ length is linear. This is the regime typically
considered relevant for experiments. (iii) For $\frac{L^{2}}{D_{1}}\ll\tau_{1}$
one finds $t^{search}\propto L^{2}$.

It is natural to ask which $\tau_{1}$ optimizes $t^{search}$ when $\tau_3$ is held fixed. Using
Eq. (\ref{ts1}) it is easy to verify that
\begin{equation}
\tau_{1}^{opt}=\tau_{3}\label{t1opt0} \;.
\end{equation}
It can be shown that this result is exact in the large $L$ limit \cite{CBVM2004}. Alternatively, one
can consider an optimal antenna size $l_{opt}=\sqrt{2D_{1}\tau_{3}}$.
When this condition is met, the total search time scales as
\begin{equation}
t_{opt}^{search}=\sqrt{\frac{\tau_{3}}{D_{1}}}L\sim\sqrt{\frac{\Lambda^{3}L}{D_{1}D_{3}}}.\label{tSM}
\end{equation}
Note that the $\sqrt{L}$ dependence is obtained by optimizing, say
$\tau_{1}$, as $L$ is varied.
This model, \textit{at the optimal} $\tau_{1}$ and assuming known
values for $D_{1}$, $L$ and $\tau_{3}$, predicts reasonable search
times \textit{in vivo} and is commonly believed to give a possible
explanation for the efficiency of the target location process in
experiments.

The combined strategy, while better than the pure three-dimensional
or one-dimensional search strategies, comes at a cost of being sensitive
to changes in the properties of either the three-dimensional or the
one-dimensional diffusive processes. Given the many constraints on the
protein to function, it is restrictive to demand an optimization
of the search process. Specifically, within the model an optimal search process
requires fine tuning of the antenna size, $l$, as a function of the
parameters $D_{1}$ and $\tau_{3}$. These parameters depend on various
cell and environmental conditions such as the size of the cell, the
DNA length, the ionic strength etc. The dependence can be quite significant:
for example, the parameter $\tau_{3}/\tau_{1}$ has been argued to have an exponential
dependence on the square root of the ionic strength \cite{LR75}.
Deviations of this parameter from the optimum value might be crucial
to the search time since $\frac{t^{search}}{t_{opt}^{search}}=\frac{1}{2}\left(\sqrt{\frac{\tau_{3}}{\tau_{1}}}+\sqrt{\frac{\tau_{1}}{\tau_{3}}}\right)$.
Indeed, a strong dependence of the search time on the ionic strength
was found in \textit{in vitro} experiments \cite{RBC70}.

Interestingly, \textit{in vivo}, when the DNA is densely packed, no
effect of the ionic strength on the efficiency of the Lac repressor
was revealed \cite{RCMKAFR87}. Other experiments also suggest that
$\tau_{1}$ is not optimized. In particular, equilibrium measurements
\cite{K-HRBOCNVH77}, as well as recent single molecule experiments
\cite{WAC2006,ELX2007}, find a value of $\tau_{1}$ for the
Lac repressor that is much larger than the predicted optimum $\tau_{3}$
\textit{in vivo}.
The lack of sensitivity to the ionic strength \textit{in vivo} and
the rapid search times found for the Lac repressor, even with very
large values of $\tau_{1}$, suggest that other processes, apart
from jumping and sliding, are involved in the search process.
These seem to be more important \textit{in vivo} than \textit{in
vitro}.
One such mechanism which was suggested to speed the search
time is intersegmental transfers (IT) \cite{HRGW75,BC75}. During
an IT the protein moves from one site to another by transiently
binding both at the same time. This mechanism is expected to be important for systems with a high DNA density \cite{Broek2008kl}.
In principle the new site can be
either close along the one-dimensional DNA sequence (or chemical
distance) or distant (see Fig. \ref{fig1ChIT}). An analysis shows that the
average search time remains similar to the combined one-dimensional and
three-dimensional diffusion described above but with $\tau_3$
which obtains a different dependence on the DNA length. This has
been discussed in detail in \cite{SK2009}.

Finally, we comment that in principle all search strategies can be
made arbitrarily fast by increasing the number of searchers. This
allows, in principle, any of the above discussed mechanisms, one-dimensional
diffusion, three-dimensional diffusion, facilitated diffusion with
or without intersegmental transfers, to be at work for different proteins \cite{Halford2009}. This statement, however, becomes more problematic when the
stability requirements discussed above are included. As stated above, it is clear that the TFs also interact
with non-target sites so that pure three dimensional searches are unlikely. This implies that facilitated diffusion is
hard to avoid. To this end, in what follows we analyze in detail the problems which arise when facilitated diffusion is combined with
the stability requirements.

\section{The speed-stability paradox and possible solutions}

It has been recognized early that there is a tight connection and
antagonism between the stability of the TF at the target and the search
speed. The conflict is commonly termed the speed-stability paradox
\cite{WBH81,SM2004}. As we have seen, the experimental data shows
that the proteins can be classified into several classes. These classes
call for different mechanisms. We begin by introducing the speed-stability
paradox and then discuss possible solutions for each class of proteins.

\subsection{The speed-stability paradox\label{The speed-stability paradox}}

Recall that a fast search of one protein (we later return to the
case of $n_{p}$ proteins and discuss it in detail) requires a fast one-dimensional
diffusion on the DNA. Then note that the binding energies between the transcription
factor and the DNA on each site $i$, $U_{i}$, is an independent
random variable with a Gaussian distribution with a variance $\sigma_{U}^{2}$.
This disorder in the binding energies of the protein to different sites
implies that this diffusion takes place on a disordered potential. On long times this leads
to an effective diffusion constant whose value is given by \cite{SM2004}
\begin{equation}
D_{1}=D_{1}\left(\sigma_{U}=0\right)\sqrt{1+\frac{\sigma_{U}^{2}}{2}}e^{-\frac{7}{4}\sigma_{U}^{2}}.\label{1ddiffdis}
\end{equation}
The important thing to note is the exponential dependence of the
diffusion coefficient on $\sigma_{U}^{2}$. It can be understood
up to prefactors by recalling that the diffusion is an activated
process so that $D_{1}\propto\int dUe^{-U}Pr(U)$ . This implies that
even for $\sigma_{U}=5.5$ (the boarder line between the small and the large disorder regimes
for an \textit{E. coli} genome) the one-dimensional diffusion constant
becomes $19$ orders of magnitude smaller than the diffusion constant
on a flat energy landscape. This in turn leads to a very slow search process. Essentially,
speed requirements prohibit the large disorder regime discussed above.
For the search to be fast $\sigma_{U}$ has to be kept small, of the
order of $0.5$, to ensure a diffusion coefficient of the same order
as that on a flat energy landscape.

On the other hand, this requirement conflicts with the stability requirements
for proteins which demand (see section \ref{sec:targetoccupation}) either a large value of $\sigma_{U}$ (for marginally gapped and gapped TFs), or a large TF length $l_p$
(for gapped TFs in the small disorder regime), to create
a significant gap between the energy at the target and the rest of
the DNA. From the analysis above, a priori only gapped TFs might satisfy both speed and stability requirements. Below we analyze in detail the speed and stability requirements for gapped and marginally gapped TFs and discuss  possible solutions of the paradox.

More puzzling are non-gapped proteins which are unstable at the target. A new possible mechanism which ensures
both speed and stability for those is discussed in later sections.

\subsection{Possible solutions of the speed-stability paradox for gapped, marginally gapped and non-gapped TFs.}

\subsubsection{Gapped TFs}

In principle both speed and stability requirements can be easily satisfied
using, for example, cooperative interactions on the target sequence and small $\sigma_{U}$
for the rest of the sequence. In this case, when the TF is clearly gapped, the binding energy at the
target can be made arbitrarily low without affecting the value $\sigma_{U}$.

However, as stated above the experimental data seem to suggest no
significant cooperative effect such that $U_{{\cal T}}=l_{p}E_{c}$
where as above $l_{p}$ is the length of the target and $E_{c}$ is
the average (over different binding sites of the protein) minimal
binding energy. Since $E_{c}$ depends on $\sigma_{U}$ it is not
clear how both requirements on speed and stability can be satisfied.
As argued above the speed requirement demands (see Eq. \ref{1ddiffdis})
\begin{equation}
e^{-\frac{7}{4}\sigma_{U}^{2}}\sim1\;,\label{Speed}
\end{equation}
which prohibits $\sigma_U$ to be in the large disorder regime. Following the discussion above this rules out the stability of marginally gapped and non gapped targets.
As before we assume that the probability of a mismatch is $3/4$ and using our
convention that $\langle U\rangle=0$ we have
$\sigma_{U}^{2}=\frac{l_{p}E_{c}^{2}}{3}$. The stability requirement in the small disorder regime is, using Eq. (\ref{Zapprox}),
\begin{equation}
e^{-l_{p}E_{c}}\simeq Ne^{\frac{\sigma_{U}^{2}}{2}}=Ne^{\frac{l_{p}E_{c}^{2}}{6}}.
\end{equation}
Thus, to ensure stability we need $E_{c}<3\left(\sqrt{1-\frac{2\ln N}{3l_{p}}}-1\right)$. Note that
a solution for $E_{c}$ exist only when $l_{p}\geq\frac{2}{3}\ln N$
-- as expected the target length has to be large enough to ensure
stability. This can be re-expressed in terms of the variance to read
\begin{equation}
\sigma_{U}^{2}\gtrsim3l_{p}\left(\sqrt{1-\frac{2\ln N}{3l_{p}}}-1\right)^{2}.
\end{equation}
 For $N\sim10^{7}$ one may check that  both the stability
criterion and the speed requirement, Eq. (\ref{Speed}), can only
be met for $l_{p}>70$. This argument suggests that both speed and
stability requirements demand a very large target size. The database
studied above does not contain proteins of that size
\footnote{However, during the process of the homologous recognition the length of
the searcher and the target may be much larger \cite{Kupiec2008}}.

\subsubsection{Two-state models\label{Two-state models}}

The previous solution relies on having a gapped TF. Another possible resolution of the speed-stability paradox, which applies also to marginally gapped TFs, lies in introducing another
conformation of the DNA-TF complex. This conformation, usually attributed to non-specific binding,
modifies the properties of the energy landscape experienced by the protein during its
one dimensional diffusion. Specifically, it was suggested \cite{WBH81,GMH2002,SM2004}
that another conformation may introduce an effective cutoff on the TF-DNA binding energy distribution
which will lower its variance and hence lead to a quick one-dimensional diffusion thus resolving the
speed-stability paradox.

The two-state model assumes that the protein (or protein-DNA complex) switches rapidly between its two conformations
so that the two can be assumed to be equilibrated. We assume that
in the first non-specific conformation the protein has a constant binding
energy, $U_{ns}$, on all sites and that in the second, specific conformation
the binding energy, $U_{i}$, is, as before, an independent random
variable with a Gaussian distribution (\ref{Pr(U)}) with a variance
$\sigma_{U}^{2}$.  The total free energy on site $i$ is then  given by
\cite{GMH2002}
\begin{equation}
G_{i}=-\ln\left(e^{-U_{ns}}+e^{-U_{i}}\right)\simeq\min\left(U_{ns},U_{i}\right) \;.
\end{equation}
Therefore the probability distribution of the total free energy has a cutoff as discussed above and is
given by
\begin{equation}
\Pr(G_{i})\simeq\left\{ \begin{array}{cc}
\frac{1}{\sqrt{2\pi\sigma_{U}^{2}}}e^{-\frac{G_{i}^{2}}{2\sigma_{U}^{2}}} & G_{i}<U_{ns}\\
0 & G_{i}\geq U_{ns}\end{array}\right.+\delta\left(G_{i}-U_{ns}\right)\int_{U_{ns}}^{\infty}\frac{e^{-\frac{G^{2}}{2\sigma_{U}^{2}}}}{\sqrt{2\pi\sigma_{U}^{2}}}dG.
\end{equation}
Clearly by tuning the value of $U_{ns}$ the resulting free energy
landscape can be made flat on most of the DNA allowing for a fast one-dimensional
diffusion. This happens roughly when the protein is mostly in the non-specific conformation, which yields a first constraint:
\begin{equation}
e^{-U_{ns}}>\left\langle e^{-U_{i}}\right\rangle =e^{\frac{\sigma_{U}^{2}}{2}}.\label{Gerland1}
\end{equation}
This procedure can not be carried out in an arbitrary manner as very low values of $U_{ns}$ might destroy the stability
of the target. To avoid this
the non-specific energy $U_{ns}$ also has to obey a second constraint
\begin{equation}
Ne^{-U_{ns}}<e^{-U_{\mathcal{T}}},\label{Gerland2}
\end{equation}
where $U_{\mathcal{T}}$, as before, is the target binding energy.

For marginally-gapped TFs these conditions are very restrictive and demand fine tuning:
\begin{equation}
\sigma_U \simeq \sqrt{2 \ln N} \;.
\end{equation}
Clearly, most marginally gapped TFs do not satisfy this constraint (see Fig. \ref{EminPlot}).

For gapped-TFs the constraint
is not as severe. It is easy to see that here we need
\begin{equation}
-U_{\cal T}  > \sigma_U^2/2 + \ln N \;.
\label{eq:crit}
\end{equation}
Within the additive binding energy model this implies that $l_p \geq c \ln N$ where $c$ is a constant which depend on $E_c$.
It is interesting to check this criterion, which is actually the same as demanding stability of a one state TF in the small disorder regime (see Eq. \ref{smalldisorder}) using the protein weight matrices. The results are shown in Fig. \ref{criter}. Note that more than
half the proteins do not satisfy the criterion. For these the two state model presented above does not seem to apply.

This gives a clear condition on when this
mechanism alone is sufficient to resolve the speed stability paradox.
Recently, such a cutoff was measured in a eukaryotic transcription
factor \cite{Quake2007}. However, the nonspecific binding energy
was estimated to be only $5k_{B}T$ larger than the target energy.
The fraction of the time that the protein would spend on the target
having such a small energy gap is of the order of $e^5/N$.
In a eukaryotic cell, where one typically has $N\sim10^{10}$, such a gap looks
inexplicably low.

Finally, we comment that all the above considerations ignore the free-energy associated with
the protein being off the DNA. At the optimal antenna size (see Sec. \ref{facdif})  this has the same contribution
as the non-specifically bound state. As mentioned above, it can only reduce the stability on the target.

\subsubsection{Multiple TFs}

In this subsection we assume a single state TF model and comment on the applicability
of the results to two-state models at the end. An easy resolution to the slow search speed is increasing the copy
number of each TF.  When $n_p$ TFs are searching for the target the mean search time is reduced by a factor
of $n_{p}$ (for cases where this does not apply see Sec. V). For facilitated diffusion the disorder increases
the mean search time by a factor of about $e^{\sigma_{U}^{2}/2}$. Therefore, $n_p \simeq e^{\sigma_{U}^{2}/2}$ TFs
can compensate for the effects of the disorder.

Multiple TFs may also fulfill the stability requirements
since the occupation probability of the target increases with $n_{p}$.
Ignoring interaction between TFs copies the occupation probability
of the target, namely the probability to find  at least one TF at the target,   is given by \begin{equation}
P^{{\cal \mathcal{T}}}\left(n_{p}\right)=1-\left[1-P^{{\cal \mathcal{T}}}\left(n_{p}=1\right)\right]^{n_{p}}.\end{equation}
Namely, by taking $n_{p}$ to be larger than $1/P^{{\cal \mathcal{T}}}\left(n_{p}=1\right)$
the occupation probability of the target becomes of order of one,
so that one satisfies the stability requirement.

However, there is a worry that by increasing the number of TF copies
the specificity will be reduced. Namely, the TFs will activate or
repress other genes by tightly binding to unwanted sequences on the DNA.
Consider the case when in addition to the target there are $N_{d}$
{}``dangerous'' sites. A site is defined as dangerous if a significant
occupation probability of this site affects the transcription of a
non-target gene. It seems reasonable to take $N_{d}$ to be of
the same order of magnitude as the total length (in base-pairs) of
all DNA promoters. We assume that the binding energy distribution
of the dangerous part of the DNA is the same as the rest of the DNA.

The lowest energy among $N_{d}$ dangerous sites on a typical
sequence is given by $-\sigma_{U}\sqrt{2\ln N_{d}}$ so that the occupation
probability of the most occupied dangerous site, for a one TF case,
is given by\begin{equation}
P_{d}\left(n_{p}=1\right)\simeq\frac{e^{\sigma_{U}\sqrt{2\ln N_{d}}}}{\overset{N}{\underset{i=1}{\sum}}e^{-U_{i}}}\label{Pd}\end{equation}
 while in the case of $n_{p}$ proteins the occupation probability
of the most occupied dangerous site is
\begin{equation}
P_{d}\left(n_{p}\right)=1-\left[1-P_{d}\left(n_{p}=1\right)\right]^{n_{p}}.\label{Pdnp}
\end{equation}
Thus to ensure that the dangerous part is not significantly occupied
$n_{p}$ has to be much smaller than $1/P_{d}(n_p=1)$.

In sum, three conditions limit the possible value of the TF copy number.

\noindent \textit{Rapid search (speed):}
\begin{equation}
n_{p}\gg e^{\sigma_{U}^{2}}.\end{equation}
 \textit{Significant occupation probability of the target (stability):}
\begin{equation}
n_{p}\gg\frac{1}{P^{{\cal \mathcal{T}}}\left(n_{p}=1\right)}.\end{equation}
 \textit{Small occupation probability of a dangerous site (specificity):}
\begin{equation}
n_{p}\ll\frac{1}{P_{d}\left(n_{p}=1\right)}.\end{equation}
Of course, these ignore the obvious but hard to quantify, cost involved in the production of the TFs.

In Appendix \ref{specificity} we analyze these conditions in detail. We find that a large copy number can resolve
the speed and stability issues without affecting specificity only in the small disorder regime. Finally, we comment that
similar considerations hold also for the two state model discussed above. In that case the specificity condition becomes more stringent for
large $N_d$ while for small $N_d$ they are less stringent. Moreover, one can check that adding $n_p$ TFs only eases the criterion given in
Eq. \ref{eq:crit} by a $\ln n_p$ term (where we assume that the criterion on the speed in the two state model is unchanged).
This can help the search process only for very large $n_p$ values.

\section{Search and recognition based on a barrier discrimination -- effects
of multiple time scales}\label{Search and recognition based on a barrier discrimination -- effects
of multiple time scales}

In the previous sections we saw that many proteins
have a low occupation probability at the target. Moreover,
demanding that the protein reaches the target quickly posed many
more constraints on, for example, the length of the protein
recognition site and the number of proteins searching for the
target. For about ten percent of the proteins the problem seems
particularly severe. Their occupation probability is so low that
they demand thousands of TFs for a high occupation probability.
In what follows we suggest a new mechanism which, in principle,
may apply to any of the classes above. In particular in the next
section we show how it applies even to TFs with a very low
occupation probability through what we call transient stability.

The model assumes that the protein-DNA complex can assume two
conformations. Namely, when the protein is bound
to the DNA, it can switch between two conformations separated by a free
energy barrier. A closely related model was introduced in previous
works in order to solve the speed-stability paradox (see Section \ref{Two-state models}
and Refs. \cite{GMH2002,SM2004,HGB2008}). In these works
the barrier between the two states of the protein was assumed to be low enough
so that the two conformations were equilibrated with each other. Moreover,
the barrier was assumed to be a constant for all DNA sites. A two state structure was demonstrated experimentally in transcription
factors \cite{Ferreiro2003,KBBLGBK2004,PW2007,TSBXA2007} and type
II restriction endonucleases (for a review see Ref.
\cite{Pingoud2001}). Furthermore, there are simple theoretical
arguments for their existence \cite{DPJBV2009}.

In contrast to the
two state model discussed above and in previous studies, here an important role is played
by  a difference in the association rates to different DNA sites.
As we show, and intuitively clear, this may supply an additional discriminating factor. Some evidence for the
possible importance of association rates is found in the purine repressor.
There it was shown
that, when activated, it changes its association
rate to the target by two orders of magnitude while the
dissociation rate changes only by one order of magnitude
\cite{Xu1998}. Note, that the
association rate may be very large with a small binding energy and
vice-versa. Differences between association rates to different DNA
sites were also observed in Refs. \cite{Shultzaberger2007} and
\cite{Ferreiro2003}. Interestingly in the latter work association
rates which correspond to very high energetic barriers of tens of
$k_BT$ were observed, albeit for eukaryotic cells.

\begin{figure}[ptb]

\begin{centering}
\includegraphics[width=8cm]{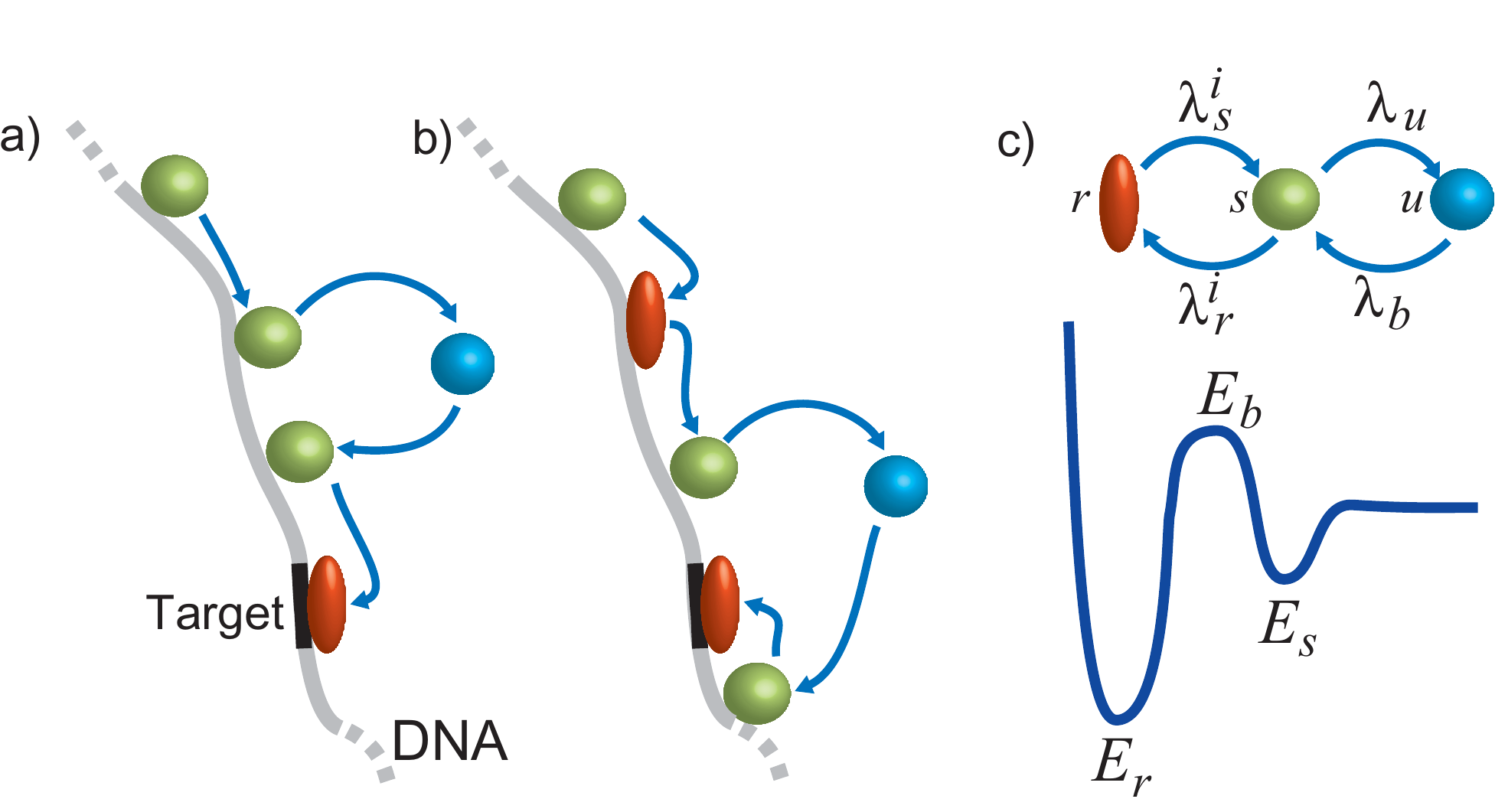}
\par\end{centering}

\caption{An illustration of the two-state model described in Sec. \ref{Search and recognition based on a barrier discrimination -- effects
of multiple time scales}. (a) A time sequence of a protein sliding
in the $s$ mode (green circle), diffusing off the DNA (blue circle)
and entering the target site in the $r$ mode (red oval). (b) A protein
finding the target after entering the $r$ state. (c) An illustration
of the rates and the energy landscape which governs them at each location,
$i=1,...,N$, along the DNA. Here $\lambda_{r}^{i}\propto e^{-(E_{b}^{i}-E_{s}^{i})/k_{B}T}$,
$\lambda_{s}^{i}\propto e^{-(E_{b}^{i}-E_{r}^{i})/k_{B}T}$ and $\lambda_{u}\propto e^{-E_{s}^{i}/k_{B}T}$,
while $\lambda_{b}$ depends on details of the three-dimensional diffusion
process.}

\label{fig1ChBarrier}
\end{figure}

Assuming two conformations, we call one the \textit{search state}.
In this conformation the protein is loosely bound to the DNA and can slide along it. In
the second, \textit{recognition state}, it is trapped in a deep
energetic well (see Fig. \ref{fig1ChBarrier}). Note that equilibrium
measurements of binding energies to the DNA are controlled by the
recognition state. To make the discussion clear, below we analyze
search processes where the recognition is only based on barrier
discrimination. This implies that equilibrium properties of the target
site are identical to those of non-target sites.

Based on a quantitative analysis of this model, we argue that due
to the occurrence of several time scales in the search process the
widely used definition of the reaction rate of a single protein as
the inverse of the average search time $t^{ave}$ \cite{HANGGI1990},
is generally irrelevant as a measure of the efficiency of target location
on DNA. When $n_{p}$ proteins are searching for the target, the relevant
quantity is the probability $\mathcal{R}_{n_{p}}(t)$ for a reaction
to occur before time $t$. We show below that $\mathcal{R}_{n_{p}}(t)$
can reach values close to one on a time scale $t_{n_{p}}^{typ}$ which
can be orders of magnitude smaller than the average search time, $t_{n_{p}}^{ave}$.
Both the typical and the average search times can be orders of magnitude
smaller than the naive approach based on a one time scale assumption
which gives $t^{ave}/n_{p}$.

Our analysis has several important merits. First, it reports a \textit{fast}
search time despite a very strong binding of the protein in the recognition
state to \textit{any site} on the DNA. This renders the question of
stability in the recognition state irrelevant. We suggest that within this model the
measured binding energies of proteins to the DNA are irrelevant to
the kinetics of the search process; the relevant quantities are transition
rates (specified below). Second, we show that with a proper
choice of parameters one may solve the speed-stability paradox without designing the target.
We make two comments. (i) While there is no equilibrium stability within this
model it will be shown that the protein is present on the target site for an extended period of time. (ii) Within this model
the kinetics are independent of the equilibrium properties. Therefore,
it is straightforward to add equilibrium stability within it.

The model consists of $n_{p}$ proteins which can each be in three
states : (i) an unbound state, $u$, in which it performs three-dimensional
diffusion (jumping), (ii) a search state, $s$, where it is weakly
bound to the DNA, performing one-dimensional diffusion (sliding) and
(iii) a recognition state, $r$, where it is tightly bound to the
DNA\footnote{In the language of enzyme-ligand interactions, the discussed model
of the protein-DNA binding has an induced fit mechanism \cite{K58}.
}. We assume, for simplicity, that in the recognition state the protein
is trapped in a deep energy well (as justified by the experimentally
measured strong binding energies) and is unable to move \cite{SM2004}.
The transition rates, $\lambda_{s}^{i}$, $\lambda_{r}^{i}$, $\lambda_{b}$
and $\lambda_{u}$, between the different states are defined in Fig.
\ref{fig1ChBarrier}. To model sliding, in the $s$ state the protein
can move with transition rate $\lambda_{0}/2$ to neighboring sites
on the DNA. Note that the transition rates $\lambda_{r}^{i}$ and
$\lambda_{s}^{i}$ are expected in general to depend on the location
$i=1\ldots N$ along the DNA. In principle $\lambda_{0}$ and $\lambda_{u}$
also have a dependence on $i$. As justified later this will have
a weaker effect on our results and we omit it for clarity. Finally,
after a jump we assume that the protein relocates to a random position
on the DNA due to its packed conformation \cite{SK2009}.

The presentation of the model gives many details of the derivations of the results.
However, we have made an effort to end each subsection with a highlight
of the main results. Furthermore, some subsection focus only on results.

\subsection{Non disordered case\label{Non disordered case}}

\begin{figure}[ptb]

\begin{centering}
\includegraphics[width=14cm]{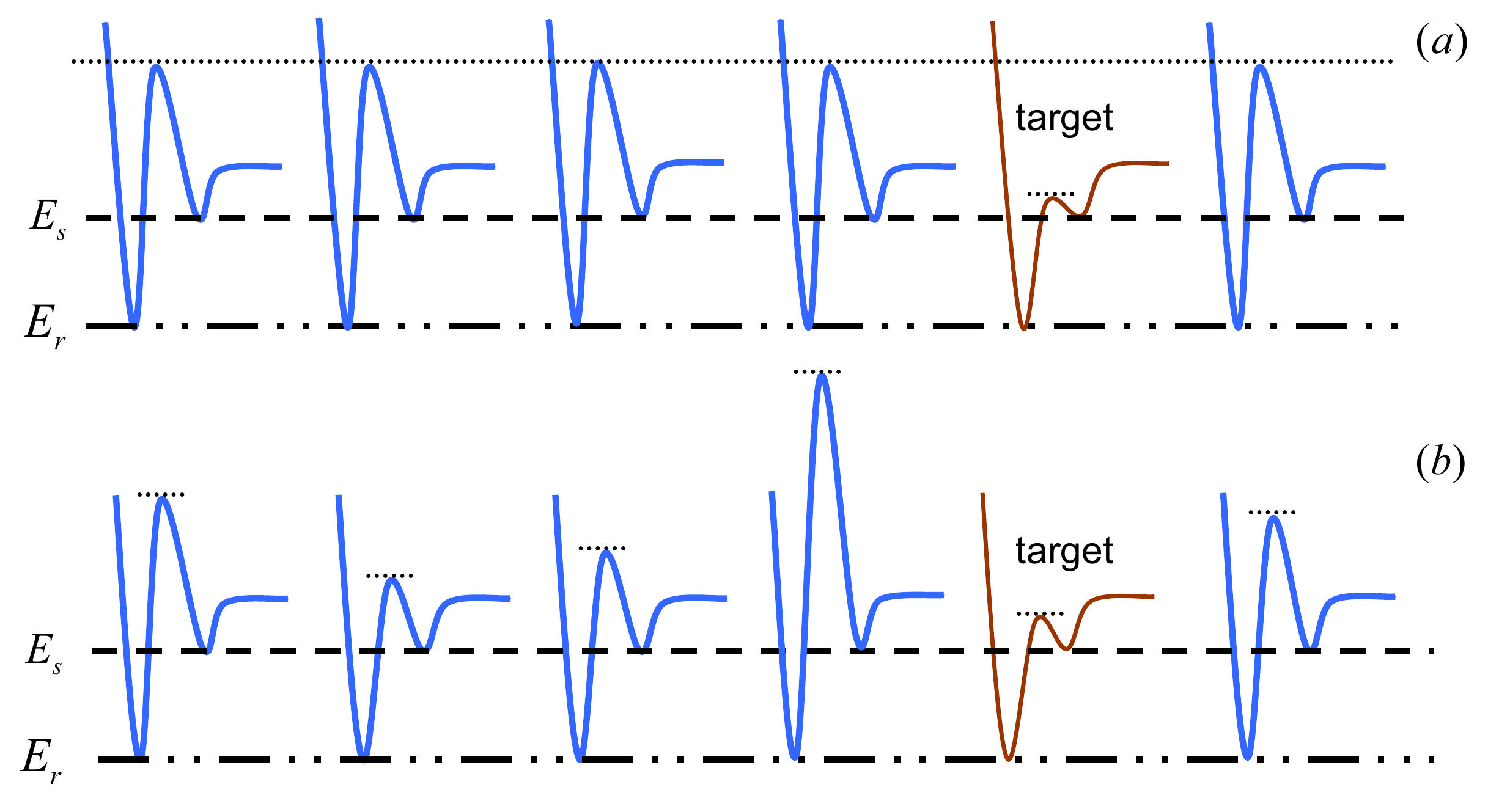}
\par\end{centering}

\caption{An illustration of the free energy landscape used in non-disordered
and disordered models. (a) Each free energy profile represents a site
on the DNA. All sites have the same profile except for the target
that is designed to have smaller barrier. (b) Each free energy
profile represents a site on the DNA. The energies of the $s$ and $r$
modes are fixed for all sites including the target. The barrier height
is drawn from a Gaussian distribution while the target is defined
as the site with the smallest barrier.}

\label{LanscapeChBarrier}
\end{figure}

To gain an understanding of the difference between the two time scales
$t_{n_{p}}^{typ}$, $t_{n_{p}}^{ave}$ \ and the naive estimation
$t^{ave}/n_{p}$\ we first consider a single searcher, $n_{p}=1$, in a simplified
model where the transition rates $\lambda_{r}^{i}=\lambda_{r}$ and
$\lambda_{s}^{i}=\lambda_{s}$ are independent of $i$ except at the
target site $\mathcal{T}$ (see Fig. \ref{LanscapeChBarrier}$(a)$).
The target site in this section is \textit{designed} such that the
transition rates on the target are different from the transition rates
on the rest of DNA. At the target site the transition rate from the $s$
state to the $r$ state is denoted by $\lambda_{r}^{\mathcal{T}}$ and
the transition rate from the $r$ state to the $s$ state is denoted by $\lambda_{s}^{\mathcal{T}}$
($\lambda_{s}^{\mathcal{T}}$ \ is irrelevant for the calculation
of the first-passage time properties). As stated above, in our considerations
we analyze a process of search and recognition based \textit{only}
on a barrier discrimination and $P^{\cal T}=1/N$.
Therefore, the relation
\begin{equation}
\frac{\lambda_{r}}{\lambda_{s}}=\frac{\lambda_{r}^{\mathcal{T}}}{\lambda_{s}^{\mathcal{T}}}
\end{equation}
holds.

To analyze the model we first consider the probability
\begin{equation}
\mathcal{R}(t)=\int_{0}^{t}P(t^{\prime})dt^{\prime}\label{RvsP}
\end{equation}
that the protein finds its target before time $t$, where $P(t)$
is the distribution of the first-passage time (FPT) \cite{R2001}
to the target (we drop the subscript when $n_{p}=1$).
The Laplace transform, \begin{equation}
{\tilde{P}}(s)=\int_{0}^{\infty}e^{-st}P(t)dt,\end{equation}
 of $P(t)$ can be obtained exactly. For simplicity we take a centered
target site (labeled $0$). Consider, first, the joint probability
density for a protein to find the target (in its $r$ state) at time
$t=t_{s}+t_{r}$ starting from a location $x_{0}$ at $t=0$ before
unbinding from the DNA. Here $t_{s}$ is the total time spent in the $s$
state and $t_{r}$ is the total time spent in the $r$ state. The
probability that exactly $n$ transitions occurred from the $s$ state
to the $r$ state is given by $\mathrm{\mathcal{P}}\left(n,\lambda_{r},t_{s}\right)$
where \begin{equation}
\mathrm{\mathcal{P}}(n,\mu,t)=(\mu t)^{n}e^{-\mu t}/n!\end{equation}
 is a Poisson distribution. The probability to spent a time $t_{r}$
in the $r$ state given that $n$ transitions occurred from the $s$
state to the $r$ state is $\lambda_{s}\mathrm{\mathcal{P}}\left(n-1,\lambda_{s},t_{r}\right)$
(with the convention $\mathrm{\mathcal{P}}(-1,\mu,t)\equiv\delta(t)/\mu$).
The probability to stay on the DNA up to time $t=t_{s}+t_{r}$ starting
at $t=0$ is given by $e^{-\lambda_{u}t_{s}}$. Finally, the probability
to cross the barrier at the target at each visit of its $s$ state
is given by\begin{equation}
p_{1}=\frac{\lambda_{r}^{\mathcal{T}}}{1+\lambda_{u}/\lambda_{0}+\lambda_{r}^{\mathcal{T}}/\lambda_{0}}.\end{equation}
 Therefore, the joint probability density for a protein to find the
target at time $t=t_{s}+t_{r}$ starting from a location $x_{0}$
at $t=0$ before unbinding from the DNA is\begin{equation}
P_{n}(t_{s},t_{r}|x_{0})=\lambda_{s}\mathrm{\mathcal{P}}\left(n-1,\lambda_{s},t_{r}\right)\mathrm{\mathcal{P}}\left(n,\lambda_{r},t_{s}\right)j_{p_{1}}\left(t_{s}|x_{0}\right)e^{-\lambda_{u}t_{s}},\label{psd}\end{equation}
 where $j_{p_{1}}\left(t|x_{0}\right)$ is the FPT density at the
target $x=0$ for a usual random walk starting from $x_{0}$ given
that the probability to cross the barrier at the target at each visit
in its $s$ state is $p_{1}$. The FPT density before unbinding starting
from $x_{0}$ then reads:
\begin{equation}
J\left(t|x_{0}\right)=\sum_{n=0}^{\infty}\int_{0}^{\infty}\!\!\int_{0}^{\infty}\!\! dt_{s}dt_{r}\delta\left(t_{s}+t_{r}-t\right)P_{n}\left(t_{s},t_{r}|x_{0}\right).\label{psd2}
\end{equation}
 After a Laplace transform and using \begin{equation}
{\tilde{\mathrm{\mathcal{P}}}}\left(n,\mu,s\right)=\frac{\mu^{n}}{\left(s+\mu\right)^{n+1}},\end{equation}
 we find \begin{equation}
{\tilde{J}}\left(s|x_{0}\right)={\tilde{j}}_{p_{1}}\left(u\left(s\right)|x_{0}\right)\label{Jp1}\end{equation}
 with \begin{equation}
u(s)=\frac{s\left(s+\lambda_{r}+\lambda_{s}+\lambda_{u}\right)+\lambda_{s}\lambda_{u}}{s+\lambda_{s}}.\end{equation}
 Following \cite{CBVM2004,pccp2008} we write the probability to
find the target, $P\left(t\right)$
as\begin{equation}
P\left(t\right)=\left\langle \sum_{m=0}^{\infty}\int_{0}^{\infty}dt_{m}{J}\left(t_{m}|x_{m}\right)\delta\left(t-\sum_{l=1}^{m-1}\left(t_{l}+\tau_{l}\right)-t_{m}\right)\underset{k=1}{\overset{m-1}{{\displaystyle \prod}}}dt_{k}d\tau_{k}\overline{{J}}\left(t_{k}|x_{k}\right)\lambda_{b}e^{-\lambda_{b}\tau_{k}}\right\rangle _{\left\{ x_{k}\right\} }\end{equation}
 where $\left\langle {}\right\rangle _{\left\{ x_{k}\right\} }$ denotes
an average over the DNA binding sites and $\overline{{J}}\left(t|x_{0}\right)$
is the probability to unbind before finding the $r$ state of the
target starting from site $x_{0}$. This is given by\begin{equation}
\overline{{J}}\left(t|x_{0}\right)=\lambda_{u}e^{-\lambda_{u}t}\left(1-\int_{0}^{t}dt^{\prime}J\left(t^{\prime}|x_{0}\right)e^{\lambda_{u}t^{\prime}}\right).\end{equation}
 We assume that each DNA binding event occurs at a random position
on the DNA. Thus\begin{equation}
P\left(t\right)=\sum_{m=0}^{\infty}\int_{0}^{\infty}dt_{m}{J}\left(t_{m}\right)\delta\left(t-\sum_{l=1}^{m-1}\left(t_{l}+\tau_{l}\right)-t_{m}\right)\underset{k=1}{\overset{m-1}{{\displaystyle \prod}}}dt_{k}d\tau_{k}\overline{{J}}\left(t_{k}\right)\lambda_{b}e^{-\lambda_{b}\tau_{k}}\end{equation}
 where ${J}\left(t\right)\equiv\left\langle {J}(t|x_{0})\right\rangle _{x_{0}}$
and $\overline{{J}}\left(t\right)\equiv\left\langle \overline{{J}}(t|x_{0})\right\rangle _{x_{0}}$.
We then obtain the Laplace transformed FPT distribution as\begin{equation}
\tilde{P}(s)={\tilde{J}}\left(s\right)\left[1-\frac{\lambda_{b}\lambda_{u}}{s+\lambda_{b}}\frac{1-{\tilde{J}}\left(s\right)}{u(s)}\right]^{-1}.\end{equation}
 Using Eq. (\ref{Jp1}) and defining $\tilde{j}_{p_{1}}\left(s\right)\equiv\left\langle {\tilde{j}}_{p_{1}}\left(s|x_{0}\right)\right\rangle _{x_{0}}$
one obtains\begin{equation}
\tilde{P}(s)=\tilde{j}_{p_{1}}\left(u\left(s\right)\right)\left[1-\frac{\lambda_{b}\lambda_{u}}{s+\lambda_{b}}\frac{1-\tilde{j}_{p_{1}}\left(u\left(s\right)\right)}{u\left(s\right)}\right]^{-1}.\label{LapTrans}\end{equation}
 Finally, $\tilde{j}_{p_{1}}\left(s\right)$ may be calculated using
that\begin{align}
j_{p_{1}}\left(t\right) & =\left\langle j_{p_{1}}\left(t|x_{0}\right)\right\rangle _{x_{0}}=\left\langle j\left(t|x_{0}\right)\right\rangle _{x_{0}}p_{1}+\left(1-p_{1}\right)\left\langle j\left(t|x_{0}\right)\right\rangle _{x_{0}}\ast j_{0}\left(t\right)p_{1}+\label{jp1Calc}\\
 & +\left(1-p_{1}\right)^{2}\left\langle j\left(t|x_{0}\right)\right\rangle _{x_{0}}\ast j_{0}\left(t\right)\ast j_{0}\left(t\right)p_{1}+...\nonumber \end{align}
 where $j\left(t|x_{0}\right)$ is the FPT density at the target $x=0$
for a usual random walk starting from $x_{0}$ and ${j}_{0}\left(t\right)$
is the generating function of the first return time to site $0$ of
a simple random walk. The $\ast$ symbol denotes a convolution. The Laplace
transform of (\ref{jp1Calc}) gives\begin{equation}
\tilde{j}_{p_{1}}\left(s\right)=\frac{p_{1}{\tilde{j}}\left(s\right)}{1-\left(1-p_{1}\right){\tilde{j}}_{0}\left(s\right)},\end{equation}
 where \begin{equation}
\tilde{j}(s)\equiv\left\langle \tilde{j}\left(s|x_{0}\right)\right\rangle _{x_{0}}\simeq\frac{1}{N}\sqrt{\frac{1+e^{-s/\lambda_{0}}}{1-e^{-s/\lambda_{0}}}}\label{j(s)}\end{equation}
 and \begin{equation}
{\tilde{j}}_{0}\left(s\right)\simeq1-\sqrt{1-e^{-2s/\lambda_{0}}}\label{j0(s)}\end{equation}
 for large $N$ \cite{montroll} (see Appendix \ref{MontrollCalc}
for details).

Applying the Laplace transform to Eq. (\ref{RvsP}) and using Eq.
(\ref{LapTrans}) one obtains\begin{equation}
\widetilde{\mathcal{R}}\left(s\right)=\frac{\tilde{P}(s)}{s}=\frac{\tilde{j}_{p_{1}}\left(u\left(s\right)\right)}{s}\left[1-\frac{\lambda_{b}\lambda_{u}}{s+\lambda_{b}}\frac{1-\tilde{j}_{p_{1}}\left(u\left(s\right)\right)}{u\left(s\right)}\right]^{-1}.\label{RLapTrans}\end{equation}

\subsubsection{Large barrier regime}

By analyzing the pole structure of Eq. (\ref{LapTrans}) (see Appendix
\ref{Poles}) one can show that in the large barrier regime \begin{equation}
\lambda_{s}\ll\lambda_{r}\ll\lambda_{u},\lambda_{b},\lambda_{0}\label{LargeBarrier}\end{equation}
 (with $\lambda_{u},\lambda_{b},\lambda_{0}$ of comparable order)
the reaction probability simplifies to \begin{equation}
\mathcal{R}\left(t\right)\simeq1-qe^{-t/\tau_{1}}-\left(1-q\right)e^{-t/\tau_{2}}\label{2exp}\end{equation}
 with\begin{align}
q & =\frac{1}{1+\frac{\lambda_{r}}{\lambda_{u}\kappa/N}},\\
\kappa & =\frac{\sqrt{\coth\left(\frac{\lambda_{u}}{2\lambda_{0}}\right)}}{1+\frac{1-p_{1}}{p_{1}}\sqrt{1-e^{-2\lambda_{u}/\lambda_{0}}}},\\
\tau_{1} & =\frac{1+\frac{\lambda_{u}}{\lambda_{b}}}{1+\frac{\lambda_{r}}{\lambda_{u}\kappa/N}}\frac{1}{\kappa\lambda_{u}/N}\end{align}
 and\begin{equation}
\tau_{2}=\frac{1}{\lambda_{s}}\left(1+\frac{\lambda_{r}}{\lambda_{u}\kappa/N}\right).\end{equation}
 Eq. (\ref{2exp}) is a central result of this Section. We show below
that a similar two exponents structure appears also in the disordered
case. The short time scale $\tau_{1}$ characterizes searches where
the protein never enters the $r$ state and is therefore independent
of the binding energy $E_{r}$ (and hence of $\lambda_{s}$). The
time scale $\tau_{2}$ characterizes searches where the protein enters
the $r$ state, and is therefore much larger than $\tau_{1}$ in the
case of strong binding ($\lambda_{s}$ small). In turn, $q$ is the
probability of an event where the target is found without falling
into a trap.

\begin{figure}[ptb]

\begin{centering}
\includegraphics[width=13cm]{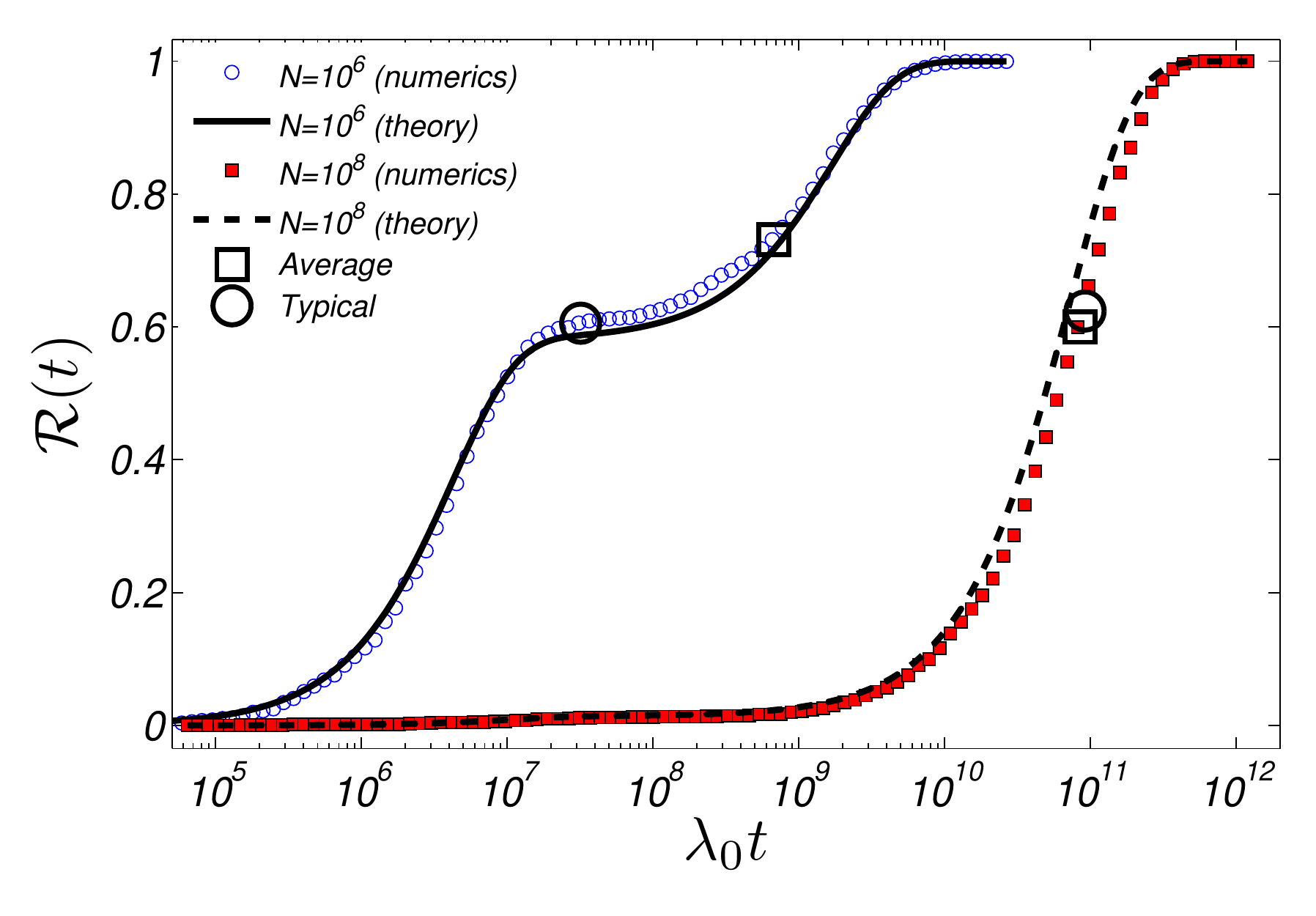}
\par\end{centering}

\caption{A plot of $\mathcal{R}(t)$ for $N=10^{6}$ (empty circles) and $N=10^{8}$
(filled squares) for non disordered case. Lines correspond to Eq.
\ref{2exp}, with $\tau_{1}$, $\tau_{2}$ and $q$ derived analytically.
Here $p_{1}=1$, $\lambda_{u}=10^{-2}\lambda_{0}$, $\lambda_{b}=0.1\lambda_{0}$,
$\lambda_{r}=10^{-7}\lambda_{0}$ and $\lambda_{s}=10^{-9}\lambda_{0}$,
in agreement with \cite{Wunderlich2009}. These correspond to energies,
measured relative to the energy of the unbound state, of $E_{s}=-4.6k_{B}T$,
$E_{b}=11.5k_{B}T$ and $E_{r}=-9.2k_{B}T$. The value of $p_{1}$
was taken to be one. Experiments suggest $\lambda_{0}\simeq10^{6}\sec^{-1}$
for the Lac repressor \cite{WAC2006}.}

\label{fig2ChBarrier}
\end{figure}

Expression (\ref{2exp}) enables an explicit determination of $t^{ave}=q\tau_{1}+(1-q)\tau_{2}$
and the typical search time $t^{typ}$. For convenience we define $t^{typ}$ through
\begin{equation}
\mathcal{R}\left(t^{typ}\right)={1-}\frac{1}{e},\label{ttyp}
\end{equation}
i.e. the time after which the target is found with probability $1-1/e \simeq 0.63$
\footnote{This choice of the typical time (in contrast to, say, the half life time
of an unoccupied target $\mathcal{R}\left(t^{1/2}\right)=\frac{1}{2}$)
has the advantage of being equal to the average time for a simple
exponential decay case.%
}. The solution for Eq. (\ref{ttyp}) in the regime when the two time
scales, $\tau_{1}$ and $\tau_{2}$ are well separated, $\tau_{1}\ll\tau_{2}$,
is given by

\begin{equation}
t_{typ}=\left\{ \begin{array}{cc}
\tau_{1}\ln\frac{q}{\frac{1}{e}+q-1} & q>1-\frac{1}{e}\\
\tau_{2}\ln\frac{1-q}{\frac{1}{e}-q} & q<1-\frac{1}{e}\end{array}\right.
\end{equation}
We stress that experimentally, the relevant time, where almost all
search processes end, is $t^{typ}$ and not $t^{ave}$. In the regime
$\lambda_{r}\gg\lambda_{u}\kappa/N$, one has $t^{typ}\simeq t^{ave}\simeq\tau_{2}$.
A difference between $t^{typ}$ and $t^{ave}$ emerges as $\lambda_{r}$
is decreased and in the limit $\lambda_{r}\ll\lambda_{u}\kappa/N$
we find that $t^{typ}\simeq\tau_{1}/(2q-1)$ (with $q\simeq1$) is
independent of $\lambda_{s}$. This shows that for DNA lengths $N\leq\lambda_{u}\kappa/\lambda_{r}$,
the typical search time is significantly smaller than the average
even in the presence of deep traps ($\lambda_{s}$ small). This is
a direct result of the competition between the two time scales.

\begin{figure}[ptb]

\begin{centering}
\includegraphics[width=13cm]{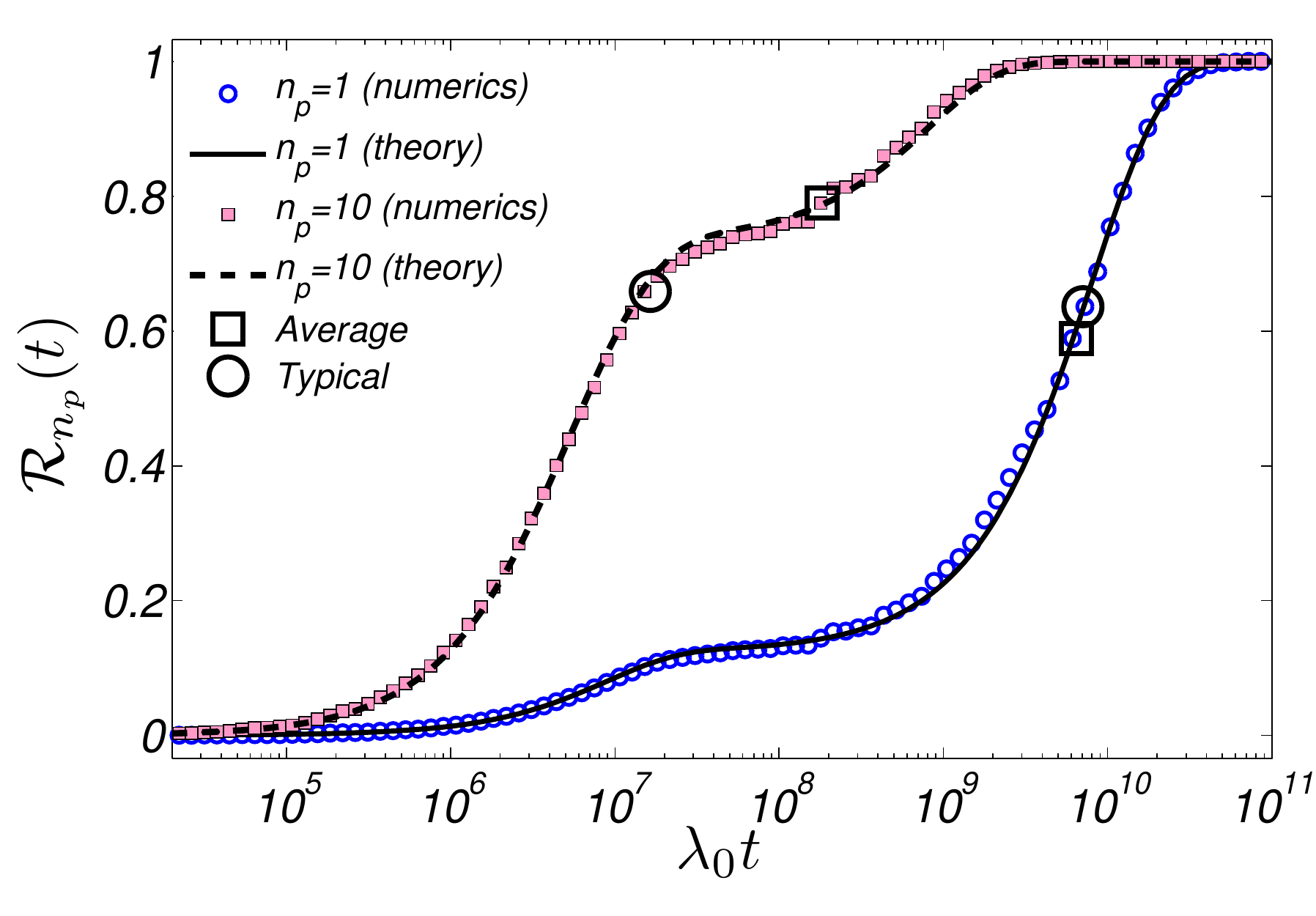}
\par\end{centering}

\caption[A plot of $\mathcal{R}_{n_{p}}(t)$ for non disordered case]{A plot of $\mathcal{R}_{n_{p}}(t)$ for $n_{p}=1$ (empty circles)
and $n_{p}=10$ (filled squares) for non disordered case. Here $N=10^{6}$,
$p_{1}=1$, $\lambda_{u}=10^{-4}\lambda_{0}$, $\lambda_{b}=0.1\lambda_{0}$,
$\lambda_{r}=10^{-7}\lambda_{0}$ and $\lambda_{s}=10^{-9}\lambda_{0}$
(see \cite{Wunderlich2009}). These correspond to energies, measured
relative to the unbound state, of $E_{s}=-9.2k_{B}T$, $E_{b}=6.9k_{B}T$
and $E_{r}=-13.8k_{B}T$. The value of $p_{1}$ was taken to be one.
Lines correspond to Eq. (\ref{2exp}) with calculated values of $\tau_{1}$,
$\tau_{2}$ and $q$. Note that here $\lambda_{u}$ is different from
Fig. \ref{fig2ChBarrier}.}
\label{fig3ChBarrier}
\end{figure}

The results, compared with numerics which were performed using a standard
continuous time Gillespie algorithm \cite{G1976} (see Appendix \ref{Numerics}
for details), are shown in Fig. \ref{fig2ChBarrier}. We use realistic
ranges of parameters (from available experimental data summarized
in \cite{Wunderlich2009}) which are specified in the caption. Since
to the best of our knowledge there are no direct measurements of the
barrier height for different DNA sequences, we assume this quantity
to be of the same order of magnitude as the experimentally measured
binding energies \cite{GMH2002}. It is found that $\mathcal{R}(t)$
reaches a plateau close to one on a typical time scale $t^{typ}$
which, for $N=10^{6}$, is smaller than the average search time $t^{ave}$
by two orders of magnitude. In next sections we show that results
of this simple model may be applied to more realistic models.

\subsection{Several searching proteins}

The interesting regime $t^{typ}\ll t^{ave}$ requires a rather
large barrier between the $s$ and $r$ state in the case of long
DNA molecules (namely, $\lambda_{r}\leq\lambda_{u}\kappa/N$). One
might argue that in general this condition may not be met by all proteins.
Despite of this we now argue that this constraint can be, to a large
extent, relaxed when $n_{p}$ proteins are searching for the target
simultaneously. In this case even when for a single protein $t^{ave}\simeq t^{typ}$
the typical search time $t_{n_{p}}^{typ}$ of $n_{p}$ proteins can
be significantly shorter than $t^{ave}/n_{p}$ even for relatively
small values $n_{p}\approx10-15$. Here, again, $t^{ave}$ is the
average search time of a single protein and $t_{n_{p}}^{typ}$ is
defined as in Eq. \ref{ttyp} where for $n_{p}$ proteins the first-passage
distribution $P_{n_{p}}(t)$ is deduced from the cumulative distribution
\begin{equation}
\mathcal{R}_{n_{p}}(t)=1-\left(1-\mathcal{R}(t)\right)^{n_{p}}\;.\label{Pnp}
\end{equation}
In Fig. \ref{fig3ChBarrier} we show the results for $\mathcal{R}_{n_{p}}(t)$
for $n_{p}=10$. Note that as claimed above $t_{n_{p}}^{typ}\ll t^{ave}/n_{p}$,
whereas $t^{typ}$ is close to $t^{ave}$ for one protein. This can
be understood as follows. Using Eqs. (\ref{2exp}) and (\ref{Pnp}),
it is obvious that when $\tau_{2}\gg\tau_{1}$, the decay of $\mathcal{R}_{n_{p}}(t)$
is dominated by $\tau_{1}$ as long as $(1-q)^{n_{p}}\ll1$. In essence
since only one protein needs to find the target, the probability of
a catastrophic event where the search time is of the order of $\tau_{2}$
is\begin{equation}
p_{cat}=(1-q)^{n_{p}}\label{pcat}\end{equation}
 which decays exponentially fast with $n_{p}$. For large enough values
of $n_{p}$ the short time scale $\tau_{1}$ controls the behavior
of $\mathcal{R}_{n_{p}}(t)$, even if it is insignificant for the
one protein search time. This implies that searches involving several
proteins strongly suppress the long time-scales induced by the traps
which control $t^{ave}$. In Section
\ref{Average and typical search time} the average and typical search
times are calculated for a given values of $\tau_{1}$, $\tau_{2}$,
$q$ and $n_{p}$.

\subsection{Calculating the average and typical search times\label{Average and typical search time}}

We showed above that the cumulative FPT distribution is given by
\begin{equation}
\mathcal{R}_{n_{p}}(t)=1-\left[qe^{-t/\tau_{1}}+\left(1-q\right)e^{-t/\tau_{2}}\right]^{n_{p}}
\end{equation}
 for the non-disordered model. In this section we calculate the typical and average search times for given values of
 $q$, $\tau_{1}$ and $\tau_{2}$. In the next section we discuss the disordered model in detail. As shown the disorder
 leaves the mathematical structure of the non-disordered case intact but with effective values of $q$, $\tau_{1}$ and $\tau_{2}$.
 Therefore all the results presented below and obtained for the non-disordered case can be easily extended to the disordered one.

\subsubsection{Typical search time}

When $\tau_{1}\ll\tau_{2}$, the
typical search time $t_{n_{p}}^{typ}$, defined through \begin{equation}
\mathcal{R}_{n_{p}}(t_{n_{p}}^{typ})={1-}\frac{1}{e}\label{typdef}\end{equation}
can be obtained by assuming $t_{n_{p}}^{typ}\gg\tau_{1}$ or $t_{n_{p}}^{typ}\ll\tau_{2}$
and checking these assumptions self-consistently. Using this method
we obtain for $n_p>1$
\begin{equation}
t_{n_{p}}^{typ}=\left\{ \begin{array}{cc}
\tau_{2}\left[\frac{1}{n_{p}}+\ln\left(1-q\right)\right] & {\rm for} \;\;\; n_{p}<\frac{1}{\ln\left(\frac{1}{1-q}\right)}\\
\tau_{1}\ln\frac{q}{e^{-1/n_{p}}+q-1}\simeq\frac{\tau_{1}}{qn_{p}} & {\rm for} \;\;\; n_{p}>\frac{1}{\ln\left(\frac{1}{1-q}\right)}\end{array}\right..\label{typ}
\end{equation}
Therefore, for large enough $n_{p}$ it is widely independent of
the binding energy in the $r$ mode.

\subsubsection{Average search time}

\begin{figure}[ptb]

\begin{centering}
\includegraphics[width=15cm]{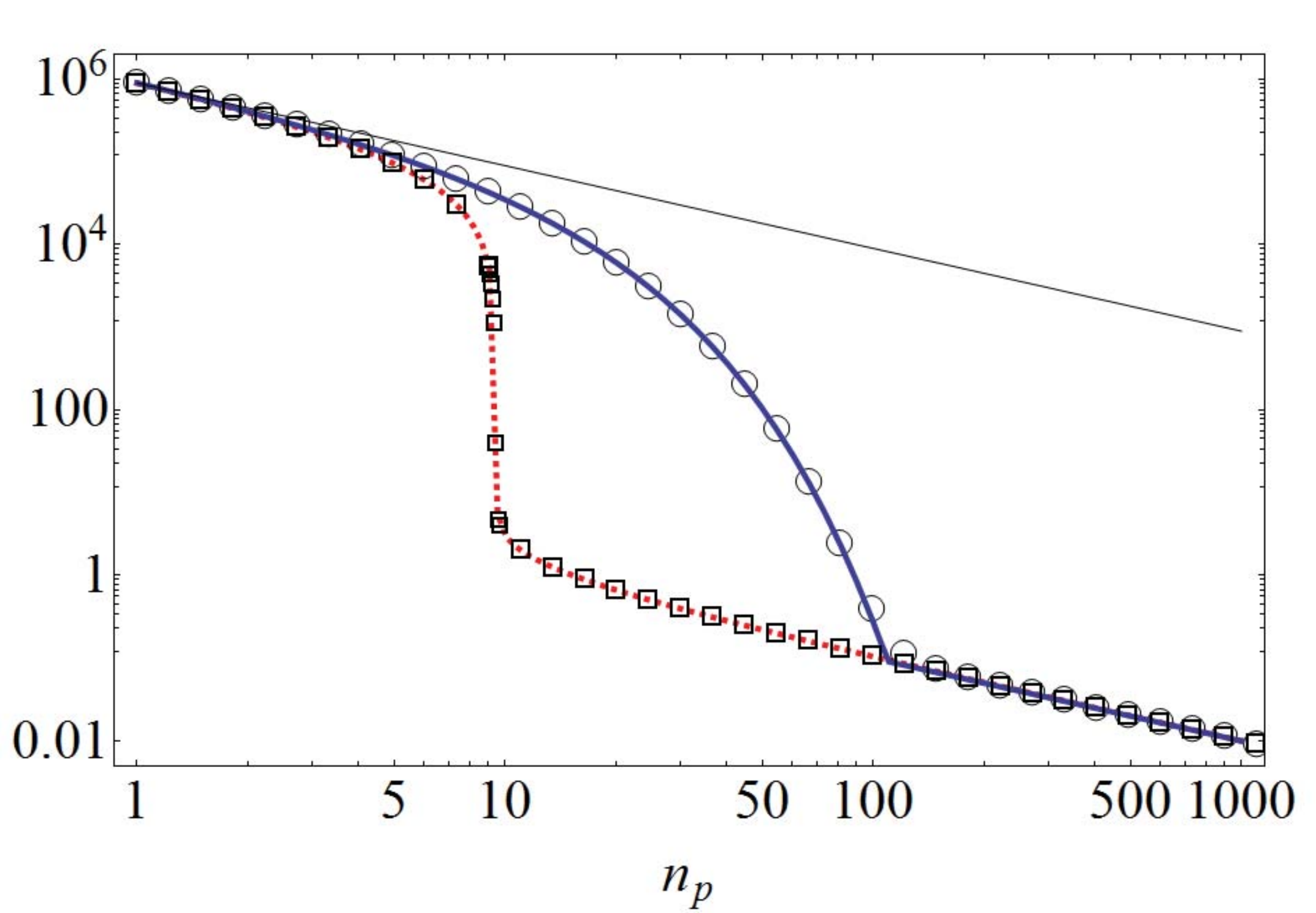}
\par\end{centering}

\caption{In this plot the average and the typical search times and their approximations
are shown as a function of the number of proteins, $n_{p}$. Circles represent
the average search time. Squares represent the typical search time,
$t_{n_{p}}^{typ}$ (\ref{typdef}). The blue, solid and thick line represents
the approximation for the average time (\ref{taveapp}), the dashed
red line represents the approximation for the typical search time
(\ref{typ}) and the thin black line represents the naive estimate
$\frac{t^{ave}}{n_{p}}$. Parameters chosen for this plot are $\tau_{1}=1$,
$\tau_{2}=10^{6}$ and $q=0.1$.}

\label{fig6ChBarrier}
\end{figure}

The average search time in the case of $n_{p}$ proteins is given
by
\begin{align}
t_{n_{p}}^{ave} & =\int_{0}^{\infty}\left(1-\mathcal{R}(t)\right)^{n_{p}}dt=\nonumber \\
& =\overset{n_{p}}{\underset{n=0}{\sum}}\frac{n_{p}!}{\left(n_{p}-n\right)!n!}\frac{q^{n}\left(1-q\right)^{n_{p}-n}}{\frac{n}{\tau_{1}}+\frac{n_{p}-n}{\tau_{2}}} \;.
 \end{align}
 This sum may be estimated using a saddle point approximation. The
saddle point is at $n^{\ast}=n_{p}q$ as expected (in the limit of
a large $\frac{\tau_{2}}{\tau_{1}}$ ratio and using the Stirling approximation).
Note that the saddle point approximation breaks when $n^{\ast}<1$.
In this case the dominant term is $n^\ast=0$. When this is not the case we find
\begin{align}
t_{n_{p}}^{ave} =\frac{\tau_{2}}{n_{p}}\left(1-q\right)^{n_{p}}+\frac{\tau_{1}}{n_{p}q}\frac{1}{1+\frac{\tau_{1}}{\tau_{2}}\frac{1-q}{q}}\;.\label{tave}
\end{align}
In the limit of $\tau_1 \ll \tau_2$ the average time is given by\begin{equation}
t_{n_{p}}^{ave}\simeq\left\{ \begin{array}{cc}
\frac{\tau_{2}}{n_{p}}\left(1-q\right)^{n_{p}} & {\rm for} \;\;\; n_{p}<\frac{\ln\frac{\tau_{2}}{\tau_{1}}}{\ln\left(\frac{1}{1-q}\right)}\\
\frac{\tau_{1}}{n_{p}q} & {\rm for} \;\;\;  n_{p}>\frac{\ln\frac{\tau_{2}}{\tau_{1}}}{\ln\left(\frac{1}{1-q}\right)}\end{array}\right.\label{taveapp}\end{equation}
 In Fig. \ref{fig6ChBarrier} the average and typical search times
are shown and compared to the approximations given by Eqs. (\ref{typ}) and (\ref{taveapp}). The data shown correspond to a choice of parameters where for $n_p=1$ the typical and average search times are roughly the same. Note that there is a large range of $n_p$ values for which $t_{n_{p}}^{typ} \ll t_{n_{p}}^{ave}$ and that they coincide again at very large values of $n_p$. The range of values of $n_p$ for which the typical and the mean search times differ scales as $\ln (\tau_2 / \tau_1)$. Remarkably, for small values of $n_p$  the average search time decreases faster than exponentially with the protein copy number.

\subsection{Disordered case\label{Disordered case}}

In this section we study a disordered version of the model. Since the barrier plays a key role
in the search we focus on effects of disorder in its height. To
account for this we consider the case where the barrier height, $E_{b}^{i}$,
is drawn from a Gaussian distribution: \begin{equation}
p(E_{b}^{i})=\frac{e^{-\frac{(E_{b}^{i}-E_{0})^{2}}{2\sigma^{2}}}}{\sqrt{2\pi\sigma^{2}}}\end{equation}
 such that the transition rate from the $s$ state to the $r$ state
at site $i$ is given by\begin{equation}
\lambda_{r}^{i}=\lambda_{0}\min\left(1,e^{-E_{b}^{i}}\right).\label{lamr}\end{equation}
 Introducing an energy difference between the $r$ state and the $s$
state (taken to be equal for all sites), $E_{r}$, the transition
rate from the $r$ state to the $s$ state at site $i$ is given
by\begin{equation}
\lambda_{s}^{i}=\lambda_{0}e^{E_{r}}\min\left(1,e^{-E_{b}^{i}}\right)=\lambda_{r}^{i}e^{E_{r}}.\label{lams}
\end{equation}
Similar to the affinity properties of the target site that is
typically very close to the highest affinity among the non-target
sites (see discussion above), we propose an intrinsic definition
of the target as the site with the \textit{lowest barrier} with no
specifically designed properties (see Fig.
\ref{LanscapeChBarrier}(b)). Indeed, our previous assumption in
Section \ref{Non disordered case} that $\lambda_{r}^{\mathcal{T}}$
is large at the target site and $\lambda_{r}^{i}$ small everywhere
else is a rather strong demand and corresponds to a designed
target. Although we show below that the barrier discrimination
mechanism may supply an efficient search even for the non-designed
target, any special design of the target may significantly
increase the search effectiveness. In the next subsection we
analyze the disordered model using a mean-field approach and check
the results using a numerical simulation in Section \ref{Numerical
results and comparison to mean-field analysis}.

\subsubsection{Mean-field analysis}

Within the mean-field approach we replace the different quantities
by their disorder average and account for the barrier at the target
site. We first compute the disorder averaged probability of crossing
the barrier at the target at each visit. The probability density of
the barrier height on the target site, $E_{b}^{\mathcal{T}}$, (the
probability density of the minimal energy among $N$ normally distributed
identical and independent random variables with a mean $E_{0}$ and
variance $\sigma^{2}$) is \cite{HF2006}\begin{equation}
\Pr\left(E_{b}^{\mathcal{T}}\right)=\frac{d}{dE_{b}^{\mathcal{T}}}\left[\int_{-\infty}^{E_{b}^{\mathcal{T}}}\frac{e^{-\frac{\left(E-E_{0}\right)^{2}}{2\sigma^{2}}}}{\sqrt{2\pi}\sigma}dE\right]^{N}=-\frac{d}{dE_{b}^{\mathcal{T}}}\left[\frac{1}{2}\operatorname{erfc}\left(\frac{E_{b}^{\mathcal{T}}-E_{0}}{\sqrt{2}\sigma}\right)\right]^{N}.\end{equation}
 For a given value of $E_{b}^{\mathcal{T}}$ the probability to pass
over a barrier to the $r$ state is $\frac{e^{-E_{b}^{\mathcal{T}}}}{1+\lambda_{u}/\lambda_{0}+e^{-E_{b}^{\mathcal{T}}}}$.
Thus the disorder averaged probability of crossing the barrier at
the target at each visit is given by \begin{equation}
\overline{p}_{1}=-\int_{-\infty}^{\infty}dE_{b}^{\mathcal{T}}\frac{e^{-E_{b}^{\mathcal{T}}}}{1+\lambda_{u}/\lambda_{0}+e^{-E_{b}^{\mathcal{T}}}}\frac{d}{dE_{b}^{\mathcal{T}}}\left[\frac{1}{2}\operatorname{erfc}\left(\frac{E_{b}^{\mathcal{T}}-E_{0}}{\sqrt{2}\sigma}\right)\right]^{N}.\end{equation}
 Here we set the time scale of the activation process across the barrier
to be $\lambda_{0}$. We finally assume that the expression for $u(s)$
of the non-disordered model (\ref{2exp}) holds with $\lambda_{r}^{i}$
replaced by its average over the barrier energy. Using Eq. (\ref{lamr})
this is given by \begin{equation}
{\bar{\lambda}}_{r}=\lambda_{0}\left[\int_{-\infty}^{0}\frac{e^{-\frac{\left(E_{b}-E_{0}\right)^{2}}{2\sigma^{2}}}}{\sqrt{2\pi}\sigma}dE_{b}+\int_{0}^{\infty}e^{-E_{b}}\frac{e^{-\frac{\left(E_{b}-E_{0}\right)^{2}}{2\sigma^{2}}}}{\sqrt{2\pi}\sigma}dE_{b}\right].\end{equation}
 Also using Eq. (\ref{lams}), within the mean-field approximation
$\lambda_{s}^{i}$ is replaced by\begin{equation}
\overline{\lambda}_{s}={\bar{\lambda}}_{r}e^{E_{r}}\end{equation}
 and $p_{1}$ replaced by $\overline{p}_{1}$. In the next Section
we check these results numerically.

\subsubsection{Numerical results and comparison to mean-field analysis\label{Numerical results and comparison to mean-field analysis}}

\begin{figure}[ptb]

\begin{centering}
\includegraphics[width=10cm]{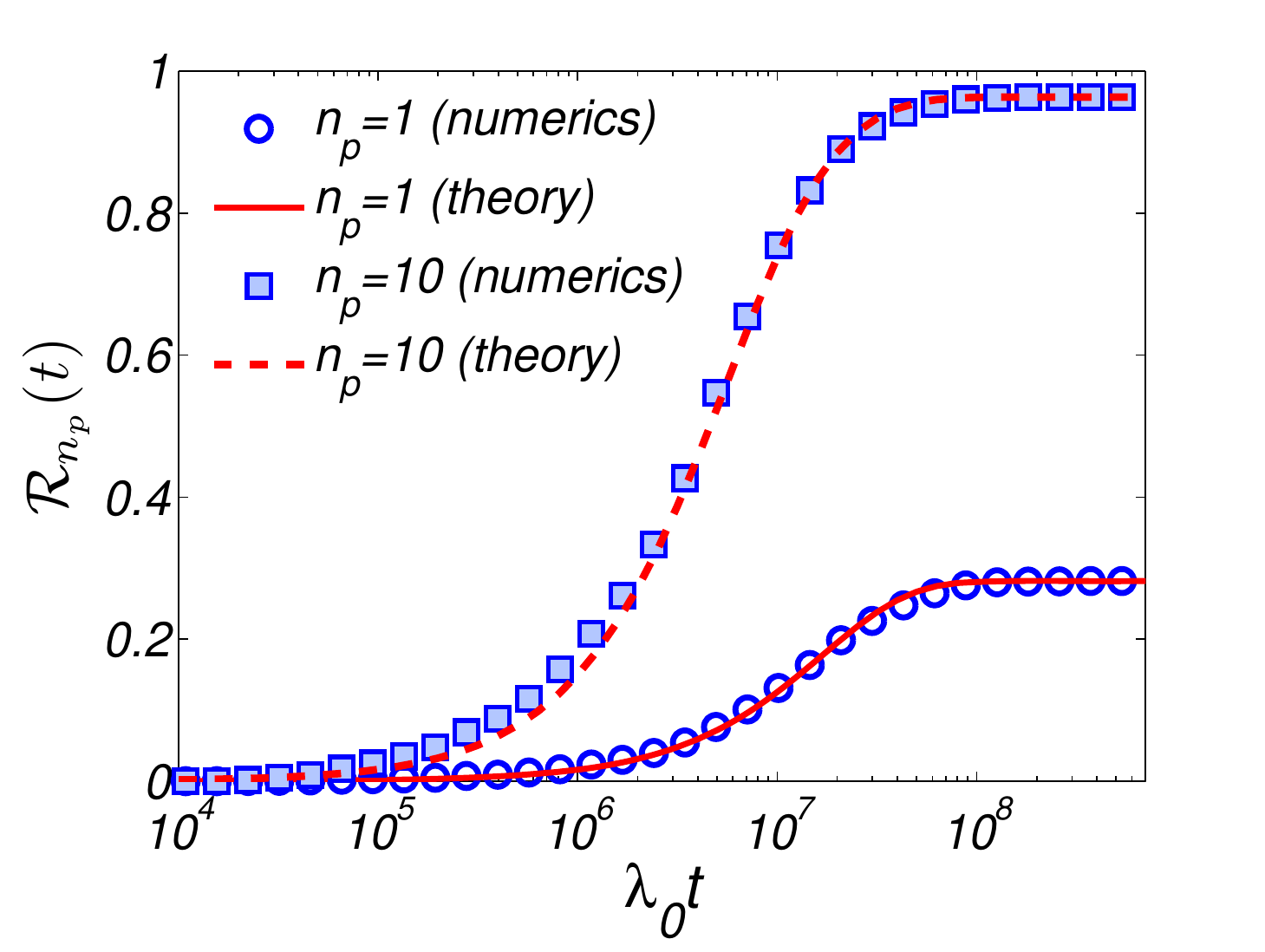}
\par\end{centering}

\caption[Plot of $\mathcal{R}_{n_{p}}(t)$ for the disordered model]{Plot of $\mathcal{R}_{n_{p}}(t)$ for $n_{p}=1$ (empty circles)
and $n_{p}=10$ (filled squares) for the disordered model. The lines
were obtained by fitting the form $1-(qe^{-t/\tau_{1}}+(1-q))^{{n}_{p}}$
to the numerical simulations with $q=0.2817$, $\lambda_{0}\tau_{1}=1.7\cdot10^{7}$
and $\tau_{2}=\infty$. These are close to the mean field prediction
$q=0.2827$, $\lambda_{0}\tau_{1}=1.1\cdot10^{7}$. Here $\lambda_{u}=10^{-2}\lambda_{0}$
($E_{s}=-4.6k_{B}T$), $\lambda_{b}=0.1\lambda_{0}$, $E_{0}=25.4k_{B}T$,
$E_{r}=-\infty$ and $\sigma=5.3k_{B}T$. Note that here the average
height of the barrier at the target site is $6.25k_{B}T$.}

\label{fig4ChBarrier}
\end{figure}

\begin{figure}[ptb]

\begin{centering}
\includegraphics[width=10cm]{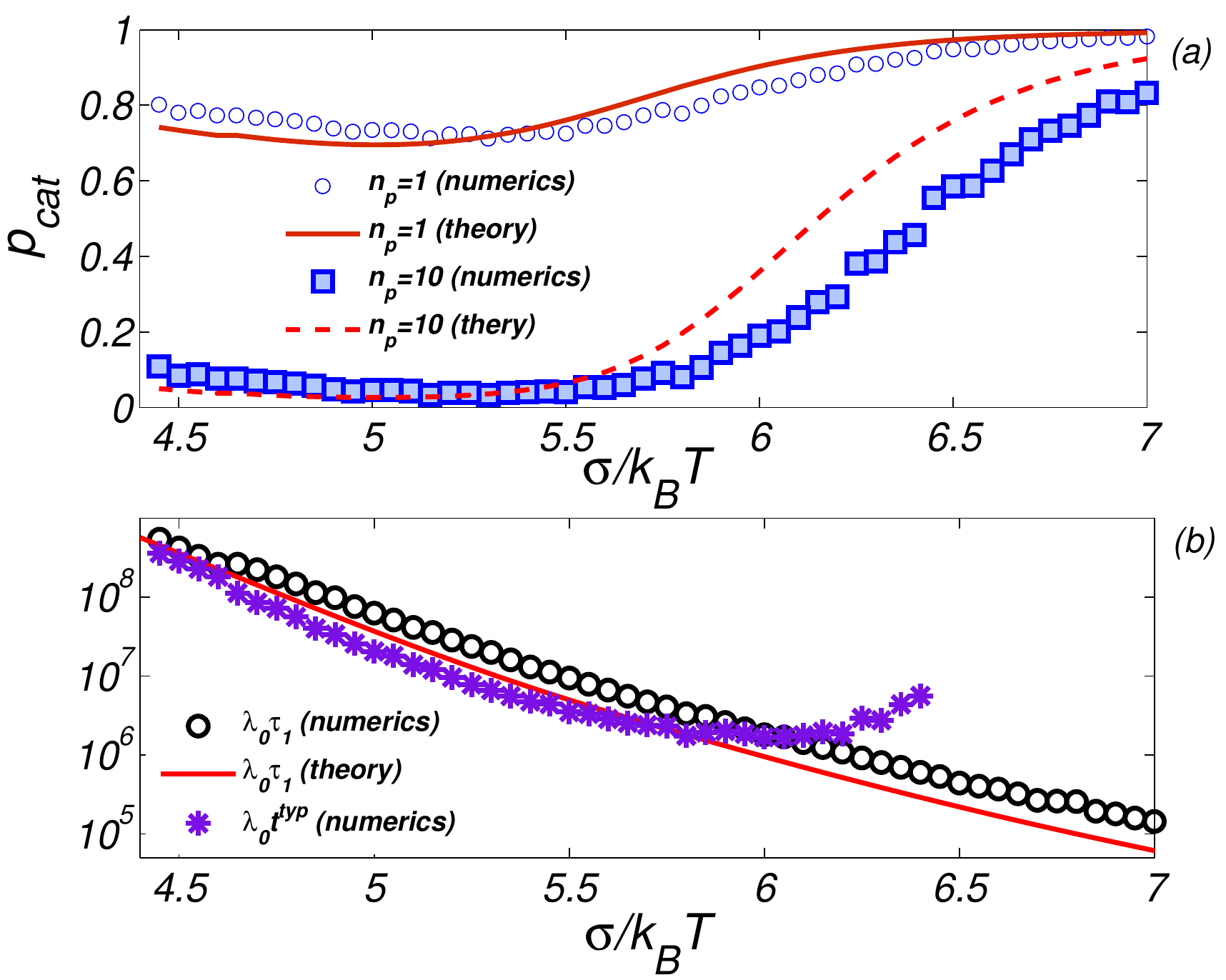}
\par\end{centering}

\caption[Results for the disordered model]{Results for the disordered model. Here $N=10^{6}$, $\lambda_{u}=10^{-2}\lambda_{0}$
($E_{s}=-4.6k_{B}T$), $\lambda_{b}=0.1\lambda_{0}$, $E_{r}=-\infty$
and $E_{0}=25.4k_{B}T$. (a) $p_{cat}$ as a function of $\sigma$
for $n_{p}=1$ and $n_{p}=10$. (b) $t^{typ}$ for $n_{p}=10$ and
$\tau_{1}$ are plotted as a function of $\sigma$. Using $\lambda_{0}=10^{6}\sec^{-1}$
\cite{WAC2006} for $n_{p}=10$ at the minimal $p_{cat}$ we find
$t^{typ}\simeq10sec$. Note that by moderate changes in $E_{0}$ similar
results can be obtained for longer DNA sequences. The parameters for
this plot are $\tau_{1}=1$, $\tau_{2}=10^{6}$, $q=0.1$.}

\label{fig5ChBarrier}
\end{figure}

We now check the mean-field results using numerics. First, we show that the two scales scenario described
above still holds. Indeed, Fig. \ref{fig4ChBarrier} shows that $\mathcal{R}(t)$
is well fitted by Eq. (\ref{2exp}) for realistic values of parameters.
Note that in Figs. \ref{fig4ChBarrier} and \ref{fig5ChBarrier} we
have chosen the worst scenario $E_{r}=-\infty$ so that $\lambda_{s}=0$
and the average search time and the value of $\tau_{2}$ are both
infinite. This implies that for $n_{p}$ large enough the only relevant
time scale is $\tau_{1}$ and the typical search time again takes
the form $t_{n_{p}}^{typ}\simeq\frac{\tau_{1}}{qn_{p}}$ (the detailed
calculation of the typical and average times is presented above in
Section \ref{Average and typical search time}). This enables a fast
search even in the presence of very deep (even infinite) traps.

The regime of a fast search with $t_{n_{p}}^{typ}$ independent of
the trap depth $E_{r}$ also requires, as above, a small catastrophe
probability, $p_{cat}=\left(1-q\right)^{n_{p}}$ (see Eq. \ref{pcat}).
We now show that this condition holds in a wide range of disorder
parameters, $E_{0}$ and $\sigma$. To illustrate this, the dependencies
(holding all other variables constant) of $p_{cat}$ and $t_{n_{p}}^{typ}$
on $\sigma$, obtained from numerics and the mean-field treatment,
are shown in Fig. \ref{fig5ChBarrier} for realistic values of parameters.
Notably, the dependence of the catastrophe probability on the disorder
strength is not monotonic so that the value of $p_{cat}$ can be minimized
as a function of $\sigma$. This reflects the fact that for small
values of $\sigma$ the DNA sequence has to be scanned many times
before the target enters in the $r$ mode. Increasing $\sigma$
lowers the barrier at the target and therefore reduces the number
of scans needed, which diminishes $p_{cat}$. For larger $\sigma$
the chance of falling into a trap increases due to lower secondary
minima of the barrier, which leads to an increase of $p_{cat}$.
As expected, $p_{cat}$ is dramatically decreased when $n_{p}$ is
increased, even by a few units, and can remain small for a wide range
of values of $\sigma$. For larger $\sigma$, $p_{cat}$ increases
and $t_{n_{p}}^{typ}$ rises quickly as it starts to depend on $\tau_{2}$.

Summarizing and using the results of Section \ref{Non disordered case},
the mean-field approach predicts that in the high barrier regime $\overline{\lambda}_{s}\ll\overline{\lambda}_{r}\ll\lambda_{u},\lambda_{b},\lambda_{0}$
(with $\lambda_{u},\lambda_{b},\lambda_{0}$ of comparable order)
the reaction probability simplifies to \begin{equation}
\mathcal{R}(t)\simeq1-qe^{-t/\tau_{1}}-(1-q)e^{-t/\tau_{2}}\end{equation}
 with \begin{align}
q & =\left(1+\frac{\overline{\lambda}_{r}}{\overline{\lambda}_{u}\kappa/N}\right)^{-1},\\
\kappa & =\frac{\sqrt{\coth\left(\frac{\overline{\lambda}_{u}}{2\overline{\lambda}_{0}}\right)}}{1+\frac{1-\overline{p}_{1}}{\overline{p}_{1}}\sqrt{1-e^{-2\overline{\lambda}_{u}/\overline{\lambda}_{0}}}},\\
\tau_{1} & =\frac{\overline{\lambda}_{b}+\overline{\lambda}_{u}}{\overline{\lambda}_{b}\left(\overline{\lambda}_{r}+\frac{\kappa\overline{\lambda}_{u}}{N}\right)}\end{align}
 and \begin{equation}
\tau_{2}=\frac{\overline{\lambda}_{r}+\kappa\overline{\lambda}_{u}/N}{\overline{\lambda}_{s}\kappa\overline{\lambda}_{u}/N}.\end{equation}

In the case of a few proteins, searchers that fall into traps tend to
occupy sites with low barriers and, therefore, increase the probability of other TFs
to reach the target. Thus, Eq. (\ref{Pnp}), in which
the searchers are assumed to be independent, provides a lower bound
on the probability to reach the target. Here and below we assume that
the number of proteins, $n_{p}$, is small enough (compared with $N$)
such that this effect does not play a role and Eq. (\ref{Pnp}) is
applicable.%

Most important, as advertised above, these results show that it is
possible to obtain relatively small values of $t_{n_{p}}^{typ}$ and
$p_{cat}$ with realistic values of the parameters (see Fig. \ref{fig5ChBarrier}).
Reasonable search times (in the range of seconds) are obtained for
a rather large range of $\sigma$ as long as $n_{p}$ is of the order
of ten or more proteins suggesting another possible resolution of the speed and stability
requirements. We stress that this mechanism can apply to any of the classes of TFs discussed above. This is
a direct consequence of the decoupling of the stability and speed requirements.
We note that by moderate changes in $E_{0}$ similar results
can be obtained for much longer DNA sequences.
In Appendix \ref{Conditions for a perfect search} we show that by increasing the disorder strength and the average barrier height such that
\begin{equation}
\frac{E_{0}}{\sigma}=\sqrt{2}\operatorname{erfc}^{-1}\left(\frac{2}{N}\right) \;,
\label{Perfect}
\end{equation}
a ``perfect'' searcher is obtained. By ``perfect'' it is implied that its search time is the same as a search on a flat (single state) model and that the target is reached with probability one.

\section{Effective model and outcomes\label{Effective model and outcomes}}

As we showed in the previous section, by only using a barrier discrimination between different
DNA sites a transcription factor may, in principle, serve as an
efficient searcher and its complex with the target can be arbitrarily
stable. Experiments show that different DNA sites are discriminated by their binding energy.
Therefore, if a barrier mechanism is at work it is likely to be combined with an energetic
discrimination between different sites.

Nonetheless, it is interesting to consider a scenario where there is only barrier discrimination. This could
apply for TFs which have a very small target occupation probability $P^{\cal T}$. As we now show a barrier
mechanism may lead to a high transient occupation probability of the target even with no energetic discrimination.
When active processes are included the occupation probability can be made large even in the long-time limit.
Furthermore, and in a more speculative manner, we show how the barrier mechanism can lead to a dynamical ordering of gene activation.
%and transcriptional bursts.

To show these we construct an effective model which uses the simple resulting mathematical structure of the previous
section. Specifically, we use the cumulative probability
\begin{equation}
\mathcal{R}\left(t\right)=1-qe^{-\frac{t}{\tau_{1}}}-\left(1-q\right)e^{-\frac{t}{\tau_{2}}}.\label{2exp2}
\end{equation}
In our discussion we concentrate on the target occupation probability.
This fact and the simplicity of expression (\ref{2exp2}) allow one
to describe our system using a three
states model. Within this approach we only consider the $r$ state
on the target ($\mathcal{T}$), the $r$ state off the target ($\mathcal{D}$)
and one state for all other configurations (including $s$ states
and the unbound state) ($\mathcal{U}$). The transition rates between
the states are defined as follows: $\lambda^{\mathcal{D}}$ is the
transition rate from $\mathcal{U}$ to $\mathcal{D}$, $\lambda_{-1}^{\mathcal{D}}$
is the transition rate from $\mathcal{D}$ to $\mathcal{U}$, $\lambda^{\mathcal{T}}$
is the transition rate from $\mathcal{U}$ to $\mathcal{T}$ and $\lambda_{-1}^{\mathcal{T}}$
is the transition rate from $\mathcal{T}$ to $\mathcal{U}$. The model is illustrated
schematically in Fig. \ref{fig7ChBarrier}. As shown below, this simplification
allows us to analyze the behavior of the system beyond its FPT properties.%
\begin{figure}[ptb]

\begin{centering}
\includegraphics[width=6cm]{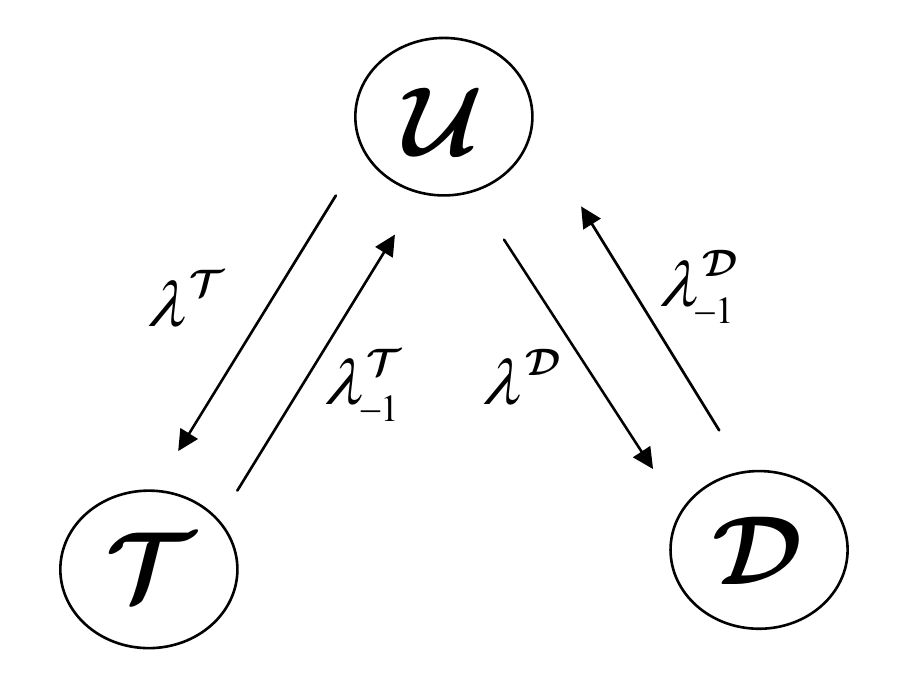}
\par\end{centering}

\caption{Schematic representation of the three-states effective system. Here
$\mathcal{T}$ denotes the $r$ state on the target, $\mathcal{D}$
denotes the $r$ state off the target and $\mathcal{U}$ denotes all
other states (including the $s$ state and the unbound state).}

\label{fig7ChBarrier}
\end{figure}

To proceed we, first, show that the effective model yields the same cumulative FPT distribution $\mathcal{R}\left(t\right)$ as the original system.
Specifically, it is straightforward to show that
\begin{equation}
\mathcal{R}(t)=1-\frac{\lambda^{+}-\lambda^{\mathcal{D}}}{\lambda^{+}-\lambda^{-}}e^{-t\lambda^{+}}-\frac{\lambda^{\mathcal{D}}-\lambda^{-}}{\lambda^{+}-\lambda^{-}}e^{-t\lambda^{-}},\label{sR}
\end{equation}
 where
 \begin{equation}
\lambda^{\pm}=\frac{\lambda_{-1}^{\mathcal{D}}+\lambda^{\mathcal{D}}+\lambda^{\mathcal{T}}\pm\sqrt{(\lambda_{-1}^{\mathcal{D}}+\lambda^{\mathcal{D}}+\lambda^{\mathcal{T}})^{2}-4\lambda^{\mathcal{T}}\lambda_{-1}^{\mathcal{D}}}}{2}.
\end{equation}
Comparing Eq. (\ref{2exp}) with Eq. (\ref{sR}) one obtains
relations between transition rates of the effective model and the
full model:
\begin{align}
\frac{\lambda^{+}-\lambda^{\mathcal{D}}}{\lambda^{+}-\lambda^{-}} & =q\nonumber \\
\lambda^{+} & =\frac{1}{\tau_{1}}\nonumber \\
\lambda^{-} & =\frac{1}{\tau_{2}}.\label{qLamPlus}
\end{align}
 The solution for the transition rates in the effective model are then
 \begin{align}
\lambda_{-1}^{\mathcal{D}} & =\frac{\tau_{1}-q\tau_{1}+q\tau_{2}-\sqrt{\left(q\tau_{1}-\tau_{1}-q\tau_{2}\right)^{2}-4\tau_{1}\tau_{2}}}{2\tau_{1}\tau_{2}}\simeq\frac{1-q}{2\tau_{2}}\nonumber \\
\lambda^{\mathcal{D}} & =\frac{q\tau_{1}+\tau_{2}-q\tau_{2}}{\tau_{1}\tau_{2}}\simeq\frac{1-q}{\tau_{1}}\nonumber \\
\lambda^{\mathcal{T}} & =\frac{q\tau_{2}+\tau_{1}-q\tau_{1}+\sqrt{\left(q\tau_{1}-\tau_{1}-q\tau_{2}\right)^{2}-4\tau_{1}\tau_{2}}}{2\tau_{1}\tau_{2}}\simeq\frac{q}{\tau_{1}},\label{transLam}
\end{align}
 where we assumed a high barrier regime, $\tau_{2}\gg\tau_{1}$. Note
that the transition rate from the target, $\lambda_{-1}^{\mathcal{T}}$,
has no influence on the FPT properties. However, it determines properties
of the target occupation probability in equilibrium. The time scale
separation in the high barrier regime, $\tau_{1}\ll\tau_{2}$, implies\begin{equation}
\lambda_{-1}^{\mathcal{D}}\ll\lambda^{\mathcal{D}}.\label{RateFromDNA}\end{equation}

As stated above we consider a case where
the binding energies of all sites (including the target) are
the same so that in equilibrium the occupation probability of the
target site is equal to the occupation probability of all other
sites on the DNA. In this case, in equilibrium the occupation probability
of the $\mathcal{D}$ state is $N$ times larger than the occupation
probability of the $\mathcal{T}$ state. This implies within the simplified
model that\begin{equation}
\frac{\lambda^{\mathcal{D}}}{\lambda_{-1}^{\mathcal{D}}}=N\frac{\lambda^{\mathcal{T}}}{\lambda_{-1}^{\mathcal{T}}},\label{tmone}\end{equation}
 so that\begin{equation}
\lambda_{-1}^{\mathcal{T}}=N\frac{\lambda_{-1}^{\mathcal{D}}\lambda^{\mathcal{T}}}{\lambda^{\mathcal{D}}}\simeq N\frac{q}{2\tau_{2}}.\label{transLam2}\end{equation}

Thus, we showed that a simple three-states effective model has the
same dynamical and equilibrium properties as the original system.
Below we use the effective model to analyze the search dynamics beyond
FPT properties. For example, we consider equilibration dynamics, the
possible existence of an active processes and temporal ordering
in the activation/repression of multiple targets.

\subsection{Transient behavior}

\begin{figure}[ptb]

\begin{centering}
\includegraphics[width=10cm]{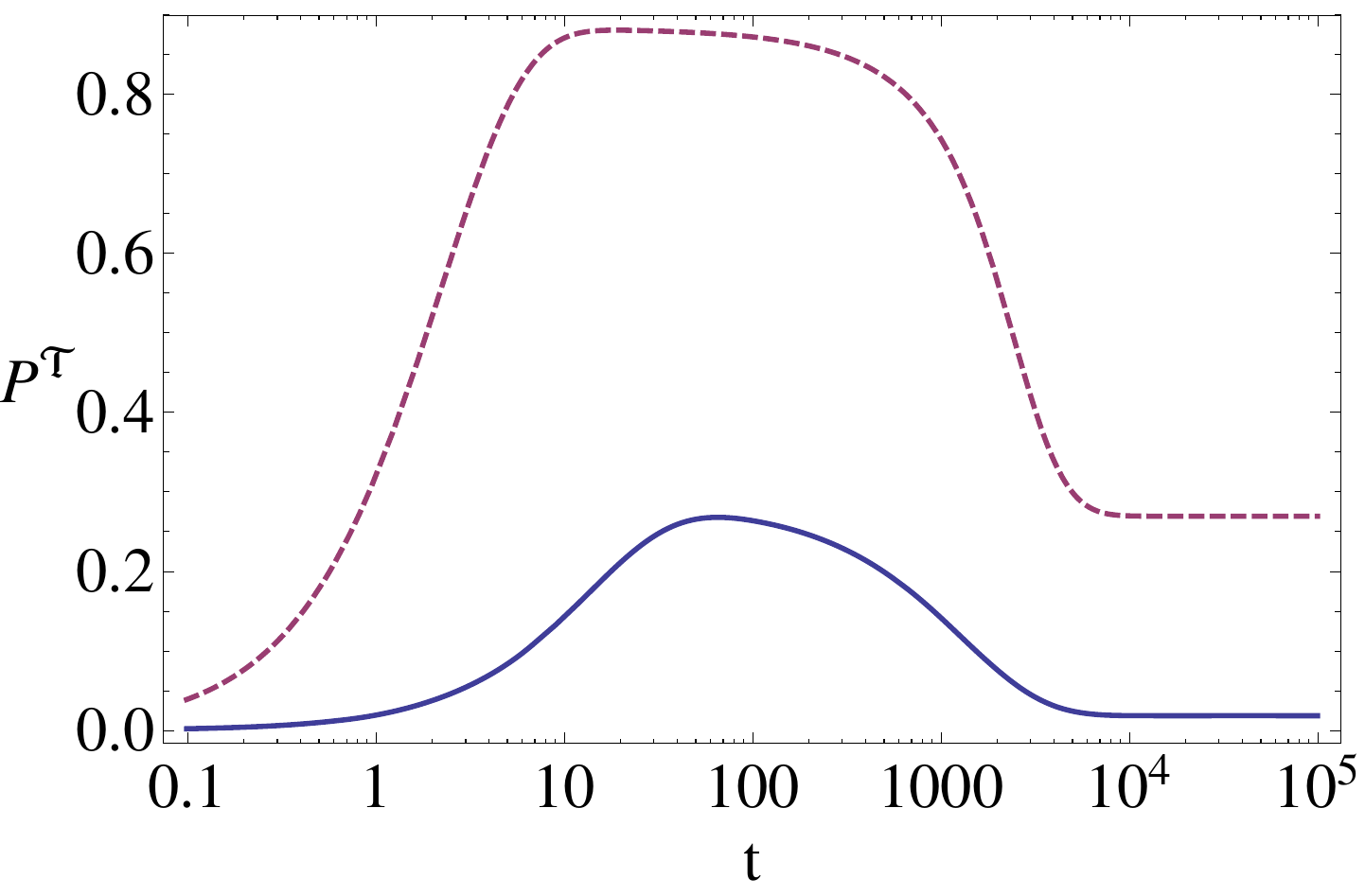}
\par\end{centering}

\caption{The occupation probability of the target site in the simplified model,
$P^{\mathcal{T}}$, is show as a function of time. The parameters
are $\lambda^{\mathcal{T}}=0.02$, $\lambda_{-1}^{\mathcal{T}}=0.05$,
$\lambda^{\mathcal{D}}=10^{-3}$ and $\lambda_{-1}^{\mathcal{D}}=5\cdot10^{-5}$.
The blue solid (red dashed) line represents the $n_{p}=1$ ($n_{p}=20$)
case.}

\label{fig8ChBarrier}
\end{figure}

Following the above the occupation probability of the target
site, $P^{\mathcal{T}}\left(t\right)$, evolves as
\begin{align}
\frac{\partial P^{\mathcal{T}}}{\partial t} & =\lambda^{\mathcal{T}}\left(1-P^{\mathcal{T}}-P^{\mathcal{D}}\right)-\lambda_{-1}^{\mathcal{T}}P^{\mathcal{T}}\nonumber \\
\frac{\partial P^{\mathcal{D}}}{\partial t} & =\lambda^{\mathcal{D}}\left(1-P^{\mathcal{T}}-P^{\mathcal{D}}\right)-\lambda_{-1}^{\mathcal{D}}P^{\mathcal{D}}\nonumber\\
P^{\mathcal{U}} & =1-P^{\mathcal{D}}-P^{\mathcal{T}}\label{PtEq}
\end{align}
 where $P^{\mathcal{D}}\left(t\right)$ is the occupation probability
of the $\mathcal{D}$ state, $P^{\mathcal{U}}\left(t\right)$ is the occupation
probability of the $\mathcal{U}$ state and the initial conditions
are $P^{\mathcal{U}}\left(t=0\right)=1$ so that $P^{\mathcal{T}}\left(t=0\right)=P^{\mathcal{D}}\left(t=0\right)=0$.
These equations may be solved exactly. However, here we analyze the
equations by noting that there are three time regimes. For $t\ll\frac{1}{\lambda^{\mathcal{T}}}$
the occupation probability of the target is close to its initial value,
i.e. $P^{\mathcal{T}}\simeq0$. For $\frac{1}{\lambda^{\mathcal{D}}}\gg t\gg\frac{1}{\lambda^{\mathcal{T}}}$
the protein equilibrated with the target but not with the rest DNA
so that $P^{\mathcal{T}}\simeq\frac{1}{1+\frac{\lambda_{-1}^{\mathcal{T}}}{\lambda^{\mathcal{T}}}}$.
Of course, this regime exist only when $\lambda^{\mathcal{T}}\gg\lambda^{\mathcal{D}}$.
For $t\gg\frac{1}{\lambda^{\mathcal{D}}}$ the system reaches thermal equilibrium and the target occupation probability
is given by  $P^{\mathcal{T}}\simeq\frac{1}{1+\frac{\lambda^{\mathcal{T}}}{\lambda_{-1}^{\mathcal{T}}}\frac{\lambda_{-1}^{\mathcal{D}}}{\lambda^{\mathcal{D}}}}$.
In Fig. \ref{fig8ChBarrier} the occupation probability of the target
site, $P^{\mathcal{T}}\left(t\right)$, is shown. Since there is no
binding energy discrimination between DNA sites
the occupation probability of the target in the long time limit is
very small, $P^{\mathcal{T}}=\frac{1}{N}$. Note, however, that there is a transient regime
where the occupation probability is large. In fact, in this regime the TF binds and unbinds many times from the target
site before the system reaches thermal equilibrium.

The above discussion may be generalized to the case of a few proteins,
$n_{p}>1$. In this case a mean-field generalization of (\ref{PtEq})
is
\begin{align}
\frac{\partial n^{\mathcal{T}}}{\partial t} & =\lambda^{\mathcal{T}}\left(n_{p}-n^{\mathcal{T}}-n^{\mathcal{D}}\right)\left(1-n^{\mathcal{T}}\right)-\lambda_{-1}^{\mathcal{T}}n^{\mathcal{T}}\nonumber \\
\frac{\partial n^{\mathcal{D}}}{\partial t} & =\lambda^{\mathcal{D}}\left(n_{p}-n^{\mathcal{T}}-P^{\mathcal{D}}\right)-\lambda_{-1}^{\mathcal{D}}n^{\mathcal{D}}\nonumber\\
n^{\mathcal{U}} & =n_{p}-n^{\mathcal{D}}-n^{\mathcal{T}}\label{PtEqnp}
\end{align}
with the initial conditions $n^{\mathcal{U}}\left(t=0\right)=n_{p}$
and $n_{\mathcal{T}}\left(t=0\right)=n^{\mathcal{D}}\left(t=0\right)=0$. Here $n^{\mathcal{T}},n^{\mathcal{D}}$ and $n^{\mathcal{U}}$ represent the mean-occupation  number at the target  in the ${\mathcal{T}}$, ${\mathcal{D}}$ and ${\mathcal{U}}$ respectively.
The numerical solution of this nonlinear equation is shown in Fig.
\ref{fig8ChBarrier}. The qualitative behavior is similar to the $n_{p}=1$
case: there are three time regimes. Using the same arguments as
the $n_{p}=1$ case we find that, for $t\ll\frac{1}{\lambda^{\mathcal{T}}n_{p}}$
we have $n^{\mathcal{T}}\simeq0$. For $\frac{1}{\lambda^{\mathcal{D}}}\gg t\gg\frac{1}{\lambda^{\mathcal{T}}n_{p}}$
we have $n^{\mathcal{T}}\simeq\frac{1}{1+\frac{1}{n_{p}}\frac{\lambda_{-1}^{\mathcal{T}}}{\lambda^{\mathcal{T}}}}$
and for $t\gg\frac{1}{\lambda^{\mathcal{D}}}$ the mean-occupation number
is given by $n^{\mathcal{T}}\simeq\frac{1}{1+\frac{1}{n_{p}}\frac{\lambda^{\mathcal{T}}}{\lambda_{-1}^{\mathcal{T}}}\frac{\lambda_{-1}^{\mathcal{D}}}{\lambda^{\mathcal{D}}}}$.
The intermediate regime, corresponding to a transient high occupation of the target, exist when $\lambda^{\mathcal{T}}n_{p}\gg\lambda^{\mathcal{D}}$.
Its easy to check that for a large enough number of proteins, $n_{p}\gg\frac{\lambda^{\mathcal{D}}}{\lambda^{\mathcal{T}}}$,
this regime exists even when $\lambda^{\mathcal{T}}<\lambda^{\mathcal{D}}$.

Using this analysis we have shown that when the target site differs
from the rest by a low barrier between the transcription factor's
$r$ and $s$ states its occupation has a transient nature. After
a change in the environment, that activates the transcription factors,
the occupation probability of the target increases exponentially with
a fast time constant and after this decreases exponentially with a
slow time constant to its final value. When the only discrimination
between sites is the barrier height the final occupation probability
of the target is very small, so that in the long time limit the system
is in the same state as it was before the activation of the protein.
By introducing a free energy binding energy discrimination between
sites, the final occupation probability of the target may be significant
such that the long time limit of the system may be different from
the initial \textquotedbl{}pre-activated\textquotedbl{} state.

In general, when subjected to a change in the environmental condition,
a cell typically responds by increasing the activity level of certain
genes and decreasing the activity level of others. In many cases,
the expression level of a certain gene changes temporarily, exhibiting
a sharp increase or decrease, and later changing again, reaching a
new steady-state (which often is similar the original\ state). This,
two-step transient behavior, is widely observed in different transcriptional
responses, from yeast \cite{Gasch2000,Braun2004} to human \cite{Ramoni2002}
and may be explained by a negative feedback of an activated protein
\cite{AlonBook}. In this Section we showed how this kind of behavior
naturally arises in a regulation system (composed of only one
transcription factor) based on a barrier discrimination between distinct
DNA sites.

\subsection{Steady-state and an existence of an active process}

\begin{figure}[ptb]

\begin{centering}
\includegraphics[width=8cm]{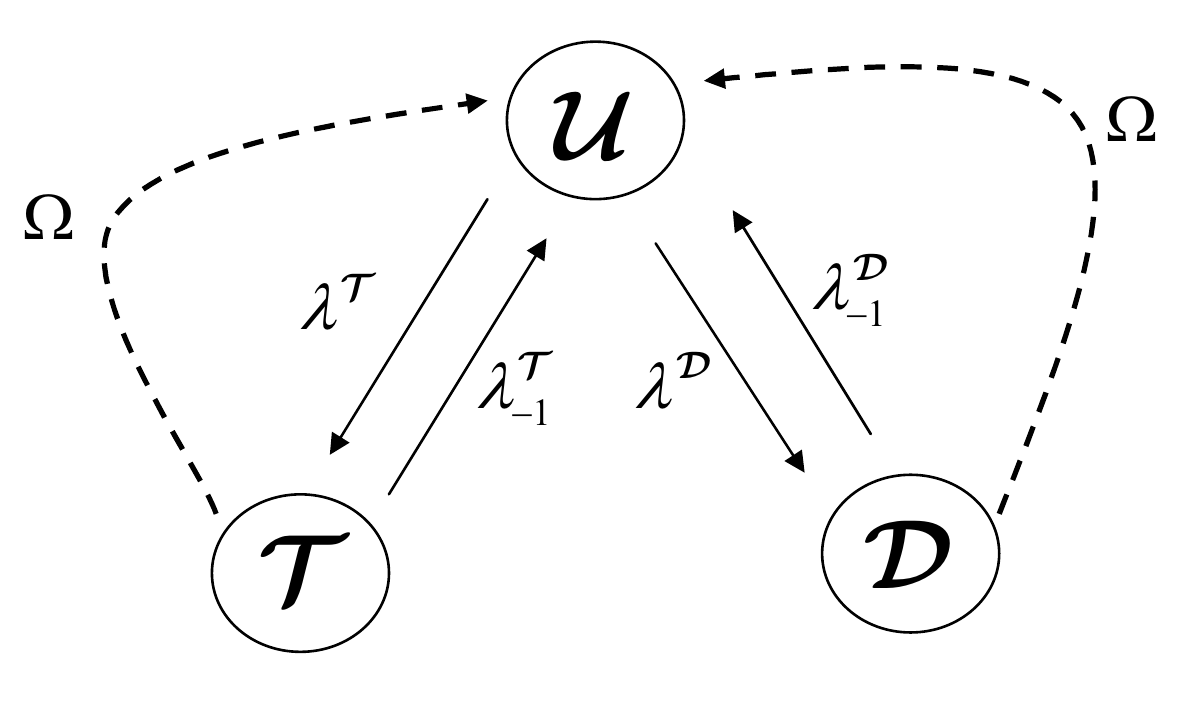}
\par\end{centering}

\caption{Schematic representation of the three-states effective system in
presence of an active process $\Omega$.}

\label{fig7ActiveChBarrier}
\end{figure}

In equilibrium the occupation probability on a DNA site depends only
on its binding energy. In cases where the only difference between
the target and non-target sites is the barrier height between the
$s$ and $r$ states, after the equilibration the probability to find
the protein on the target site is very small. As we now show, by introducing an active process that returns the searcher
to the initial state, $u$, with a transition rate, $\Omega$ (see
Fig. \ref{fig7ActiveChBarrier}) from \textit{any} state, one may
obtain a high occupation probability of the target site even at steady-state. This active
process may be loosely thought of as cell division or degradation and
production of the protein.

In this case, for $n_{p}=1$ the equation for the occupation probabilities
are given by
\begin{align}
\frac{\partial P^{\mathcal{T}}}{\partial t} & =\lambda^{\mathcal{T}}\left(1-P^{\mathcal{T}}-P^{\mathcal{D}}\right)-\left(\lambda_{-1}^{\mathcal{T}}+\Omega\right)P^{\mathcal{T}}\nonumber \\
\frac{\partial P^{\mathcal{D}}}{\partial t} & =\lambda^{\mathcal{D}}\left(1-P^{\mathcal{T}}-P^{\mathcal{D}}\right)-\left(\lambda_{-1}^{\mathcal{D}}+\Omega\right)P^{\mathcal{D}}.
\end{align}
In the steady-state ($\partial P^{\mathcal{T}}/\partial t=0$)
the target site occupation probability is, therefore,
\begin{equation}
P^{\mathcal{T}}=\lambda^{\mathcal{T}}\frac{1+\frac{\lambda_{-1}^{\mathcal{D}}}{\Omega}}{\Omega+\frac{\lambda_{-1}^{\mathcal{T}}\lambda^{\mathcal{D}}+\lambda_{-1}^{\mathcal{D}}\left(\lambda^{\mathcal{T}}+\lambda_{-1}^{\mathcal{T}}\right)}{\Omega}+\left(\lambda^{\mathcal{D}}+\lambda_{-1}^{\mathcal{D}}+\lambda^{\mathcal{T}}+\lambda_{-1}^{\mathcal{T}}\right)}.
\end{equation}
If in absence of an active process ($\Omega=0$) the steady-state
occupation of the unbound state is small, the \textquotedbl{}on\textquotedbl{}
rates, $\lambda^{\mathcal{D}}$ and $\lambda^{\mathcal{T}}$ are much
larger than the \textquotedbl{}off\textquotedbl{} rates, $\lambda_{-1}^{\mathcal{D}}$
and $\lambda_{-1}^{\mathcal{T}}$. In this case one obtains three
regimes depending on the value of $\Omega$
\begin{equation}
P^{\mathcal{T}}\simeq\left\{ \begin{array}{cc}
\frac{\lambda^{\mathcal{T}}}{\frac{\lambda_{-1}^{\mathcal{T}}\lambda^{\mathcal{D}}}{\lambda_{-1}^{\mathcal{D}}}+\lambda^{\mathcal{T}}+\lambda_{-1}^{\mathcal{T}}} & \Omega\ll\lambda_{-1}^{\mathcal{D}},\lambda_{-1}^{\mathcal{T}}\\
\frac{1}{1+\frac{\lambda^{\mathcal{D}}}{\lambda^{\mathcal{T}}}} & \lambda^{\mathcal{D}},\lambda^{\mathcal{T}}\gg\Omega\gg\lambda_{-1}^{\mathcal{D}},\lambda_{-1}^{\mathcal{T}}\\
\frac{\lambda^{\mathcal{T}}}{\Omega} & \Omega\gg\lambda^{\mathcal{D}},\lambda^{\mathcal{T}}\end{array}\right..
\end{equation}
Note that in the second regime the occupation of the target site can by significant.

\begin{figure}[ptb]

\begin{centering}
\includegraphics[width=10cm]{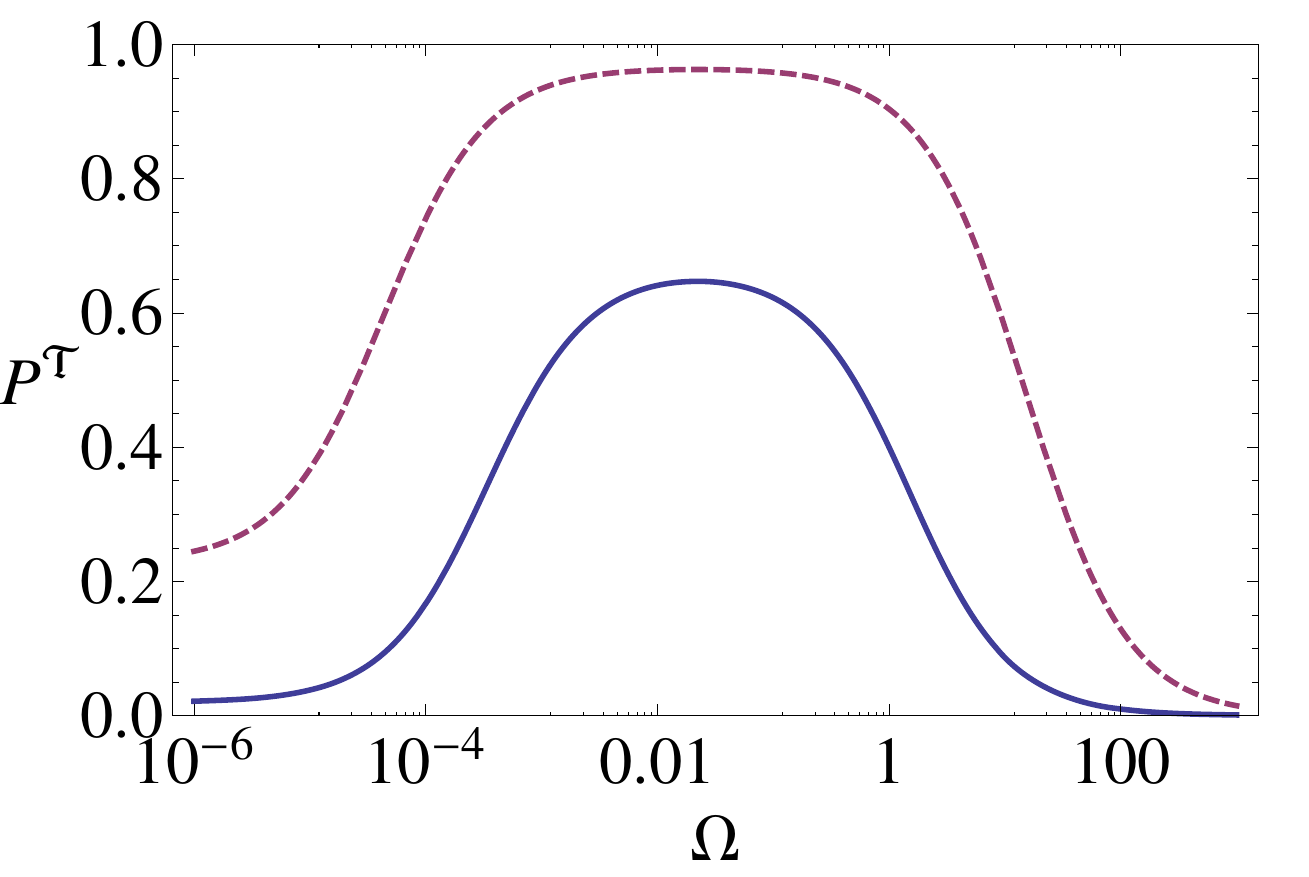}
\par\end{centering}

\caption{In this figure the occupation probability of the target site in the
steady state is shown as a function of $\Omega$. The parameters are
$\lambda^{\mathcal{T}}=1$, $\lambda_{-1}^{\mathcal{T}}=10^{-3}$,
$\lambda^{\mathcal{D}}=0.5$ and $\lambda_{-1}^{\mathcal{D}}=10^{-5}$.
The solid blue (dashed red) line represents the $n_{p}=1$
($n_{p}=15$) case.}

\label{fig9ChBarrier}
\end{figure}

Similar to above, the approach can be generalized to several proteins (we use the same notation as in the previous subsection). When a few proteins act together the mean-field equations for the occupation
probabilities are given by
\begin{align}
\frac{\partial n^{\mathcal{T}}}{\partial t} & =\lambda^{\mathcal{T}}\left(n_{p}-n^{\mathcal{T}}-n^{\mathcal{D}}\right)\left(1-n^{\mathcal{T}}\right)-\left(\lambda_{-1}^{\mathcal{T}}+\Omega\right)n^{\mathcal{T}}\nonumber \\
\frac{\partial n^{\mathcal{D}}}{\partial t} & =\lambda^{\mathcal{D}}\left(n_{p}-n^{\mathcal{T}}-n^{\mathcal{D}}\right)-\left(\lambda_{-1}^{\mathcal{D}}+\Omega\right)n^{\mathcal{D}}.
\end{align}
Here $n^{\mathcal{T}}$and $n^{\mathcal{D}}$ are the mean occupations numbers in the $\mathcal{T}$ and $\mathcal{D}$ state respectively.
Assuming, as before, that without an active process the protein in
equilibrium spends most of its time bound to the DNA we obtain in steady-state
\begin{equation}
n^{\mathcal{T}}\simeq\left\{ \begin{array}{cc}
\frac{\lambda^{\mathcal{T}}n_{p}}{\frac{\lambda_{-1}^{\mathcal{T}}\lambda^{\mathcal{D}}}{\lambda_{-1}^{\mathcal{D}}}+\lambda^{\mathcal{T}}n_{p}+\lambda_{-1}^{\mathcal{T}}} & \Omega\ll\lambda_{-1}^{\mathcal{D}},\lambda_{-1}^{\mathcal{T}}\\
\frac{1}{1+\frac{\lambda^{\mathcal{D}}}{\lambda^{\mathcal{T}}n_{p}}} & n_{p}\lambda^{\mathcal{D}},n_{p}\lambda^{\mathcal{T}}\gg\Omega\gg\lambda_{-1}^{\mathcal{D}},\lambda_{-1}^{\mathcal{T}}\\
\frac{\lambda^{\mathcal{T}}}{\Omega}n_{p} & \Omega\gg n_{p}\lambda^{\mathcal{D}},n_{p}\lambda^{\mathcal{T}}\end{array}\right..
\end{equation}
 The optimal $\Omega$ (that maximizes the steady-state occupation
of the target, $n^{\mathcal{T}}$) is independent of $n_{p}$ and
given by \begin{equation}
\Omega_{opt}=\sqrt{\lambda_{-1}^{\mathcal{T}}\lambda^{\mathcal{D}}-\lambda_{-1}^{\mathcal{D}}\lambda^{\mathcal{D}}}-\lambda_{-1}^{\mathcal{D}}.\label{OmOpt}\end{equation}
 In Fig. \ref{fig9ChBarrier} the steady-state probability of the
target site as a function of the rate of the active process, $\Omega$, is shown.

Summarizing, non-equilibrium effects of the barrier discrimination
between DNA sites may lead to a high target occupation probability even at steady-state.

\subsection{The possibility of the genetic temporal ordering}

It is often the case that each TF activates more than one gene  \cite{A2007}. For example, in \textit{E. coli}
there are $68$ transcription factors which individually regulate
more than $13$ operons \cite{RMC98,SMMA2002}. In some cases the activation of different genes, regulated by the same TF, are
temporally ordered \cite{KMPSRLSA2001,RRSA2002,ZMRBSTSA2004}.
In these systems it seems that the temporal ordering is not caused by
the transcriptional network (for example, by a genetic cascade). It
was suggested \cite{AlonBook,GMH2002} that different genes have different
activation thresholds. In this case a temporally increased concentration
of the transcription factor activates them one-by-one. Different thresholds
arise from non-linear effects, such as cooperativity between the transcription
factors. Recently \cite{KWLGM2007,WM2008,BLV2008} it was proposed that genetic temporal
ordering may be influenced by different distances between the production location of the TF and its target
(this mechanism seems plausible only for prokaryotic cells).

\begin{figure}[ptb]

\begin{centering}
\includegraphics[width=6cm]{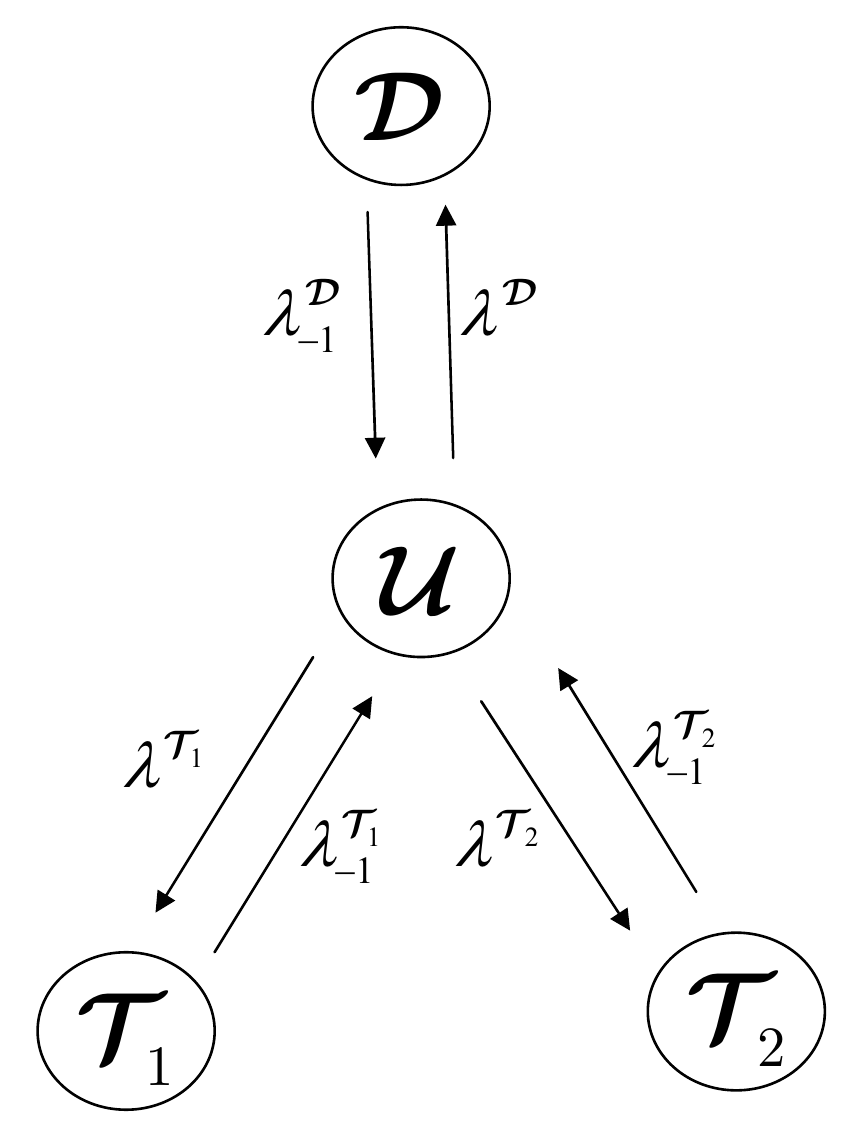}
\par\end{centering}

\caption{Schematic representation of the four-states effective system for
temporal ordering. Here $\mathcal{T}_{1}$ denotes the $r$ state
of the first target, $\mathcal{T}_{2}$ denotes the $r$ state of
the second target, $\mathcal{D}$ denotes the $r$ state out of both
targets and $\mathcal{U}$ denotes the remaining states (including the
$s$ state and the unbound state).}

\label{fig12ChBarrier}
\end{figure}

Here we show that a search mechanism based on barrier discrimination can also
lead to temporal ordering. This does not rely on cooperativity
and appears even for a TF with a constant concentration. To
show this we generalize the effective three states model to four states
by adding an additional target site (see Fig. \ref{fig12ChBarrier}).
Now the states of the model are the $r$ state of the first target ($\mathcal{T}_{1}$),
the $r$ state of the second target ($\mathcal{T}_{2}$), the $r$
state out of both targets ($\mathcal{D}$) and the $\mathcal{U}$
states (including the $s$ states and the unbound state).

\begin{figure}[ptb]
\begin{centering}
\includegraphics[width=13cm]{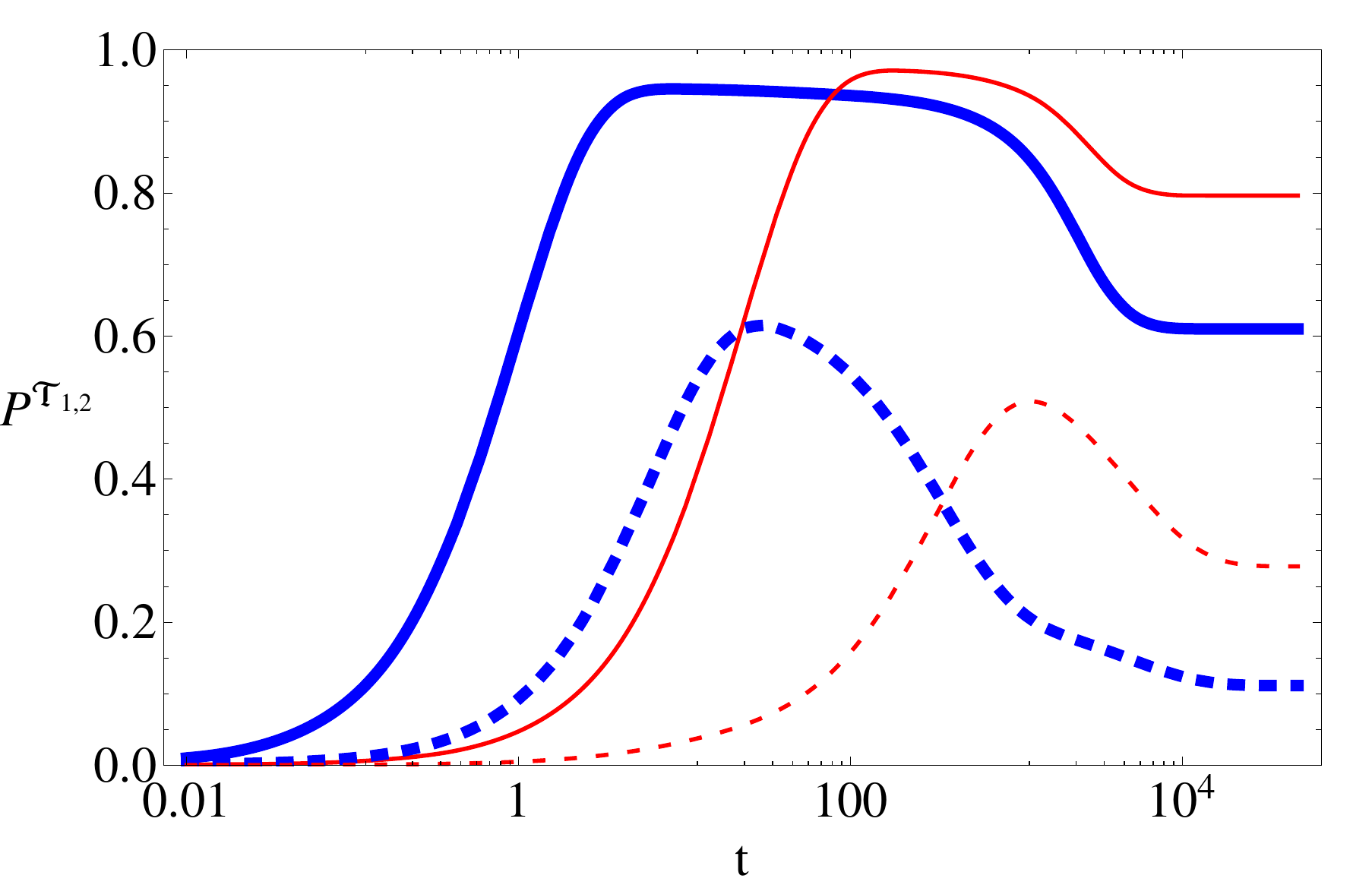}
\end{centering}
\caption{On this graph the occupation probabilities of two target sites, $P^{\mathcal{T}_{1}}$
(blue, thick lines) and $P^{\mathcal{T}_{2}}$ (red, thin lines), are shown as
a function of time. The parameters are $\lambda^{\mathcal{T}_{1}}=0.05$,
$\lambda_{-1}^{\mathcal{T}_{1}}=0.005$, $\lambda^{\mathcal{T}_{2}}=5\cdot10^{-3}$,
$\lambda_{-1}^{\mathcal{T}_{2}}=10^{-3}$, $\lambda^{\mathcal{D}}=10^{-3}$
and $\lambda_{-1}^{\mathcal{D}}=10^{-4}$. The solid (dashed) lines
represent the $n_{p}=10$ ($n_{p}=1$) case.}
\label{fig11ChBarrier}
\end{figure}

For $n_{p}=1$ the evolution equations for the occupation probability
of the first target, $P^{\mathcal{T}_{1}}$, the second target, $P^{\mathcal{T}_{2}}$,
and the rest of the DNA, $P^{\mathcal{D}}$ are given by
\begin{align}
\frac{\partial P^{\mathcal{T}_{1}}}{\partial t} & =\lambda^{\mathcal{T}_{1}}\left(1-P^{\mathcal{T}_{1}}-P^{\mathcal{T}_{2}}-P^{\mathcal{D}}\right)-\lambda_{-1}^{\mathcal{T}_{1}}P^{\mathcal{T}_{1}}\nonumber \\
\frac{\partial P^{\mathcal{T}_{2}}}{\partial t} & =\lambda^{\mathcal{T}_{2}}\left(1-P^{\mathcal{T}_{1}}-P^{\mathcal{T}_{2}}-P^{\mathcal{D}}\right)-\lambda_{-1}^{\mathcal{T}_{2}}P^{\mathcal{T}_{2}}\nonumber \\
\frac{\partial P^{\mathcal{D}}}{\partial t} & =\lambda^{\mathcal{D}}\left(1-P^{\mathcal{T}_{1}}-P^{\mathcal{T}_{2}}-P^{\mathcal{D}}\right)-\lambda_{-1}^{\mathcal{D}}P^{\mathcal{D}}
\end{align}
 while the occupation probability of the $\mathcal{U}$ state is determined
by
\begin{equation}
P^{\mathcal{U}}=1-P^{\mathcal{T}_{1}}-P^{\mathcal{T}_{2}}-P^{\mathcal{D}}.
\end{equation}
In the  case of a few proteins the mean-field equations for the evolution of the occupation
probability are\begin{align}
\frac{\partial n^{\mathcal{T}_{1}}}{\partial t} & =\lambda^{\mathcal{T}_{1}}\left(n_{p}-n^{\mathcal{T}_{1}}-n^{\mathcal{T}_{2}}-n^{\mathcal{D}}\right)\left(1-n^{\mathcal{T}_{1}}\right)-\lambda_{-1}^{\mathcal{T}_{1}}n^{\mathcal{T}_{1}}\nonumber \\
\frac{\partial n^{\mathcal{T}_{2}}}{\partial t} & =\lambda^{\mathcal{T}_{2}}\left(n_{p}-n^{\mathcal{T}_{1}}-n^{\mathcal{T}_{2}}-n^{\mathcal{D}}\right)\left(1-n^{\mathcal{T}_{2}}\right)-\lambda_{-1}^{\mathcal{T}_{2}}n^{\mathcal{T}_{2}}\nonumber \\
\frac{\partial n^{\mathcal{D}}}{\partial t} & =\lambda^{\mathcal{D}}\left(n_{p}-n^{\mathcal{T}_{1}}-n^{\mathcal{T}_{2}}-n^{\mathcal{D}}\right)-\lambda_{-1}^{\mathcal{D}}n^{\mathcal{D}}\end{align}
 with
 \begin{equation}
n^{\mathcal{U}}=n_{p}-n^{\mathcal{T}_{1}}-n^{\mathcal{T}_{2}}-n^{\mathcal{D}}.
\end{equation}
Here $n^{\mathcal{U}}, n^{\mathcal{T}_1},n^{\mathcal{T}_2}$ and $n^{\mathcal{D}}$ are the mean occupation numbers in the $\mathcal{U},\mathcal{T}_1,\mathcal{T}_2$ and $\mathcal{D}$ states respectively.
These equations may be solved analytically for the $n_{p}=1$ case
and for the $n_{p}\gg1$ case. In Fig. \ref{fig11ChBarrier} one may
see that by the tuning transition rates it is possible to obtain a temporal
ordering of gene activation or/and repression for a single protein
and few ($n_{p}=10$) proteins. Genes $\mathcal{T}_{1}$ and $\mathcal{T}_{2}$
are activated or repressed at different times depending on their association
rates. The subsequent\ deactivation or/and repression of genes
$\mathcal{T}_{1}$ and $\mathcal{T}_{2}$ take place if $\frac{\lambda^{\mathcal{T}_{1,2}}}{\lambda_{-1}^{\mathcal{T}_{1,2}}}\frac{\lambda_{-1}^{\mathcal{D}}}{\lambda^{\mathcal{D}}}\ll1$
and occurs also at different times depending on their dissociation
rates. This results, in principle, can be generalized to any
number of genes.

\section{Summary}

Search and recognition problems appear in many contexts in
biological systems. Organisms activate/repress many processes by
specifically designed proteins. This may be an enzyme that
catalyzes some chemical reaction by binding to specific molecules,
a transcription factor that changes the transcriptional activity
by binding to some specific locations on the DNA, etc. In fact,
the flow of information, from a genotype to a phenotype and vice
versa is regulated and implemented by searchers for specific
DNA/RNA sequences. Needless to say, every particular biological
searcher \textquotedbl{}invented\textquotedbl{} its own search
strategy. However, there is hope that it is possible to divide all
search strategies to a few classes similarly to the division of
the transcription factors' binding domains to a few DNA binding
motifs \cite{Luscombe2000}. Different aspects of a transcription
factor may dictate its strategy: the number of copies in the cell,
its structure and function, its interactions with other proteins, the number of its
targets  and many more.
The beginning of this review suggested a possible classification
of TFs. It is certainly of great interest to extend it and test it on
large databases. A particular simple classification arose when
the length of the binding site of the protein to the DNA was
considered.

The bulk part of the review deals with a possible scenario where
TFs locate their target using barrier discrimination. This is
different from  other mechanisms which assume a local equilibration
of the TF with its environment before the target is located. A
more detailed discussion of  those can be found in many other reviews
\cite{HB89,HM2004,pccp2008,Mirny2009,kolo2010,rmp2011}. When assuming equilibration, each site on
the DNA can be characterized by a single parameter - the binding
energy at the site. According to the equilibrium assumption, this
is the only parameter that may discriminate between different
sites. On the contrary in the barrier model the search kinetics
can become widely independent of the binding energy, and controlled by the barrier height. This gives a
particularly simple resolution to the speed-stability paradox
described in detail in this review. Moreover, it suggests that TFs
can be weakly bound to the target site for extended period. We
termed this transient activation. The end of the review dealt with more speculative processes that may occur in a barrier mechanism. These include  time ordering in the  activation of genes.

Under the conditions described above the search process in the barrier model is not characterized by
a single time scale but by two. One short and one long. This leads to a distinction between typical search times,
which are relevant to experiments, and average search times. The latter are dominated by rare events.

A clear indication that the barrier mechanism is at work in-vivo would lie  in  measurements of the full FPT distribution of a TF at a target gene,  and the observation of more than one time scale.  Even if the FPT distribution shows only a single time scale,  the model predicts that the appearance of a new short time scale as $n_p$, number of searchers, is increased. Therefore, the appearance of the new time scale would be a clear indiction for the presence of the mechanism.  In this case the typical search time will generally not scale as $1/n_p$. Such experimental data  are not yet available but in principle accessible, and will hopefully provide in the near future  a better understanding of the kinetics of gene activation.

\vspace{2cm}
{\bf Acknowledgments}: We are grateful for many stimulating and critical discussions with E. Braun, A. Finkelstein, K. Keren and D. Levine.

\appendix
%dummy comment inserted by tex2lyx to ensure that this paragraph is not empty
%dummy comment inserted by tex2lyx to ensure that this paragraph is not empty
%dummy comment inserted by tex2lyx to ensure that this paragraph is not empty
%dummy comment inserted by tex2lyx to ensure that this paragraph is not empty
%dummy comment inserted by tex2lyx to ensure that this paragraph is not empty

\section{An information theoretic approach to the calculation of disorder parameters
and binding energies\label{An information theoretic approach}}

In Sections \ref{Protein-DNA energetics} we characterized
the ability of a protein to recognize its target by the free
energy gap between the target and the rest of the DNA sites and the equilibrium target occupation probability. As an
alternative approach one often uses the information content of a protein,
denoted by $IC$ \cite{SSGE86,HS99}. While our personal preference is to the presentation used in the main text
the two can be used interchangeably. In this appendix we provide a brief overview of the relations between the information
theory quantities (the information content and the sequence score) and the physical
(the disorder strength and the binding energy of a sequence) quantities.

\subsection{The information content and the disorder strength\label{The information content and the disorder strength}}

Before turning to the information content of a protein of length $l_p$ we first discuss the information content associated with a single binding site $i$. Assuming that the frequency of each nucleotide
in the genome is close to $1/4$ \cite{EcoliGenome}, this quantity, denoted by $IC_{i}$,
is given by its maximal amount of information minus its Shannon entropy.
Note that we consider the $IC$ of the ``specific'' protein conformation.
The non-specific one, presumably, does not contain any information. With this in mind we have
\begin{equation}
IC_{i}=-\log_2(1/4)+\underset{s=\left\{ A,T,C,G\right\} }{\sum}\Pr\left(s,i\right)\log_{2}\Pr\left(s,i\right)\label{IC}
\end{equation}
 where $\Pr\left(s,i\right)$ is defined in Eq. (\ref{PrCI}). This quantity is maximal when only one nucleotide
type can bind to site $i$. In this case $\Pr\left(s,i\right)=\delta_{s,s^{\prime}}$
so that $IC_{i}=2bits$. If two types of nucleotides can bind with equal
probability ($\Pr\left(s,i\right)=\frac{1}{2}\delta_{s,s^{\prime}}+\frac{1}{2}\delta_{s,s^{\prime\prime}}$)
the information content is reduced to one bit. In a case when each of the four
types of nucleotides can bind with an equal probability of $\frac{1}{4}$
the information content is zero\footnote{Remarkably the average information content of one binding site of a TF is $1.054bits$ for TFs from the RegulonDB database \cite{regulonDB}.}.

With these definitions the total information content of the protein is given by a sum of information contents of all protein
binding sites:
\begin{equation}
IC=\underset{i=1}{\overset{l_{p}}{\sum}}IC_{i}.\label{ICtot}
\end{equation}
To identify the target this has to be larger than $\log_{2}N$. Using
Eqs. (\ref{Ueps}),(\ref{PrCI}) and (\ref{ICtot}) we obtain
\begin{align}
IC & =2l_{p}+\underset{\textbf{s}}{\sum}\frac{e^{-U\left(\textbf{s}\right)}}{\underset{\textbf{s}}{\sum}e^{-U\left(\textbf{s}\right)}}\log_{2}\frac{e^{-U\left(\textbf{s}\right)}}{\underset{\textbf{s}}{\sum}e^{-U\left(\textbf{s}\right)}}=\nonumber \\
 & =2l_{p}-\underset{\textbf{s}}{\sum}\frac{\frac{U\left(\textbf{s}\right)}{\ln2}e^{-U\left(\textbf{s}\right)}}{\underset{\textbf{s}}{\sum}e^{-U\left(\textbf{s}\right)}}-\log_{2}\underset{\textbf{s}}{\sum}e^{-U\left(\textbf{s}\right)}=\nonumber \\
 & =2l_{p}-\left(1-\frac{\partial}{\partial\beta}\right)\left(\log_{2}\underset{\textbf{s}}{\sum}e^{-\beta U\left(\textbf{s}\right)}\right)_{\beta=1}.\end{align}
Since, as we mentioned above, $U\left(\textbf{s}\right)$
behaves to a good approximation as a Gaussian random variable the
information content is given by
\begin{equation}
IC=2l_{p}-\left(1-\frac{\partial}{\partial\beta}\right)\left\langle \log_{2}\underset{i=1}{\overset{4^{l_{p}}}{\sum}}e^{-\beta U_{i}}\right\rangle _{\beta=1}\label{ICave}
\end{equation}
where $\left\{ U_{i}\right\} $ is a set of Gaussian random variables
with a probability density
\begin{equation}
\Pr(U_{i})=\frac{e^{-\frac{U_{i}^{2}}{2\sigma_{U}^{2}}}}{\sqrt{2\pi\sigma_{U}^{2}}}
\end{equation}
 and the angular brackets denote an average over realizations
of disorder. The expression $\left\langle \log_{2}\underset{i=1}{\overset{4^{l_{p}}}{\sum}}e^{-\beta U_{i}}\right\rangle $
is similar to the free energy of the Random Energy Model and in the
limit of a long protein ($l_{p}\gg1$) may be solved using ideas developed in
\cite{D1981}. This gives
\begin{equation}
\left\langle \log_{2}\underset{i=1}{\overset{4^{l_{p}}}{\sum}}e^{-\beta U_{i}}\right\rangle =2l_{p}+\left\langle \log_{2}\frac{\underset{U_{\min}}{\overset{\infty}{\int}}e^{-\frac{U^{2}}{2\sigma_{U}^{2}}}e^{-\beta U}dU}{\underset{U_{\min}}{\overset{\infty}{\int}}e^{-\frac{U^{2}}{2\sigma_{U}^{2}}}dU}\right\rangle
\end{equation}
where $U_{\min}$ is the minimal observed energy which is well approximated
by
\begin{equation}
\underset{-\infty}{\overset{U_{\min}}{\int}}\frac{e^{-\frac{U^{2}}{2\sigma_{U}^{2}}}}{\sqrt{2\pi\sigma_{U}^{2}}}dU=\frac{1}{4^{l_{p}}}
\end{equation}
 or
 \begin{equation}
U_{\min}=-\sigma_{U}\sqrt{2}\operatorname{erfc}^{-1}\left(\frac{2}{4^{l_{p}}}\right)\simeq-2\sigma_{U}\sqrt{l_{p}\ln2}.\end{equation}
 In the limit of a long protein one obtains\begin{equation}
\left\langle \log_{2}\underset{i=1}{\overset{4^{l_{p}}}{\sum}}e^{-\beta U_{i}}\right\rangle =\left\{ \begin{array}{cc}
0 & \sigma_{U}\geq2\sqrt{l_{p}\ln2}\\
2l_{p}+\frac{\beta^{2}\sigma_{U}^{2}}{2\ln2} & \sigma_{U}<2\sqrt{l_{p}\ln2}\end{array}\right..\end{equation}
 Using Eq. (\ref{ICave}) the information content is finally given by\begin{equation}
IC=\left\{ \begin{array}{cc}
2l_{p} & \sigma_{U}\geq2\sqrt{l_{p}\ln2}\\
\frac{\sigma_{U}^{2}}{2\ln2} & \sigma_{U}<2\sqrt{l_{p}\ln2}\end{array}\right..\label{ICsigma}\end{equation}

Schneider \textit{et al.} \cite{SSGE86} suggested, and showed for
a few transcription factors, that the information content is just
sufficient for the target to be distinguished from the rest of the
genome. Specifically, with $N$ potential binders, where $N$ is twice the genome length ($N\simeq10^{7}bp$ for \textit{E. coli}), the amount of the information needed to distinguish a single binder is $\log_{2}N$. Therefore, we would expect
\begin{equation}
IC\gtrsim\log_{2}N\label{ICest}
\end{equation}
If this is not fulfilled one expects a wrong sequence to be incorrectly recognized.
Comparing Eqs. (\ref{ICsigma}) and (\ref{ICest}) this translated to a condition on the length of the protein
\begin{equation}
l_{p}\gtrsim\frac{\log_{2}N}{2} \label{l_pCond}
\end{equation}
and
\begin{equation}
\sigma_{U}\gtrsim\sqrt{2\ln N}.\label{sigmaN}
\end{equation}
This is identical to results of the simple arguments presented in Sec. \ref{Protein-DNA energetics}. This condition is based on the assumption that the maximal information content per TF's length is two \emph{bits}. However, we find that the average information content per TF's length is actually close to one \emph{bit}. This fact increases the minimal protein's length from $11$ to $22$ basepairs. Similar conclusion may be obtained from Fig. \ref{Ptlp}.

\begin{figure}[ptb]

\begin{centering}
\includegraphics[width=10cm]{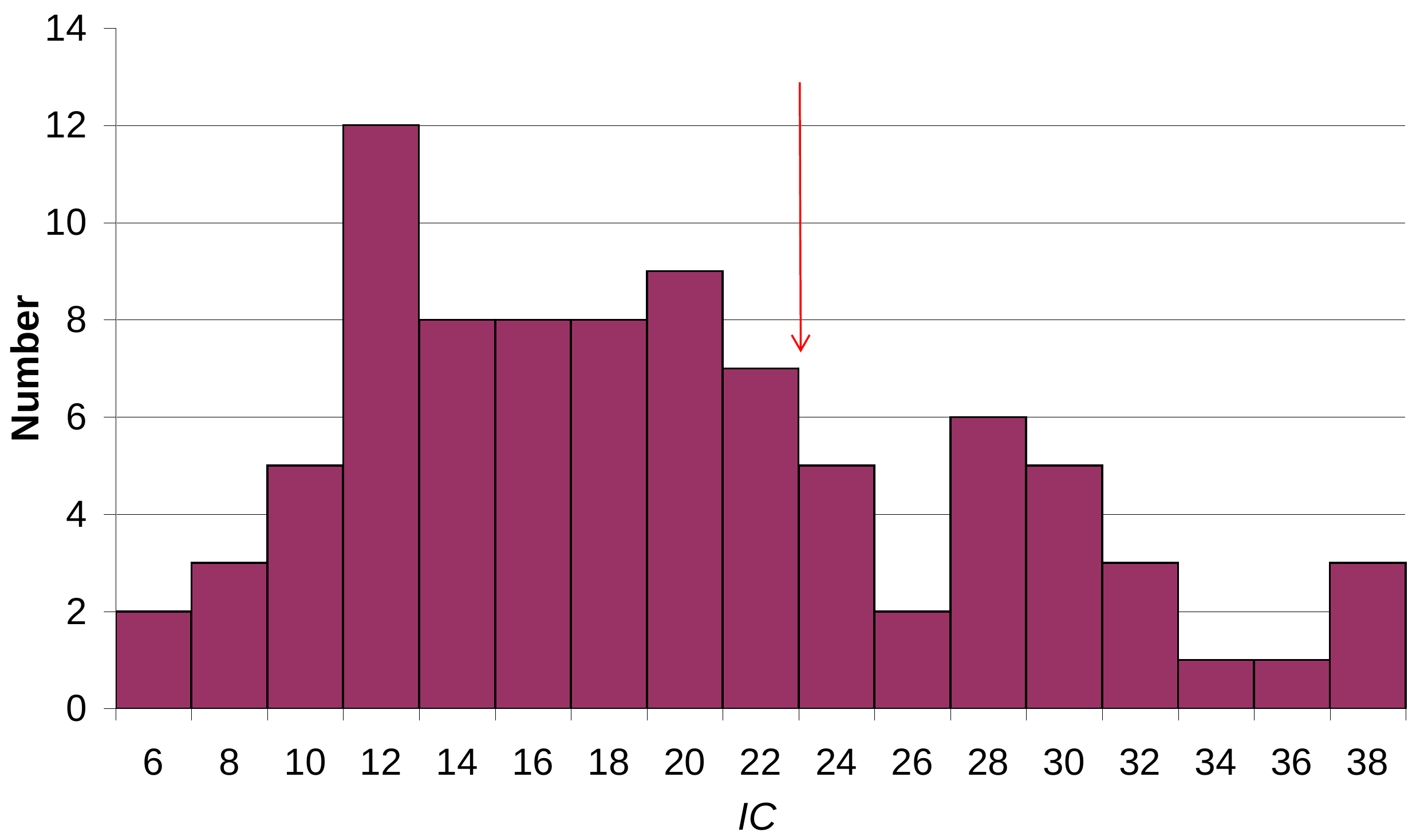}
\par\end{centering}

\caption{A histogram of the information content, $IC$.
The data is based on the $89$ weight matrices of \textit{E. coli} DNA-binding
proteins from RegulonDB database \cite{regulonDB}.
The prediction of Eq. (\ref{ICest}) is $23bits$ and is represented
on the figure by the red arrow.}

\label{figCImyChBarrier}
\end{figure}

One can see that the condition for a broad line information content
(\ref{l_pCond}) and (\ref{sigmaN}) are identical to the conditions for a marginally gapped target with a significant occupation probability (see Sec. \ref{Protein-DNA energetics}). In the
limit of $N\rightarrow\infty$ both conditions represent the freezing
point of the Random Energy Model \cite{D1981}.
In Fig. \ref{figCImyChBarrier}
a histogram of the information content of several DNA-binding proteins is
presented. Note that many proteins have significantly less
information about the target than predicted by Eq. (\ref{ICest}).
This conclusion is independent of any assumption on the
binding energy distribution. More detailed study which includes eukaryotic TFs and other databases was done in Ref. \cite{Wunderlich2009}. In this work authors found that for an eukaryotic TFs the problem of the not sufficient information content is much more severe.

\subsection{The sequence score and the binding energy\label{The sequence score and its binding energy}}

We showed above that the information content of many proteins is not
sufficient for efficient target location. In Sec. \ref{Protein-DNA energetics} we
present the same results, in particular, by comparing the binding
energy of the target to the rest of the DNA. To this end we had to obtain the binding energy of a given sequence. In an information theory context this may be evaluated
using a sequence score \cite{Staden1984}. The score of a sequence
$\textbf{s}=\left(s_{1},s_{2},...,s_{l_{p}}\right)$ is defined
as\begin{equation}
Sc\left(\textbf{s}\right)=\underset{i=1}{\overset{l_{p}}{\sum}}\ln\left[4\Pr\left(s_{i},i\right)\right]=l_{p}\ln4+\ln\left[\Pr\left(\textbf{s},i\right)\right]\end{equation}
 where $\Pr\left(s_{i},i\right)$ is defined in Eq. (\ref{PrCI})
and\begin{equation}
\Pr\left(\textbf{s},i\right)=\underset{i=1}{\overset{l_{p}}{{\displaystyle \prod}}}\Pr\left(s_{i},i\right).\end{equation}
 The probability that a sequence $\textbf{s}$ is bound is
proportional to the Boltzmann factor $e^{-U\left(\textbf{s}\right)}$.
Therefore, the binding energy, $U\left(\textbf{s}\right)$, is
equal to the score of sequence $\textbf{s}$. Note that this convention differs from the one used in the text by a constant.
The constant may be calculated by recalling that we defined the binding
energy so that its average is zero. Therefore,
\begin{align}
U\left(\textbf{s}\right) & =-Sc\left(\textbf{s}\right)+l_{p}\ln4+\frac{1}{4}\ln\left[\underset{i=1}{\overset{l_{p}}{{\displaystyle \prod}}}\underset{s_{i}^{\prime}=\left\{ A,T,C,G\right\} }{{\displaystyle \prod}}\Pr\left(s_{i}^{\prime},i\right)\right]=\nonumber\\
 & =-\underset{i=1}{\overset{l_{p}}{\sum}}\ln\frac{\Pr\left(s_{i},i\right)}{\sqrt[4]{\underset{s_{i}^{\prime}=\left\{ A,T,C,G\right\} }{{\displaystyle \prod}}\Pr\left(s_{i}^{\prime},i\right)}}=\underset{i=1}{\overset{l_{p}}{\sum}}u_{i}\left(s_{i}\right).\label{Umin}
 \end{align}
where
\begin{equation}
u_{i}\left(s_{i}\right)=-\ln\frac{\Pr\left(s_{i},i\right)}{\sqrt[4]{\underset{s_{i}^{\prime}=\left\{ A,T,C,G\right\} }{{\displaystyle \prod}}\Pr\left(s_{i}^{\prime},i\right)}}\label{Emin}\end{equation}
is the contribution of site $i$ on the protein to the total binding
energy of a sequence $\textbf{s}$. Eqs. (\ref{Umin},\ref{Emin}) are used
in Section \ref{Experimental data} to calculate
(for each transcription factor) the binding energy of a given sequence.

\section{A derivation of Eqs. (\ref{j(s)}) and (\ref{j0(s)})\label{MontrollCalc}}

In this appendix we analyze the FPT properties of a simple random
walk and derive Eqs. (\ref{j(s)}) and (\ref{j0(s)}). A similar calculation can be found in Ref. \cite{montroll} and
is presented here for completeness. The Laplace transformed
FPT probability density of the discrete space and continuum time random walk,
\begin{equation}
\widetilde{j}\left(x|x_{0};s\right)=\int_{0}^{\infty}e^{-st}j\left(x|x_{0};t\right)dt,
\end{equation}
may be expressed in terms of the $z$-transformed FPT probability
density of the discrete time random walk,
\begin{equation}
\widetilde{j}_{d}\left(x|x_{0};z\right)=\underset{t=1}{\overset{\infty}{\sum}}z^{t}j_{d}\left(x|x_{0};t\right),
\end{equation}
by \cite{H95Volume1}
\begin{equation}
\widetilde{j}\left(x|x_{0};s\right)=\widetilde{j}_{d}\left(x|x_{0};\frac{1}{1+s/\lambda_{0}}\right).\label{DisCon}
\end{equation}
In discrete space and discrete time the occupation probability
at origin starting from the site $x\geq0$ at $t=0$\ is given by
\begin{equation}
\widetilde{P}\left(x|x_{0};z\right)=\frac{\left(\frac{1-\sqrt{1-z^{2}}}{z}\right)^{\left\vert x-x_{0}\right\vert }}{\sqrt{1-z^{2}}}.
\end{equation}
 Using relation \cite{MW1965} that connects the FPT to the origin
with the occupation probabilities
\begin{equation}
\widetilde{j}_{d}\left(x|x_{0};z\right)=\frac{\widetilde{P}\left(x|x_{0};z\right)-\delta_{x,x_{0}}}{\widetilde{P}\left(x|x;z\right)},
\end{equation}
the $z$-transformed probability density of the first return time
to site $0$ of a discrete time random walk is given by
\begin{equation}
\widetilde{j}_{d}\left(0|0;z\right)=1-\frac{1}{\widetilde{P}\left(0|0;z\right)}=1-\sqrt{1-z^{2}}
\end{equation}
With Eq. (\ref{DisCon}) we obtain the Laplace transform of the
first return time of a continuous time random walk (we denote this
quantity in the text by ${\tilde{j}}_{0}\left(s\right)$)
\begin{equation}
\widetilde{j}\left(x|x_{0};s\right)=1-\sqrt{1-\left(\frac{1}{1+s/\lambda_{0}}\right)^{2}}\simeq1-\sqrt{1-e^{-2s/\lambda_{0}}}.
\end{equation}
 The $z$-transformed probability density of the first-passage time
to site $0$ from site $x$ of a discrete time random walk is given
by
\begin{equation}
\widetilde{j}_{d}\left(0|x_{0};z\right)=\frac{\widetilde{P}\left(0|x_{0};z\right)}{\widetilde{P}\left(0|0;z\right)}=\left(\frac{1-\sqrt{1-z^{2}}}{z}\right)^{\left\vert x_{0}\right\vert }.\end{equation}
 Its average over all possible starting sites in the large $N$ limit
gives\begin{equation}
\left\langle \widetilde{j}_{d}\left(0|x_{0};z\right)\right\rangle _{x_{0}}=\frac{1}{N}\underset{x_{0}=-N/2}{\overset{N/2}{\sum}}\left(\frac{1-\sqrt{1-z^{2}}}{z}\right)^{\left\vert x_{0}\right\vert }\simeq\frac{2}{N}\underset{x_{0}=1}{\overset{N/2}{\sum}}\left(\frac{1-\sqrt{1-z^{2}}}{z}\right)^{x_{0}}\simeq\frac{2}{N}\frac{1}{1-\frac{1-\sqrt{1-z^{2}}}{z}} \;.
\end{equation}
Using Eq. (\ref{DisCon}) we obtain the Laplace transformed first
passage time to site $0$ averaged over the initial sites of a discrete
time random walk (we denote this quantity in the text by $\tilde{j}(s)\equiv\left\langle \tilde{j}\left(s|x_{0}\right)\right\rangle _{x_{0}}$)
\begin{equation}
\left\langle \widetilde{j}\left(0|x_{0};s\right)\right\rangle _{x_{0}}=\frac{2}{N}\frac{1}{1-\frac{1-\sqrt{1-\left(\frac{1}{1+s/\lambda_{0}}\right)^{2}}}{1+s/\lambda_{0}}}\simeq\frac{1}{N}\sqrt{\frac{1+e^{-s/\lambda_{0}}}{1-e^{-s/\lambda_{0}}}}.
\end{equation}

\section{Specificity}
\label{specificity}

The conditions discussed in Sec. IV B and Sec. IV C imply that
\begin{equation}
P_{d}\left(n_{p}=1\right)\ll e^{-\sigma_{U}^{2}}.\end{equation}
and
\begin{equation}
P^{{\cal \mathcal{T}}}\left(n_{p}=1\right)\gg P_{d}\left(n_{p}=1\right)\label{SpecificityCond}
\end{equation}
have to hold.

It is useful to separate the discussion into two cases.

\paragraph{Small disorder $\sigma_{U}\ll\sqrt{2\ln N}$}:

In this case\begin{equation}
P^{{\cal \mathcal{T}}}\left(n_{p}=1\right)\simeq\frac{1}{1+Ne^{\sigma_{U}^{2}/2+U_{\mathcal{T}}}}\end{equation}
 \begin{equation}
P_{d}\left(n_{p}=1\right)\simeq\frac{e^{\sigma_{U}\sqrt{2\ln N_{d}}}}{Ne^{\sigma_{U}^{2}/2}}\;,\end{equation}
 so that the conditions (\ref{SpecificityCond}) take the form

\begin{equation}
\sqrt{2\ln N}\gg\sigma_{U}\sqrt{1+\frac{2\sqrt{2\ln N_{d}}}{\sigma_{U}}}\label{1}
\end{equation}

\begin{equation}
U_{\mathcal{T}}\ll-\sigma_{U}\sqrt{2\ln N_{d}}.\label{2}
\end{equation}
Condition (\ref{1}) can be satisfied for $2\sqrt{2\ln N_{d}}\ll\sigma_{U}\ll\sqrt{2\ln N}$ or
for $\sigma_{U}\ll\min\left(\sqrt{2\ln N},2\sqrt{2\ln N_{d}},\frac{\ln N}{4\ln N_{d}}\right)$.
The second condition (\ref{2}) is satisfied automatically in the case
of a non-designed TF (so that $U_{\mathcal{T}}=-\sigma_{U}\sqrt{2\ln N}$),
in the case of designed TF with an additive binding energy (so that $U_{\mathcal{T}}=l_{p}E_{c}=-\sigma_{U}\sqrt{3l_{p}}$
where, as stated above, $l_{p}>\frac{2}{3}\ln N$) and, obviously,
in the case of a gapped TF with cooperative binding energy (so that
the value of $U_{\mathcal{T}}$ is not bound). Thus, in the small
disorder regime the speed-stability paradox can be resolved by increasing
the copy number of a TF without destroying the specificity. For a
small number of dangerous sites, $N_{d}\ll N^{1/4}$, one may do so
with a relatively large disorder while in the opposite case the upper
bound on the disorder strength decreases. The number TF copies is given by

\begin{equation}
n_{p}\sim\max\left(e^{\sigma_{U}^{2}},Ne^{\sigma_{U}^{2}/2+U_{\mathcal{T}}}\right).\end{equation}

\paragraph{Large disorder $\sigma_{U}\gg\sqrt{2\ln N}$:}

In this case\begin{equation}
P^{{\cal \mathcal{T}}}\left(n_{p}=1\right)\simeq1\end{equation}
 \begin{equation}
P_{d}\left(n_{p}=1\right)=\frac{e^{\sigma_{U}\sqrt{2\ln N_{d}}}}{e^{\sigma_{U}\sqrt{2\ln N}}}.\end{equation}
 so that the conditions (\ref{SpecificityCond}) take the form

\begin{equation}
\frac{e^{\sigma_{U}\sqrt{2\ln N_{d}}}}{e^{\sigma_{U}\sqrt{2\ln N}}}\ll e^{-\sigma_{U}^{2}}\end{equation}
 or\begin{equation}
\sqrt{2\ln N}-\sqrt{2\ln N_{d}}\gg\sigma_{U}.\end{equation}
 This condition cannot be satisfied simultaneously with the regime
assumption $\sqrt{2\ln N}\ll\sigma_{U}$. Thus, in the large disorder
regime the speed-stability paradox cannot be resolved by increasing
the copy number of a TF without destroying the specificity.

\section{Details of the simulations \label{Numerics}}

The model is simulated using a standard continuous time Gillespie
algorithm \cite{G1976}. The protein on a given site, $i$, in the
$s$ mode can perform four possible moves: it can move in one
of the possible directions along the DNA with probability $\frac{\lambda_{0}/2}{\lambda_{0}+\lambda_{r}^{i}+\lambda_{u}}$,
go to the $r$ mode with probability $\frac{\lambda_{r}^{i}}{\lambda_{0}+\lambda_{r}^{i}+\lambda_{u}}$
or dissociate from the DNA and reassociate on a randomly chosen site
with probability $\frac{\lambda_{u}}{\lambda_{0}+\lambda_{r}^{i}+\lambda_{u}}$.
In the first two cases time is advanced by an amount drawn from a
Poisson distribution with an average of $\frac{1}{\lambda_{0}+\lambda_{r}^{i}+\lambda_{u}}$.
When the protein dissociates, the time is advanced by first, drawing
a time from a Poissonian distribution with an average time of $\frac{1}{\lambda_{0}+\lambda_{r}^{i}+\lambda_{u}}$,
and adding to it a time, drawn from a Poissonian distribution with
an average $\frac{1}{\lambda_{b}}$. This corresponds to the
time needed for a relocation to a new site. The $r$ mode can transform
only into the $s$ mode. The time for this step is drawn from a Poisson
distribution with an average time $\frac{1}{\lambda_{s}}$.

\section{Pole structure analysis and derivation of Eq. (\ref{2exp})\label{Poles}}

In this appendix we show that Eq. (\ref{2exp}) holds when there is a sufficient
time scale separation in the problem. Following standard practice we perform
the inverse Laplace transform by studying the poles of $\widetilde{\mathcal{R}}\left(s\right)$.
In our case $\widetilde{\mathcal{R}}\left(s\right)$ has no poles
in the region $\operatorname{Re}\left\{ s\right\} >0$ of the complex
plane so that \begin{equation}
\mathcal{R}\left(t\right)=\frac{1}{2\pi i}\int_{-i\infty}^{i\infty}e^{st}\widetilde{\mathcal{R}}\left(s\right)ds=\underset{s_{i}\in poles}{\sum}e^{s_{i}t}\operatorname*{Res}\left[\widetilde{\mathcal{R}}\left(s\right),s_{i}\right].\label{RvsPoles}\end{equation}
 As we showed above (see Eq. (\ref{RLapTrans}))\begin{equation}
\widetilde{\mathcal{R}}(s)=\frac{\tilde{j}_{p_{1}}\left(u\left(s\right)\right)}{s}\left\{ 1-\frac{\lambda_{b}\lambda_{u}}{s+\lambda_{b}}\frac{1-\tilde{j}_{p_{1}}\left(u\left(s\right)\right)}{u\left(s\right)}\right\} ^{-1},\label{Rpoles}\end{equation}
 where
 \begin{equation}
u(s)=\frac{s\left(s+\lambda_{r}+\lambda_{s}+\lambda_{u}\right)+\lambda_{s}\lambda_{u}}{s+\lambda_{s}}.
\end{equation}
 and
 \begin{equation}
\tilde{j}_{p_{1}}\left(s\right)=\frac{p_{1}{\tilde{j}}(s)}{1-(1-p_{1}){\tilde{j}}_{0}(s)},
\end{equation}
 where
 \begin{equation}
\tilde{j}(s)\equiv\left\langle \tilde{j}(s|x)\right\rangle _{x}\simeq\frac{1}{N}\sqrt{\frac{1+e^{-s/\lambda_{0}}}{1-e^{-s/\lambda_{0}}}}
\end{equation}
 {and}
 \begin{equation}
{\tilde{j}}_{0}(s)=1-\sqrt{1-e^{-2s/\lambda_{0}}}
\end{equation}
 for large $N$ \cite{montroll}. Finally, $p_{1}$ is probability
of crossing the barrier at the target at each visit of its $s$ state,
\begin{equation}
p_{1}=\frac{\lambda_{r}^{\mathcal{T}}}{1+\lambda_{u}/\lambda_{0}+\lambda_{r}^{\mathcal{T}}/\lambda_{0}}.
\end{equation}
 The trivial pole of $\widetilde{\mathcal{R}}(s)$ is easily found (from
Eq. (\ref{Rpoles})) to be\begin{equation}
s_{0}=0\label{pol0}\end{equation}
 and its residue is\begin{equation}
\mathrm{Res}_{0}=1.\label{res0}\end{equation}
 Note that a pole in $\widetilde{j}_{p_{1}}\left(u\left(s\right)\right)$
would not lead to a pole in $\widetilde{\mathcal{R}}\left(s\right)$
due to the occurrence of $\widetilde{j}_{p_{1}}\left(u\left(s\right)\right)$
in the numerator and denominator. Thus, the equation for other poles
is given by
\begin{equation}
\left(s+\lambda_{b}\right)u\left(s\right)=\lambda_{b}\lambda_{u}\left[1-\widetilde{j}_{p_{1}}\left(u\left(s\right)\right)\right].\label{PolesEq1}\end{equation}
 Next we assume that $\lambda_{s}\ll\lambda_{r},\lambda_{u},\lambda_{b},\lambda_{0}$
with $\lambda_{u},\lambda_{b},\lambda_{0}$ of comparable order. The
order of $\lambda_{r}$ will be discussed below in more details. We
focus on the interesting regime $s/\lambda_{0}\ll1$ in which the
target is not found immediately. The analysis is carried out by considering
the pole equation at different regimes. Using \begin{equation}
\frac{u\left(s\right)}{\lambda_{0}}=\frac{1}{\lambda_{0}}\frac{s\left(s+\lambda_{r}+\lambda_{s}+\lambda_{u}\right)+\lambda_{s}\lambda_{u}}{s+\lambda_{s}}=\frac{s+\lambda_{r}+\lambda_{u}}{\lambda_{0}}-\frac{\lambda_{s}\lambda_{r}}{\lambda_{0}\left[\lambda_{s}-\left(-s\right)\right]}\end{equation}
 and that $\left(-s\right)\ll\lambda_{0}$ and $\lambda_{r}\ll\lambda_{0}$
one may see that for \begin{equation}
\left\vert \lambda_{s}-\left(-s\right)\right\vert \gg\frac{\lambda_{r}\lambda_{s}}{\lambda_{0}}\label{Cond}\end{equation}
 $\widetilde{j}\left(u\left(s\right)\right)\ $is well approximated
by $\frac{1}{N}\sqrt{\frac{1+e^{-\lambda_{u}/\lambda_{0}}}{1-e^{-\lambda_{u}/\lambda_{0}}}}$
and $\widetilde{j}_{0}\left(u\left(s\right)\right)\ $is well approximated
by $1-\sqrt{1-e^{-2\lambda_{u}/\lambda_{0}}}$. Below, for each solution
we check the condition (\ref{Cond}) to be self consistent.

\textit{Regime I}: Here we consider $-s\gg\lambda_{s}$ so that to
leading order $u\left(s\right)=s+\lambda_{r}+\lambda_{u}$ and
\begin{eqnarray*}
\widetilde{j}_{p_{1}}\left(u\left(s\right)\right) & = & \frac{p_{1}{\tilde{j}}(u\left(s\right))}{1-(1-p_{1}){\tilde{j}}_{0}(s)}\simeq\\
 & \simeq & \frac{1}{N}\sqrt{\frac{1+e^{-\lambda_{u}/\lambda_{0}}}{1-e^{-\lambda_{u}/\lambda_{0}}}}\frac{1}{1+\frac{1-p_{1}}{p_{1}}\sqrt{1-e^{-2\lambda_{u}/\lambda_{0}}}}=
 \frac{1}{N}\frac{\sqrt{\coth\left(\frac{\lambda_{u}}{2\lambda_{0}}\right)}}{1+\frac{1-p_{1}}{p_{1}}\sqrt{1-e^{-2\lambda_{u}/\lambda_{0}}}}\equiv\frac{\kappa}{N}
 \end{eqnarray*}
in this regime (this is verified self-consistently below). Eq.
(\ref{PolesEq1}) then reduces to \begin{equation}
(s+\lambda_{b})\left(s+\lambda_{r}+\lambda_{u}\right)=\lambda_{b}\lambda_{u}\left(1-\frac{\kappa}{N}\right).\end{equation}
 This equation has one pole of the order of $\lambda_{u},\lambda_{b}$,
which corresponds to trajectories finding the target within the first
sliding event. This pole can be discarded in the large $N$ limit
since its residue scales as $1/N$. Its second pole reads, to leading
order in $\lambda_{r}/\lambda_{u}$ and $1/N$: \begin{equation}
\tau_{1}^{-1}\equiv-s_{1}\simeq\frac{\lambda_{b}\left(\lambda_{r}+\frac{\kappa\lambda_{u}}{N}\right)}{\lambda_{b}+\lambda_{u}}.\label{pol1}\end{equation}
 To ensure that $-s\gg\lambda_{s}$, as was assumed one should satisfy\begin{equation}
\lambda_{s}\ll\frac{\lambda_{b}\left(\lambda_{r}+\frac{\kappa\lambda_{u}}{N}\right)}{\lambda_{b}+\lambda_{u}}.\end{equation}
 The corresponding residue of $\widetilde{\mathcal{R}}\left(s\right)$
then reads: \begin{equation}
\mathrm{Res}_{1}\simeq-q=-\frac{1}{1+\frac{\lambda_{r}}{\lambda_{u}\kappa/N}}.\label{res1}
\end{equation}
This second pole corresponds to trajectories which find the target before
crossing of the barrier. In the limit $\lambda_{r}\ll\frac{\kappa\lambda_{u}}{N}$,
such events occur with a high probability $q\simeq1$ and are characterized
by a time scale $\tau_{1}\simeq\frac{N}{\kappa}\left(\frac{1}{\lambda_{b}}+\frac{1}{\lambda_{u}}\right)$.
In the limit $\lambda_{r}\gg\frac{\kappa\lambda_{u}}{N}$ this second
pole corresponds to processes which find the target before the typical
time which characterizes a fall into the trap $\lambda_{r}{}^{-1}$
and without scanning the whole length. Such events are unlikely as
shown by $q\ll1$. To check the self-consistency we calculate the
condition (\ref{Cond}) for $s=s_{1}$. \begin{equation}
\frac{\lambda_{b}\left(\lambda_{r}+\frac{\kappa\lambda_{u}}{N}\right)}{\lambda_{s}\left(\lambda_{b}+\lambda_{u}\right)}-1\gg\frac{\lambda_{r}}{\lambda_{0}}\end{equation}
 but\begin{equation}
\frac{\lambda_{b}\left(\lambda_{r}+\frac{\kappa\lambda_{u}}{N}\right)}{\lambda_{s}\left(\lambda_{b}+\lambda_{u}\right)}-1>\frac{\lambda_{r}}{\lambda_{s}}\end{equation}
 so one may see that condition (\ref{Cond}) for the first pole
holds easily.

\noindent \textit{Regime II}: Here we consider $-s\ll\lambda_{b}$.
To proceed we take \begin{equation}
\frac{p_{1}{\tilde{j}}(u\left(s\right))}{1-(1-p_{1}){\tilde{j}}_{0}(s)}\simeq\frac{1}{N}\sqrt{\frac{1+e^{-\lambda_{u}/\lambda_{0}}}{1-e^{-\lambda_{u}/\lambda_{0}}}}\frac{1}{1+\frac{1-p_{1}}{p_{1}}\sqrt{1-e^{-2\lambda_{u}/\lambda_{0}}}}=\frac{1}{N}\frac{\sqrt{\coth\left(\frac{\lambda_{u}}{2\lambda_{0}}\right)}}{1+\frac{1-p_{1}}{p_{1}}\sqrt{1-e^{-2\lambda_{u}/\lambda_{0}}}}\equiv\frac{\kappa}{N}\end{equation}
 as before (this is verified self-consistently below). The equation
becomes\begin{equation}
s+\lambda_{s}+\lambda_{r}+\lambda_{u}=\lambda_{u}\left(s+\lambda_{s}\right)\left(1-\frac{\kappa}{N}\right).\end{equation}
 The interesting pole is given by \begin{equation}
\tau_{2}^{-1}\equiv-s_{2}\simeq\frac{\lambda_{s}\frac{\kappa\lambda_{u}}{N}}{\lambda_{r}+\frac{\kappa\lambda_{u}}{N}},\label{pol2}\end{equation}
 and the corresponding residue of $\widetilde{\mathcal{R}}\left(s\right)$
reads \begin{equation}
\mathrm{Res}_{2}\simeq-\left(1-q\right)=-\left(1-\frac{1}{1+\frac{\lambda_{r}}{\lambda_{u}\kappa/N}}\right).\label{res2}\end{equation}
 Similar to the case discussed above when $\lambda_{r}\gg\lambda_{u}\kappa/N$
the search involves a high chance of many entrances and exists from
the $s$ state, as shown by $1-q\simeq1$. In the opposite limit trajectories
entering the $s$ state are very unlikely ($q\simeq1$) and the search
time is dominated by the trapping time $1/\lambda_{s}$. To check
the self-consistency we calculate the condition (\ref{Cond}) for
$s=s_{2}$:
\begin{equation}
1-\frac{\frac{\kappa\lambda_{u}}{N}}{\lambda_{r}+\frac{\kappa\lambda_{u}}{N}}=1-\frac{1}{\frac{N\lambda_{r}}{\kappa\lambda_{u}}+1}\gg\frac{\lambda_{r}}{\lambda_{0}}\end{equation}
 but\begin{equation}
\frac{\kappa\lambda_{u}}{N}\ll\lambda_{0}\;.
\end{equation}
This condition is also easily met. Applying the poles (\ref{pol0}),(\ref{pol1}), and (\ref{pol2}) with the corresponding residues (\ref{res0}),(\ref{res1}), and (\ref{res2}) to
Eq. (\ref{RvsPoles}) one obtains Eq. (\ref{2exp}):
\begin{equation}
\mathcal{R}(t)\simeq1-qe^{-t/\tau_{1}}-(1-q)e^{-t/\tau_{2}},\end{equation}
 where
 \begin{align}
\kappa & =\frac{\sqrt{\coth\left(\frac{\lambda_{u}}{2\lambda_{0}}\right)}}{1+\frac{1-p_{1}}{p_{1}}\sqrt{1-e^{-2\lambda_{u}/\lambda_{0}}}}\nonumber \\
q & =\frac{1}{1+\frac{\lambda_{r}}{\lambda_{u}\kappa/N}}\nonumber \\
\tau_{1} & =\frac{\lambda_{b}+\lambda_{u}}{\lambda_{b}\left(\lambda_{r}+\frac{\kappa\lambda_{u}}{N}\right)}\nonumber \\
\tau_{2} & =\frac{\lambda_{r}+\frac{\kappa\lambda_{u}}{N}}{\lambda_{s}\frac{\kappa\lambda_{u}}{N}}.
\end{align}

 \section{Conditions for a perfect search\label{Conditions for a perfect search}}

In this appendix we show that using the search and recognition strategy
based on a disorder in a barrier height one may in principle achieve
a "perfect" search without any design of the target and using only one
searcher by optimizing the values of $E_{0}$ and $\sigma$. We define
a "perfect" search as one where one searcher goes to the $r$ state
of the target site with probability one ($q=1$) within a typical
search time\footnote{This is the time to reach a designed target on a flat energy landscape.} $t^{typ}\simeq\frac{N}{\sqrt{\lambda_{0}/\lambda_{u}}}\left(\frac{1}{\lambda_{u}}+\frac{1}{\lambda_{b}}\right)$%
. The last demand may be satisfied by ensuring that the barrier height
on the target site, $E_{b}^{\mathcal{T}}$, is not positive with probability
one. In this case the number of scans of the DNA that the searcher
performs before a transition to the $r$ state is of order one. Below
we show that this perfectly fast search can be achieved with a perfect
recognition of the target, $q=1$. Namely we show that the occupation
probability of the target in the infinite time limit may arbitrarily
close to one. To this end we calculate below the occupation probability
of each site on the DNA, $P_{i}$, and in particular, the occupation
probability of the target site, $P^{\mathcal{T}}$. We take here $\lambda_{s}^{i}=0$
($E_{r}=-\infty$) to ensure the stability of the protein-DNA complex
after the target is located. We assume now (and check this assumptions self consistently) that $q=1$. This implies that the protein scans
the whole DNA in the $s$ state (and, therefore, equilibrates in the
$s$ states, but not in the $r$ state, on all DNA\ sites) before
it passes over the barrier on the target. Thus, the transition rate
to the $r$ state of site $i$ is given by $\frac{\lambda_{b}}{\lambda_{b}+\lambda_{u}}\lambda_{r}^{i}$.
The time evolution of the occupation probability of the $r$ state
at site $i$ is then given by\begin{equation}
\frac{dP_{i}}{dt}=\frac{\lambda_{b}}{\lambda_{b}+\lambda_{u}}\lambda_{r}^{i}\left(1-\overset{N}{\underset{j=1}{\sum}}P_{j}\right).\end{equation}
 At steady-state the occupation probability of the $r$ state at site
$i$ is\begin{equation}
P_{i}=\frac{\lambda_{r}^{i}/\lambda_{0}}{\overset{N}{\underset{j=1}{\sum}}\lambda_{r}^{i}/\lambda_{0}}.\label{Pi}\end{equation}
 The transition rate $\lambda_{r}^{i}$ is given by Eq. (\ref{lamr}).
As stated above, the barrier height, $E_{b}^{i}$, is drawn from a
Gaussian distribution: \begin{equation}
\Pr(E_{b}^{i})=\frac{e^{-\frac{\left(E_{b}^{i}-E_{0}\right)^{2}}{2\sigma^{2}}}}{\sqrt{2\pi\sigma^{2}}}.\end{equation}
 The occupation probability of the site with the lowest barrier (a
non-designed target) for a given realization of disorder is then given
by\begin{equation}
P^{\mathcal{T}}=\frac{e^{-E_{b}^{\mathcal{T}}}}{\overset{N}{\underset{i=1}{\sum}}e^{-E_{b}^{i}}}=\frac{1}{\overset{N}{\underset{i=1}{\sum}}e^{-E_{b}^{i}+E_{b}^{\mathcal{T}}}}=\frac{1}{1+\underset{i\neq\mathcal{T}}{\sum}e^{-E_{b}^{i}+E_{b}^{\mathcal{T}}}}\label{PT}\end{equation}
 where $E_{b}^{\mathcal{T}}=\min\left\{ E_{b}^{i}\right\} $ is the
barrier height on the target site and the sum over $i\neq\mathcal{T}$
does not include the target site. Note that in the \textquotedbl{}thermodynamic\textquotedbl{}
limit $N\rightarrow\infty$ and $\sigma\rightarrow\infty$ holding\begin{equation}
\frac{\sigma}{\sqrt{2}\operatorname{erfc}^{-1}\left(\frac{2}{N}\right)}=const\equiv J,\label{Jdef}\end{equation}
 where $\operatorname{erfc}^{-1}$ is the inverse complementary error
function \cite{AS}, this model is similar to the Random Energy Model
and may be solved using the same approach \cite{D1981}.

The barrier height on the target site may be well estimated using\begin{equation}
\int_{-\infty}^{E_{b}^{\mathcal{T}}}\frac{e^{-\frac{\left(E-E_{0}\right)^{2}}{2\sigma^{2}}}}{\sqrt{2\pi\sigma^{2}}}dE=\frac{1}{N}.\end{equation}
 \ This gives\begin{equation}
E_{b}^{\mathcal{T}}\simeq E_{0}-\sigma\sqrt{2}\operatorname{erfc}^{-1}\left(\frac{2}{N}\right),\end{equation}
 As was mentioned above, we assume that for almost each realization
of the disorder $E_{b}^{i}>0$ for every $i$. To make the search
as fast as possible one should decrease the barrier of the target
site. These two restrictions lead to the choice\begin{equation}
E_{0}=\sigma\sqrt{2}\operatorname{erfc}^{-1}\left(\frac{2}{N}\right).\label{E0}\end{equation}
 In this case $E_{b}^{\mathcal{T}}=0$ so that the probability distribution
of the non-target sites may be well approximated in the large $N$
limit by
\begin{equation}
\Pr(E_{b}^{i\neq\mathcal{T}})=\left\{ \begin{array}{cc}
{\cal N}^{-1} e^{-\frac{\left(E_{b}^{i\neq\mathcal{T}}-E_{0}\right)^{2}}{2\sigma^{2}}} & E_{b}^{i\neq\mathcal{T}}>0\\
0 & E_{b}^{i\neq\mathcal{T}}<0\end{array}\right.,
\end{equation}
where ${\cal N}$ is a normalization constant which can be easily obtained.
 Since for almost all realizations of the disorder the minimal barrier
height is close to zero, Eq. (\ref{lamr}) simplifies to
\begin{equation}
\frac{\lambda_{r}^{i}}{\lambda_{0}}=e^{-E_{b}^{i}}.\label{lamrS}
\end{equation}
 At steady-state the average occupation probability of the site with
lowest barrier, $q$, is given by $\left\langle P^{\mathcal{T}}\right\rangle $.
Using Eqs. (\ref{PT}), (\ref{lamrS}) with Jensen's inequality,
\begin{equation}
\left\langle P^{\mathcal{T}}\right\rangle \geq\frac{1}{\left\langle \frac{1}{P^{\mathcal{T}}}\right\rangle },
\end{equation}
 one gets \begin{equation}
q=\left\langle P^{\mathcal{T}}\right\rangle \gtrsim\frac{1}{\left\langle \frac{1}{P^{\mathcal{T}}}\right\rangle }\simeq\frac{1}{1+N\frac{\int_{0}^{\infty}e^{-\frac{\left(E-E_{0}\right)^{2}}{2\sigma^{2}}}e^{-E}dE}{\int_{0}^{\infty}e^{-\frac{\left(E-E_{0}\right)^{2}}{2\sigma^{2}}}dE}}=\left[1+N\frac{e^{-E_{0}+\frac{\sigma^{2}}{2}}\operatorname{erfc}\left(\frac{-E_{0}+\sigma^{2}}{\sqrt{2}\sigma}\right)}{1+\operatorname{erf}\left(\frac{E_{0}}{\sqrt{2}\sigma}\right)}\right]^{-1}.\end{equation}
 Using Eqs. (\ref{Jdef}), (\ref{E0}) and taking the leading order
in $\frac{1}{N}$ we obtain
\begin{align}
q & \gtrsim\left[1+\sqrt{2\pi\ln\left(\frac{N^{2}}{2\pi}\right)}\frac{\exp\left[\left(J-1\right)\left[\operatorname{erfc}^{-1}\left(\frac{2}{N}\right)\right]^{2}\right]
\operatorname{erfc}\left[\left(J-1\right)\operatorname{erfc}^{-1}\left(\frac{2}{N}\right)\right]}{2}\right]^{-1}\simeq\nonumber\\
 & \simeq\left\{
\begin{array}{cc}
\left[1+\frac{2}{J-1}\left(1-\frac{\ln\ln\frac{N^{2}}{2\pi}}{\ln\frac{N^{2}}{2\pi}}\right)^{-\frac{1}{2}}\right]^{-1} & J>1\\
\left[1+2\sqrt{2\pi\ln\frac{N^{2}}{2\pi}}e^{\left(J-1\right)^{2}\left[\operatorname{erfc}^{-1}\left(\frac{2}{N}\right)\right]^{2}}\right]^{-1} & J\leq1.
\end{array}\right.
\end{align}
 In the limit $N\rightarrow\infty$ one gets a behavior similar to
the usual second order phase transition of the Random Energy Model:\begin{equation}
q=\left\{ \begin{array}{cc}
\frac{J-1}{J+1} & J\geq1\\
0 & J<1\end{array}\right.\end{equation}
 Therefore, for large enough $J$ the searcher finds its target with a
probability close to one so that all our assumptions in this Section
based on $q=1$ are self consistent. Also, since the typical barrier
between the $s$ and $r$ states on the target is zero, the searcher
finds its target within the facilitated diffusion limit. Note that,
although the search and the recognition are perfect, the \textit{average}
time is infinite since there is a finite, $1-q$, probability to be
trapped on a non-target site.

Summarizing, to ensure a perfect search one should set\footnote{This choice of finite values of $N$, $\sigma$ and $E_{0}$ provides
a good approximation to the optimal $E_{0}$ for a given $\sigma$
(and vice versa). For example, for the case shown on Fig. \ref{fig5ChBarrier}(a)
Eq. (\ref{E0}) predicts that the optimal value of $\sigma$ is $5.34$
which is close to the numerically obtained value of $5.25$ (see Fig.
\ref{fig5ChBarrier}(a)).%
}\begin{equation}
\sigma=J\sqrt{2}\operatorname{erfc}^{-1}\left(\frac{2}{N}\right)\end{equation}
 with some constant $J\gg1$ and
 \begin{equation}
E_{0}=\sigma\sqrt{2}\operatorname{erfc}^{-1}\left(\frac{2}{N}\right).
\end{equation}
In this case the probability to find the target is
\begin{equation}
q=\frac{J-1}{J+1}\simeq1-\frac{2}{J}
\end{equation}
 and the typical search time is comparable to the facilitated diffusion
limit. Therefore, $J$ should be as large as possible. For the case
of $n_{p}$ searchers (assuming that they are independent) the condition
on $J$ becomes\begin{equation}
p_{cat}=\left(1-q\right)^{n_{p}}=\left(\frac{2}{J+1}\right)^{n_{p}}\ll1.\end{equation}

In fact, a perfect search may be impossible from practical reasons.
For example, for $N=10^{6}$ our results suggest that to ensure $q=0.5$
one should take $E_{0}\simeq67.7$ and $\sigma\simeq14.3$. Such large energies may be difficult to achieve.
Nevertheless, as we showed in Section \ref{Disordered case} the proposed
mechanism of a barrier discrimination is very efficient even when
parameters are far from perfect search conditions and the mean and
the variance of the barrier energy are comparable to the mean and
the variance of the experimentally found binding energy distribution.

 \bibliographystyle{unsrt} \bibliographystyle{unsrt} \bibliographystyle{unsrt}
\bibliographystyle{unsrt} \bibliographystyle{unsrt} \bibliographystyle{unsrt}
\bibliography{BibNoTitle}

\end{document}